\documentclass[11pt]{article}
\usepackage{cite}
\usepackage{amssymb}
\usepackage{amsmath}
\usepackage[all]{xy}
\input{epsf}
\usepackage{epsfig}
\usepackage{hyperref}
\numberwithin{equation}{section}

\usepackage{color}

\newcommand{\bea}{\begin{eqnarray}}
\newcommand{\eea}{\end{eqnarray}}
\def\be{\begin{equation}}
\def\ee{\end{equation}}
\def\ba{\begin{align}}
\def\ea{\end{align}}
\def\beq{\begin{eqnarray}}
\def\eeq{\end{eqnarray}}

\def\p{\partial}

\def\im{{\rm Im}}

\newcommand{\half}{\frac{1}{2}}

\def\A{{\cal A}}
\def\B{{\cal B}}

\begin{document}

\title{\Large{\bf Geometry of open strings ending on backreacting D3-branes}}
\author{Raphael Benichou $^\flat$ and John Estes $^\natural$}

\maketitle

\begin{center}
$^\flat$ Theoretische Natuurkunde, Vrije Universiteit Brussel and \\
The International Solvay Institutes,\\
Pleinlaan 2, B-1050 Brussels, Belgium \\
\textsl{raphael.benichou@vub.ac.be}
\end{center}

\vskip 3mm

\begin{center}
$^\natural$ Instituut voor Theoretische Fysica, Katholieke Universiteit Leuven, \\
Celestijnenlaan 200D B-3001 Leuven, Belgium \\
\textsl{johnalondestes@gmail.com}
\end{center}

\begin{abstract}
We investigate open string theory on backreacting D3-branes using a spacetime approach.  We study in detail the half-BPS supergravity solutions describing open strings ending on D3-branes, in the near horizon of the D3-branes.  We recover quantitatively several non-trivial features of open string physics including the appearance of D3-brane spikes, the polarization of fundamental strings into D5-branes, and the Hanany-Witten effect.  Finally we detail the computation of the gravitational potential between two open strings, and contrast it with the holographic computation of Wilson lines.  We argue that the D-brane backreaction has a large influence on the low-energy gravity, which may lead to experimental tests for string theory brane-world scenarios.
\end{abstract}

\vfill\eject

\tableofcontents
\vfill\eject

\section{Introduction}

D-branes have played an important role in the recent developments of string theory.
It is well known that the elementary degrees of freedom of D-branes are open strings with endpoints attached to the D-branes \cite{hep-th/9510017,hep-th/9611050}.
The open string theory can be defined in terms of a two-dimensional conformal field theory living on worldsheets with boundaries.
It is equally well-known that D-branes are sources for the (super)gravity fields. In particular D-branes backreact on the surrounding geometry, and the resulting supergravity backgrounds are known at least in some particular cases (see e.g. \cite{UCSB-TH-90-66,Stelle:1996tz} and references therein).  In many cases, the near-horizon geometry of the D-branes is strongly deformed even when the number of branes is taken to be small.  Usually people work strictly in either the probe brane limit or in the supergravity limit where the open and closed string theories are decoupled.  In this paper we are interested in the open string physics on backreacting D-branes.

The open string description of D-branes has been extensively studied in the limit where the D-branes are treated as probes.  Strictly speaking this requires the string coupling $g_s$ to be equal to zero.
At non-zero string coupling, one can also consider D-branes on top of orientifold-planes so that the net backreaction is null (see \cite{CALT-68-1194,PUPT-1045,UTTG-09-87,hep-th/0204089}). Such constructions are often referred to as type I string theory.  In the cases where the D-brane backreaction is non-trivial, one can still define an open/closed string theory.  However, the theory has closed string tadpoles which induce an evolution into the background fields.
On short scales, the effects of the tadpoles may be ignored, but to understand the long distance physics, one has to take them fully into account.  In this paper we will only consider the more general case where the backreaction is non-vanishing.  

The supergravity backgrounds associated to backreacting D-branes have also been extensively studied.  However when considering such backgrounds, the interactions between the D-branes and the bulk degrees of freedom are usually not considered.
This is particularly true in the context of the AdS/CFT correspondence where a limit is taken so that the open and closed string theories decouple \cite{hep-th/9711200}.

In this paper we investigate the open string theory living on backreacting D-branes.
This is an important part of string theory about which surprisingly little is understood at the present time.
Given the importance of D-branes in current research in string theory, progress in this direction is clearly important.  One specific motivation for us is to understand better the gravitational sector of string theory brane-world scenarios.  In most type II phenomenological models, our visible universe is localized on a configuration of intersecting D-branes embedded in a higher-dimensional spacetime. The open string theory living on the D-branes realizes (a supersymmetric extension of) the Standard Model of particles physics, while gravity is mediated by closed strings living in the bulk.  Usually the D-branes are treated as probes in a four-dimensional compactification and the gravitational theory reduces at large distances to four-dimensional Einstein gravity.  However it is legitimate to wonder whether the gravitational backreaction of the D-branes alters this picture.  Indeed, we will argue that the backreaction of the D-branes has visible consequence on the effective gravitational physics even when the open strings are taken very far apart.  These corrections are further discussed in a companion letter \cite{BE}.

In this paper we give the detailed computation of the gravitational potential between open strings ending on a stack of $N$ backreacting D3-branes.  For a large separation between the strings, the resulting gravitation potential energy behaves as
\be E(r) \sim \frac{1}{r} \frac{g_s}{N}  + \mathrm{subleading}, \ee
where $r$ is the separation distance between the two strings and $g_s$ is the string coupling.  To be precise, the above is valid when the backreacting D3-branes are placed in a spacetime with five or more non-compact dimensions.  In a four-dimensional compactification, there is an additional contribution coming from the zero-mode of the graviton which gives an additional $1/r$ contribution reproducing the usual Newton's law \cite{hep-th/9906064,hep-th/9906182,Benichou:2010sj,BE}.
Since gravity propagates in more than four dimensions,
one would naively expect that the gravitational potential falls of faster than $1/r$.  However, there are important stringy effects which enhance the potential.  Finally, we note that in the usual AdS/CFT limit where $N \rightarrow \infty$ with $g_s N$ fixed, the potential we compute vanishes.  This is just the usual statement that the closed and open string theory decouple in this limit.

Other long-term motivations to study open string theory on backreacting D-branes include for instance the microscopic counting of black-hole entropy \cite{hep-th/9601029} at non-zero string coupling.  Another interesting question is whether and how the holographic principle \cite{'tHooft:1993gx, Susskind:1994vu} is realized in string theory beyond the decoupling limit of AdS/CFT. This question can presumably be studied by starting form an AdS/CFT set-up and turning on a small coupling between the open string theory and the bulk closed string theory.

\paragraph{Strategy and content of the paper.}
In this paper we take a spacetime approach to describe the open string theory living on D3-branes.
We limit ourselves to the classical theory, and we describe the open strings as extended objects that stretch up to the horizon created by the D-brane backreaction.  More precisely we will study in detail a family of exact half-BPS supergravity solutions describing stacks of open strings ending on D3-branes, in the near-horizon limit. These solutions were found in \cite{D'Hoker:2007fq}.  The configurations of open strings we study have been discussed before in the context of the holographic description of Wilson lines.\footnote{It is important to note that a fundamental string and the gravitational dual of a Wilson line are not exactly the same object.  In particular, they satisfy different boundary conditions \cite{hep-th/9904191} and their actions differ by boundary terms. For the static configurations we consider in this paper, their bulk geometry is identical.}

In section \ref{sec:probe} we discuss the probe brane description of fundamental strings ending on D3-branes.
In section \ref{sec:sugraSol} we describe the supergravity solutions of \cite{D'Hoker:2007fq}. We perform a careful computations of the charges in these background, and obtain a one-to-one matching between these solutions and the dual Wilson lines. We also find a bulk description of the Hanany-Witten effect.
In section \ref{sec:probeLim} we study the various small-charge limits of the supergravity solutions.
This allows us to describe quantitatively from the bulk two features of open string theory: the appearance of D3-bane spikes and the polarization of fundamental strings into D5-branes in the presence of fluxes.
In section \ref{sec:potential} we compute the gravitational potential between open strings ending on backreacting D3-branes.  The differences between the gravitational potential and the holographic computation of the gauge theory potential are discussed.
Section \ref{sec:discussion} contains our final remarks.
To keep the paper readable we gathered many technical details in the appendices.

\paragraph{Other approaches to the open string theory on backreacting D-branes.}
Before we begin let us briefly discuss some other possible strategies to describe the open string theory on backreacting D-branes.
\begin{itemize}
\item On probe D-branes, the open string theory is defined in terms of a two-dimensional conformal field theory living on worldsheets with boundaries.
The corrections due to the D-brane backreaction can be accounted for with the insertion of additional boundaries on the worldsheet, with proper boundary conditions.
This perturbative treatment of the D-brane backreaction is legitimate when we work far away from the D-branes, and the backreaction can be seen as a small perturbation of the initial geometry.
However in the neighborhood close to the D-branes, the backreaction is never small.
The open strings attached to the backreacting D-branes certainly evolve close to the D-brane.
It is not clear that the perturbative treatment of the D-brane backreaction is useful to study this open string theory.  A better way to deal with the closed string tadpoles is the Fischler-Susskind mechanism \cite{Fischler:1986ci, Fischler:1986tb, Keller:2007nd}.  The tadpoles are canceled by counter-term insertions on worldsheets of lesser genus. These counter-terms trigger a renormalisation group flow, whose IR fixed point is the worldsheet CFT that should be used to define string theory in the presence of backreacting D-branes. Following this RG flow is a difficult problem.
\item The backreaction of D-branes leads to supergravity backgrounds that are known at least in some cases.
String theory in these backgrounds can be defined using a two-dimensional non-linear sigma-model on the background.
Then one can try to describe the open string sector using for instance the boundary state formalism (see e.g. \cite{hep-th/0209241} for an introduction).
A first technical difficulty is that the backreacting D-branes source RR-fluxes, and string theory is still ill-understood in this context.
A second conceptual difficulty is that a generic boundary state would describe a probe D-brane in the background of the backreacting D-branes. One can try to tune the boundary state parameters so that the probe D-brane lies on top of the backreacting D-branes: in this case the boundary state may capture correctly the open string theory on the backreacting D-branes.
We will leave the study of this program for future work.
\item A popular approach is to use an effective field theory to describe the open string theory at low-energy.
This field theory is defined by the DBI action, whose non-abelian version is still not completely understood.
The main issue with this approach is that it is difficult to couple the field theory living on the worldvolume of the D-branes to the bulk degrees of freedom, when the D-brane backreaction is taken into account.
In the simplest case of extremal D-branes, the worldvolume coincides with the horizon created by the D-brane backreaction.  Because of the infinite redshift at the horizon, the couplings between the brane and bulk degrees of freedom vanish.  Even if this obstacle were overcome, the stringy gravitational effects we discuss in this article and in \cite{BE} would probably be missed in this approach.
Let us mention that recent progress has been made in understanding the coupling between a brane-worldvolume field theory and the bulk degrees of freedom (see e.g. \cite{arXiv:0705.3212,arXiv:0812.3820,arXiv:0912.3039}).
Among other interesting results, let us mention that the D-brane backreaction seems to facilitate the construction of de-Sitter vacua \cite{arXiv:1109.0532}.
\item Finally a holography-inspired approach would be to describe the D-brane worldvolume theory using the closed string theory living in the near-horizon of the D-branes.
However the AdS/CFT correspondence states that the closed string theory is exactly equivalent to a gauge theory, which in turn differs from the open string theory by string-length ($\alpha'$) corrections.  The gravitational potential energy between open strings that we want to compute is exactly given by these types of corrections and is not computable using holography.  The holographic closed string theory also misses all the massive excitations of the open string theory.
\end{itemize}

%%%%%%%%%%%%%%%%%%%%%%%%%%%%%%%%%%%%%%%%%%

\section{Fundamental strings, probe branes and Wilson lines}\label{sec:probe}

In this paper we study in detail the physics of fundamental strings ending on a stack of D3-branes, in the near-horizon limit.
This configuration is relevant for the holographic description of half-BPS Wilson lines in $\mathcal{N}=4$ SYM.
In the literature it has mostly been discussed in this context (see for instance \cite{Rey:1998ik,Maldacena:1998im,Drukker:2005kx,Yamaguchi:2006tq,Gomis:2006im,Gomis:2006sb}).
In this section we summarize what is known about the probe-string (and, as it turns out, probe branes) description of these holographic Wilson lines.

As was argued in \cite{hep-th/9904191}, there is a subtle difference between proper fundamental strings and supersymmetric Wilson lines.  More precisely the supersymmetric Wilson line is dual to a fundamental string which satisfies the reverse boundary conditions of the usual open strings ending on D3-branes, namely Dirichlet boundary conditions along the brane and Neumann boundary conditions for directions perpendicular to the brane.  This can be seen by starting with a Wilson line in 10-dimensions, which satisfies Dirichlet boundary conditions in all spatial directions.  Making use of T-duality then gives a string which satisfies Dirichlet boundary conditions along the D3-brane and Neumann boundary conditions in the six perpendicular directions.  This subtlety will not be crucial for us, since the static configurations we consider here are described by the same bulk geometries and in the following we will not distinguish between Wilson lines and fundamental strings.

We consider straight Wilson lines that break half of the Poincar\'e supersymmetries as well as half of the superconformal supersymmetries.
Such a Wilson line breaks the four-dimensional conformal symmetry $SU(2,2)$ down to $SU(1,1) \times SU(2)$ and the five-sphere $SO(6)$ isometry down to $SO(5)$.
Including the supersymmetry generators, the supergroup preserved by the Wilson line is $OSp(4^*|4)$.
In \cite{Rey:1998ik,Maldacena:1998im}, it was proposed that such a Wilson line in the fundamental representation is described by a fundamental string in the bulk which extends up to the boundary.  The intersection of the fundamental string with the $AdS$ boundary gives the path of the Wilson line.  For a half-BPS Wilson line, the fundamental string also extends all the way to the Poincar\'e horizon.  In fact, its worldvolume spans an $AdS_2$ geometry inside $AdS_5$, which preserves the bosonic symmetry $SU(1,1) \times SU(2) \times SO(5)$.

For higher-dimensional representations, one expects that we should consider multiple fundamental strings.  But this picture is not entirely satisfactory for several reasons.
First fundamental strings ending on D3-branes are best understood as spiky deformations of the D3-branes \cite{Callan:1997kz}. The presence of many fundamental strings may result in a large D3-brane spike.
Secondly coincident fundamental strings tend to polarize in the presence of certain fluxes \cite{Myers:1999ps,Emparan:1997rt,hep-th/0111156}.
In our case D5-branes with fundamental string charge dissolved in them can be generated in this way.
To clarify this point, we first discuss the holographic description of Wilson lines in totally anti-symmetric and totally symmetric representations. These are represented by probe D5- and D3-branes respectively.  We then discuss the probe brane description of Wilson lines in generic irreducible representations.

%%%%%

\subsection{Anti-symmetic representations: D5-brane description}\label{subprobed5}

\begin{figure}[t]
\centering
\includegraphics[width=1\linewidth]{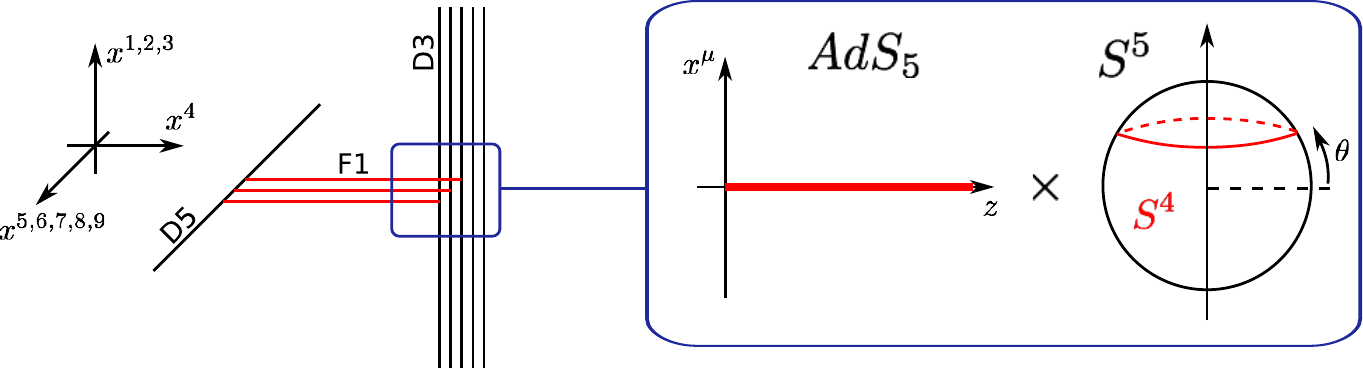}
\caption{Probe brane realization of a Wilson line in an anti-symmetric representation. In the near-horizon of the D3-brane stack, the fundamental strings polarize into a D5-brane that wraps a $S^4$ in the $S^5$. The angle $\theta$ is given by \eqref{thetaD5}.}
\label{probed5}
\end{figure}

First let us consider a Wilson line in an anti-symmetric representation. The Young tableau is a single column with $N_{F1}$ boxes.  Such a Wilson line is realized by stretching $N_{F1}$ fundamental strings between the stack of D3-branes and a single D5-brane (see Figure \ref{probed5}). Each fundamental string ends on a different D3-brane, in agreement with the $s$-rule \cite{Hanany:1996ie}.
Consequently the number of fundamental string $N_{F1}$ is bounded by the number $N$ of D3-branes in the stack.
Indeed Young tableau for $SU(N)$ cannot have columns with more than $N$ boxes.

From a worldsheet analysis \cite{Yamaguchi:2006tq}, one finds that the ground states of the individual fundamental strings behave as fermions (there is a unique ground state in the Ramond sector).  Correspondingly, when we consider multiple strings stretched between the D3-brane stack and the D5-brane, the Chan-Paton indices are anti-symmetrized.  Thus this setup is associated with an anti-symmetric representation.  In \cite{Gomis:2006sb}, it was shown that integrating out the low energy D5-brane degrees of freedom generates an insertion of a Wilson line operator into the path integral.

Let us now focus on the near-horizon of the D3-branes.
Because of the presence of the RR 5-form flux, the stack of fundamental strings polarizes into a D5-brane. This is a manifestation of the Myer's effect \cite{Myers:1999ps,Emparan:1997rt,hep-th/0111156}. The resulting D5-brane wraps a $S^4$ in the $S^5$ and an $AdS_2$ slice of Poincar\'e $AdS_5$ \cite{Pawelczyk:2000hy,Camino:2001at}.  Thus it preserves the same bosonic symmetry as the fundamental strings.
In the usual Poincare coordinates
\begin{align}
\label{PoincareAdSp}
ds^2_{AdS_5} = L^2 \frac{dz^2 - dt^2 + dr^2 + r^2 d\Omega_{(2)}^2}{z^2}
\end{align}
the D5-brane extends along time and the $z$ direction, and is located at $r=0$. Equivalently in the $AdS_2 \times S^2$ coordinate system \eqref{SU11SU2AdS} the D5-brane sits at $\eta = 0$.
The five-sphere metric can be written as
\begin{align}
ds^2_{S^5} = d \theta^2 + \cos^2 \theta\ ds_{S^4}^2.
\end{align} The D5-brane sits at a constant latitude $\theta$ that is determined by the number of fundamental strings $N_{F1}$ dissolved into the D5-brane.
In the probe limit, one can use the DBI description of the D5-brane to obtain a relation between the latitude $\theta$ and fundamental string charge \cite{Camino:2001at}\footnote{The coordinate $\bar \theta$ of \cite{Camino:2001at} is related to our $\theta$ as $\bar \theta = \frac{\pi}{2}-\theta$}:
\begin{align}\label{thetaD5}
\theta + \sin \theta \, \cos \theta = \frac{\pi}{2} - \pi \frac{N_{F1}}{N}.
\end{align}

%%%%%%

\subsection{Symmetric representations: D3-brane description}\label{subprobed3}

\begin{figure}[t]
\centering
\includegraphics[width=1\linewidth]{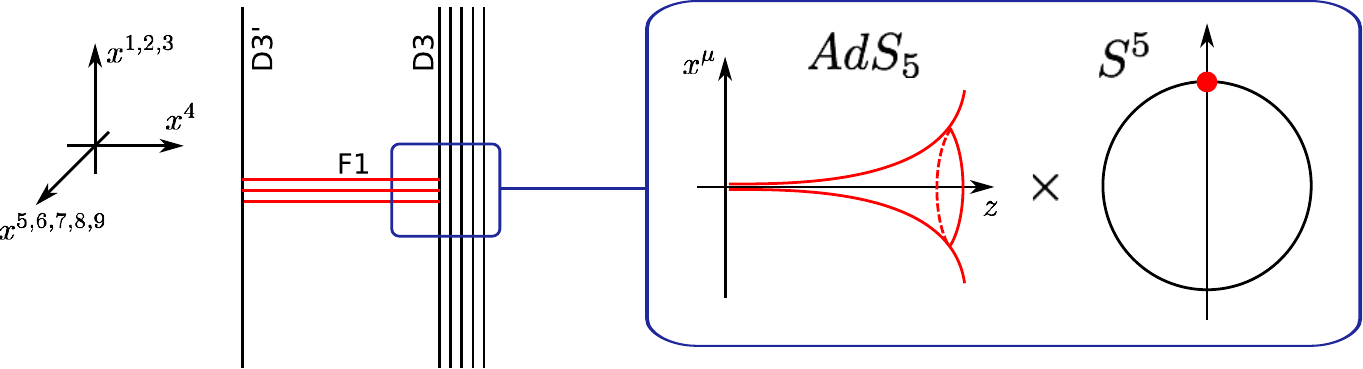}
\caption{Probe-brane realization of a Wilson line in a symmetric representation. In the near-horizon of the stack of D3's, the fundamental strings become a D3-brane spike that wraps an $AdS_2 \times S^2$ in $AdS_5$.}
\label{probed3}
\end{figure}

Now let us discuss the case of a Wilson line in a symmetric representation.
The corresponding Young tableau is a single horizontal line with $N_{F1}$ boxes.
This Wilson line is realized by putting one extra D3-brane parallel to the stack of D3-branes, and stretching $N_{F1}$ fundamental strings between the single D3-brane and the D3-brane stack \cite{Drukker:2005kx,Gomis:2006im} (see Figure \ref{probed3}).  In the ground state the fundamental strings behave as bosons and are thus symmetrized with respect to each other.  Correspondingly, the Chan-Paton indices are also symmetrized and the ground state gives rise to a symmetric representation.  Note that the number of fundamental strings $N_{F1}$ is not bounded in this case.
Indeed in $SU(N)$ Young tableau the lines can be arbitrarily long.
In \cite{Gomis:2006im}, it was shown that integrating out the low energy D3-brane degrees of freedom generates an insertion of a Wilson line operator into the path integral.

The fundamental strings pull on the D3-brane stack creating D3-brane spikes \cite{Callan:1997kz}.  In the near-horizon geometry of the D3-brane stack, this results in fundamental strings dissolved into a D3-brane which wraps an $AdS_2 \times S^2$ slice of $AdS_5$.  The location of the slice is determined by the number of fundamental strings $N_{F1}$ dissolved into the D3-brane, which in turn is related to the curvature radius of the induced $AdS_2$ and $S^2$ metrics.
More precisely in the $AdS_2 \times S^2$ parametrization of $AdS_5$ \eqref{SU11SU2AdS}  the probe D3-brane is located at a constant $\eta$ given by \cite{Drukker:2005kx}:
\be\label{DBIProbeD3}
\sinh \eta = N_{F1} \frac{\sqrt{\pi}}{2} \sqrt{\frac{g_s}{N}}
\ee
where $g_s$ is the closed string coupling and $N$ is the number of D3-branes in the stack.
In the usual Poincare coordinates \eqref{PoincareAdSp} this reads $r/z= N_{F1} \sqrt{\pi}/2 \sqrt{g_s/N}$.

%%%%%

\subsection{Generic irreducible representations}
\begin{figure}[t]
\centering
\includegraphics[width=1\linewidth]{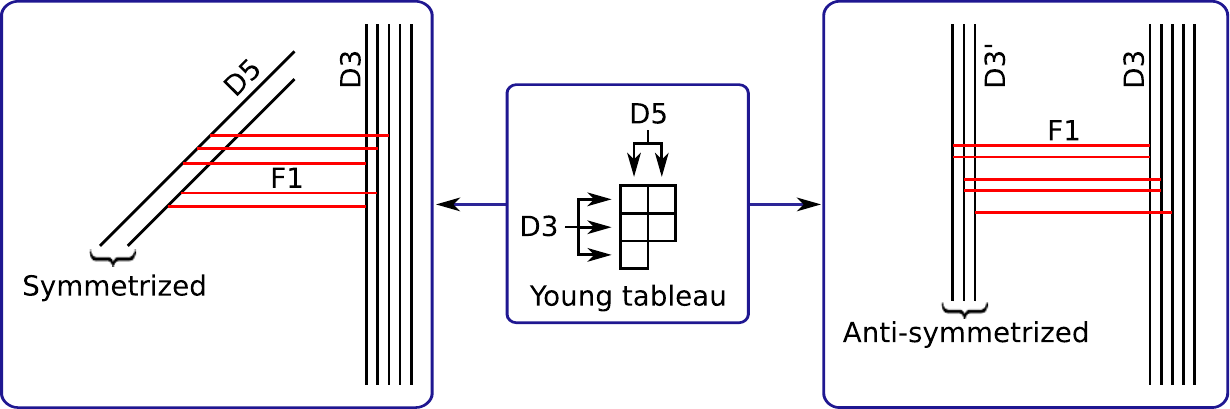}
\caption{Two different probe brane realizations of a Wilson line in a generic representation. The lines and the columns of the Young tableau can be associated respectively to D3- and D5-branes. Each box of the Young tableau is associated to a fundamental string.}
\label{probed3d5}
\end{figure}

A Wilson line associated to a generic irreducible representation of the gauge group $SU(N)$ is conveniently labeled by a Young tableau\footnote{We discuss Wilson lines only in irreducible representations as reducible representations of $SU(N)$ can always be decomposed on a basis of irreducible representations.}.  There are two different probe brane descriptions of such generic Wilson lines in terms of fundamental strings attached either to D3- or D5-branes \cite{Gomis:2006sb} (see Figure \ref{probed3d5}).

First, let us discuss the D5-brane realization.
We associate to each column of the Young tableau one probe D5-brane. For each box in this column, we stretch one fundamental string between the probe D5-brane and the stack of D3-branes.  Finally we symmetrize all the D5-branes.

Second, let us discuss the D3-brane realization.
We associate to each line of the Young tableau one probe D3-brane (denoted by D3' in Figure \ref{probed3d5}). For each box in this line, we stretch one fundamental string between the probe D3-brane and one of the D3-branes in the stack.
For each probe D3-brane we have to pick a different D3-brane in the stack.  Consequently there cannot be more than $N$ lines in the Young tableau, as expected from representation theory.
Finally we anti-symmetrize all the probe D3-branes.

In the near-horizon of the stack of D3-branes, the fundamental string polarizes in a complicated way that we will discuss in more details in the next section. However in some limits we expect to recover the probe-D5 and probe-D3 picture presented respectively in sections \ref{subprobed5} and \ref{subprobed3}.
More precisely, when the Young tableau is almost vertical and the number of columns is of order one, we expect the fundamental strings to polarize into D5-branes wrapping a $S^4$ in the $S^5$. There should be one such D5 for each column in the Young tableau. The amount of fundamental string charge dissolved in each D5 is given by the number of boxes in the column.  In this case, the D3-brane spike effect is reduced.
On the other hand, if the Young tableau is almost horizontal and the number of lines is of order one, we expect the fundamental strings to become D3-brane spikes in the near-horizon of the stack. There should be one spike for each line in the Young tableau. The amount of fundamental string charge dissolved in each D3-brane spike is given by the number of boxes in the line.  In this case the polarization effect is reduced.

The situation here is very reminiscent of the story of half-BPS local operators in $N=4$ SYM. Spinning gravitons in $AdS_5 \times S^5$ have two probe brane descriptions. One in terms of a giant graviton \cite{McGreevy:2000cw}: the string state blows up into a D3-brane that wraps an $S^3$ in the $S^5$. Another one is in terms of a dual giant graviton \cite{Grisaru:2000zn}: the string state blows up into a D3-brane that wraps an $S^3$ in the $AdS_5$.
The classification of all half-BPS supergravity solutions associated to local operators \cite{Lin:2004nb} shows that these probes brane descriptions arise as some limits of more generic configurations.
For half-BPS Wilson lines, the story is similar.  The classification of half-BPS geometries dual to Wilson lines was performed in \cite{D'Hoker:2007fq}.  In section \ref{sec:sugraSol} we will describe the matching between Young tableau and these supergravity solutions.
In section \ref{sec:probeLim} we will show that the solutions of \cite{D'Hoker:2007fq} reduce to probe D3-branes (D5-branes) in the horizontal (vertical) Young tableau limit.

%%%%%

\subsection{The Hanany-Witten effect}\label{sub:probeHW}

When a D5-branes crosses a D3-brane, a fundamental string stretching between the two is created or annihilated. This process is known as the Hanany-Witten effect \cite{Hanany:1996ie}.
In this paragraph we discuss this effect in the context of the probe D5-brane realization of the half-BPS Wilson lines.

\begin{figure}[t]
\centering
\includegraphics[width=1\linewidth]{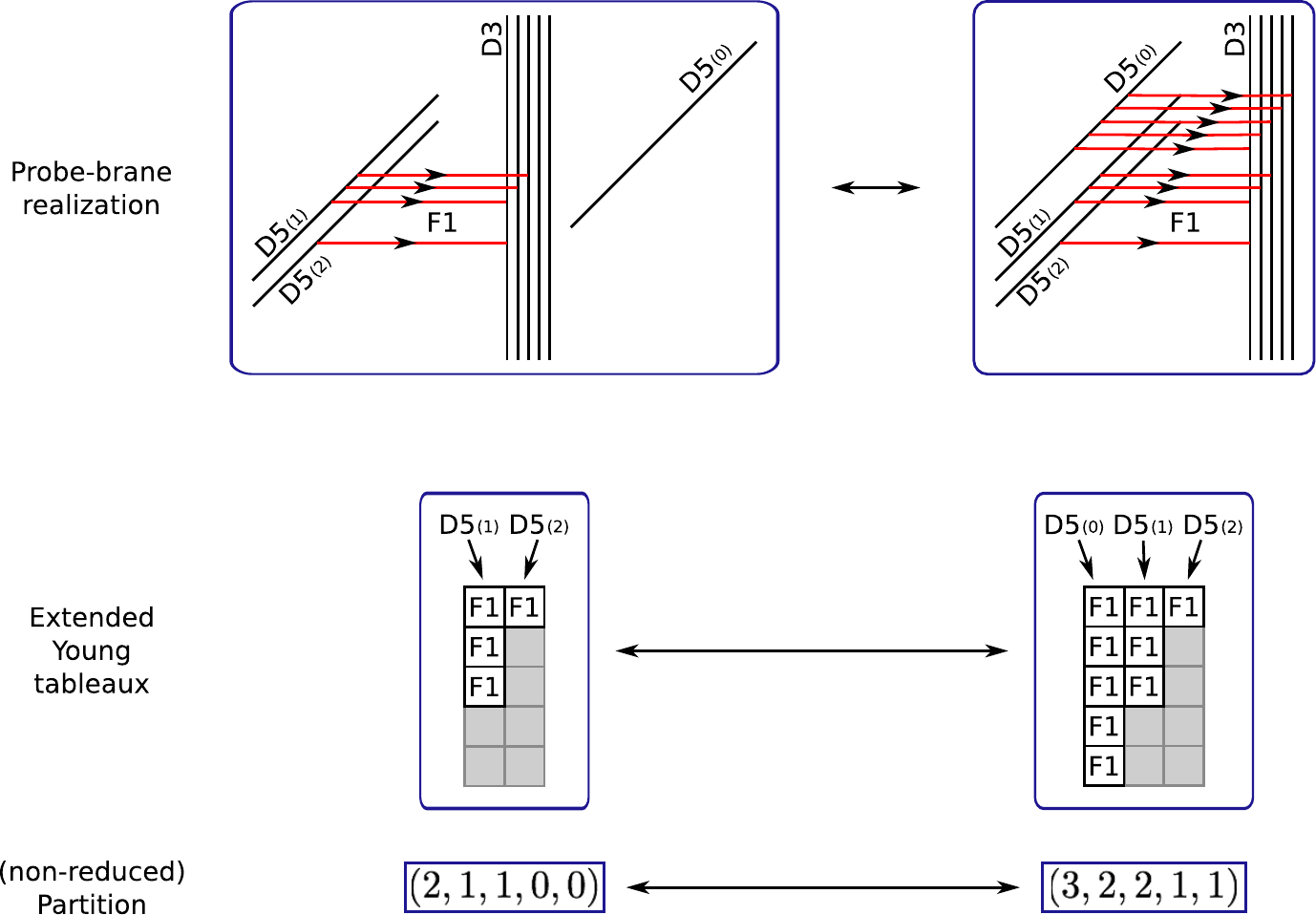}
\caption{The Hanany-Witten effect in the probe-D5 realization of the half-BPS Wilson lines. Taking a D5-brane from infinity across the stack of D3-branes induce the creation of $N$ fundamental strings.  This effect can be described in terms of Young tableau, as the addition of a column of $N$ boxes to the tableau. Equivalently in terms of partitions with $N$ entries, this amounts to adding one unit to each entry.}
\label{probeHWb}
\end{figure}

Let us consider a Wilson line in a generic irreducible representation realized as fundamental strings stretched between probe D5-branes and a stack of $N$ D3-branes (see Figure \ref{probeHWb}), as explained in the previous paragraph.
We consider one additional D5-brane that we bring from infinity across the stack of D3-branes. This results in the creation of $N$ fundamental strings stretching between the stack and this additional D5-brane.

We can conveniently keep track of this process at the level of the Young tableau.
Since we add a new D5-brane to the configuration, we add a column to the Young tableau.
Since the new D5-brane has $N$ fundamental strings ending on it, the new column we add to the Young tableau has $N$ boxes.
Adding columns of $N$ boxes to a Young tableau of $SU(N)$ does not modify the representation it describes\footnote{This would not be the case is we were to consider the full $U(N)$ gauge theory living on the D3-branes, instead of the $SU(N)$ gauge theory that is relevant in the context of AdS/CFT.}.
Equivalently, the Young tableau can be described by a partition of $N$ entries counting the number of boxes in each line.
Usual Young tableau have no columns of $N$ boxes,  and the $N$-th entry of the partition is zero.
To keep track of the Hanany-Witten effect we allow for Young tableau with columns of $N$ boxes, or equivalently we consider partitions with a non-zero $N$-th entry.

\begin{figure}[t]
\centering
\includegraphics[width=1\linewidth]{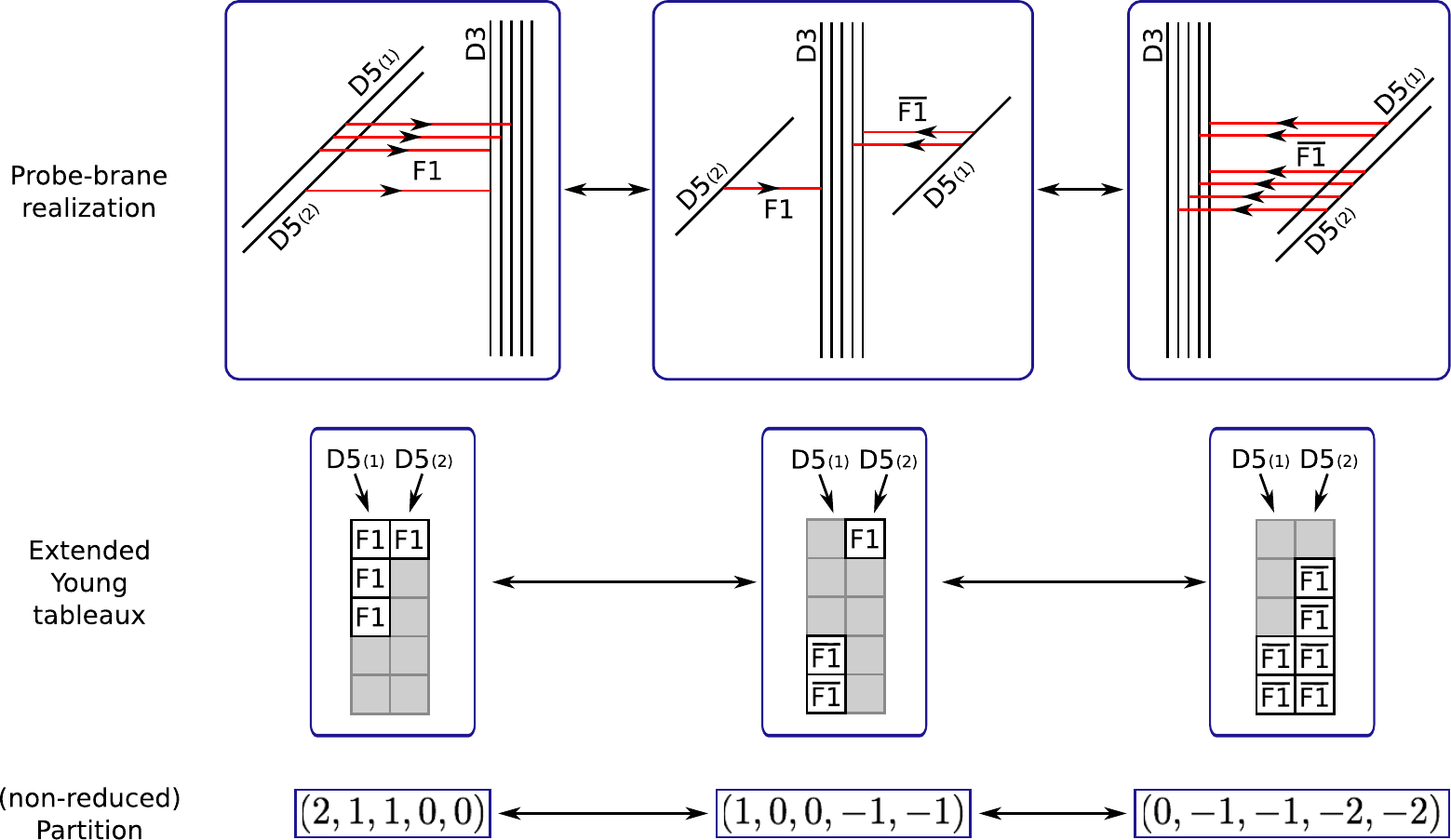}
\caption{As a D5-brane crosses the stack of D3-branes, fundamental stings disappear and anti-fundamental strings with the opposite orientation are created. At the level of the Young tableau this amounts to removing one column of $N$ boxes.
The boxes that have to be counted negatively are labeled by $\bar{F_1}$.
In terms of a partition of $N$ entries, we remove one unit for each entry.}
\label{probeHW}
\end{figure}

Instead of bringing an additional D5-brane from infinity, we can also take one of the D5-branes that are already present and bring it across the stack of D3-branes.
This process is depicted in Figure \ref{probeHW}, where we kept track of the orientation of the fundamental strings.
If we bring one of the D5-branes across the stack of D3-branes, all the fundamental string stretching between the D5-branes and the D3-branes are annihilated. Additionally there is one fundamental string created with the opposite orientation for each D3-brane in the stack that was not previously linked to the D5-brane by a fundamental string.  In Figure \ref{probeHW} the initial fundamental strings are denoted by $F_1$, and the strings created with opposite orientations are denoted by $\bar{F_1}$.
We can repeat this process until all the D5-branes have been brought across the stack of D3-branes.

This process can also be conveniently described with extended Young tableaux.
To this end we need to allow for tableaux that may have a negative number of columns of $N$ boxes.
We describe these tableaux in the following way.
We write down ``$F_1$" in all boxes of the Young tableau.
Then we complete all columns with empty boxes so that their length is equal to $N$.
When one D5-brane crosses the stack of D3-branes, we delete in the associated column all the $F_1$ and write down a $\bar{F_1}$ in the boxes that were previously empty. When all the D5-branes have crosses the stack of D3-branes, we obtain the (rotation of the) usual Young tableau for the conjugate of the initial representation. This is represented in Figure \ref{probeHW}.  In terms of partitions, we allow the $N$ entries of the partition to be negative integers.

Finally let us mention that instead of bringing the D5-branes one by one across the stack of D3-branes, one can also bring the D3-branes one by one across the D5-branes. In this process the creation and annihilations of fundamental strings can also be described using generalized Young tableau: as a D3-brane crosses the D5-branes, the empty boxes in the associated line are filled with $\bar F_1$'s, whereas the boxes with $F_1$'s are emptied.

%%%%%%%%%%%%%%%%%%%%%%%%%%%%%%%%%%%%%%%%%%%%%%%%%%

\section{The half-BPS supergravity solutions for open strings in the near-horizon of D3-branes}\label{sec:sugraSol}

The classification of smooth half-BPS solutions of type IIB supergravity with isometry $SO(2,1)\times SO(3) \times SO(5)$ was found in \cite{D'Hoker:2007fq}.
These solutions describe the backreaction of a stack of open strings ending on D3-branes, in the near-horizon limit (see also \cite{Lunin:2006xr,Lunin:2007mj} for similar work).
Equivalently they also describe the gravity dual of straight half-BPS Wilson lines.
In this section we first summarize the results of  \cite{D'Hoker:2007fq}. Then we perform a careful computation of the charges in these backgrounds, which allows for an unambiguous identification of the sources.
This leads to a one-to-one identification between the solutions of \cite{D'Hoker:2007fq} and the $SU(N)$ Young tableaux.
Finally we argue that the Hanany-Witten effect is realized at the level of the supergravity solution as large gauge transformation for the RR-potentials.

\subsection{Presentation of the solutions}\label{sub:summarySols}
\begin{figure}[t]
\centering
\includegraphics[width=1\linewidth]{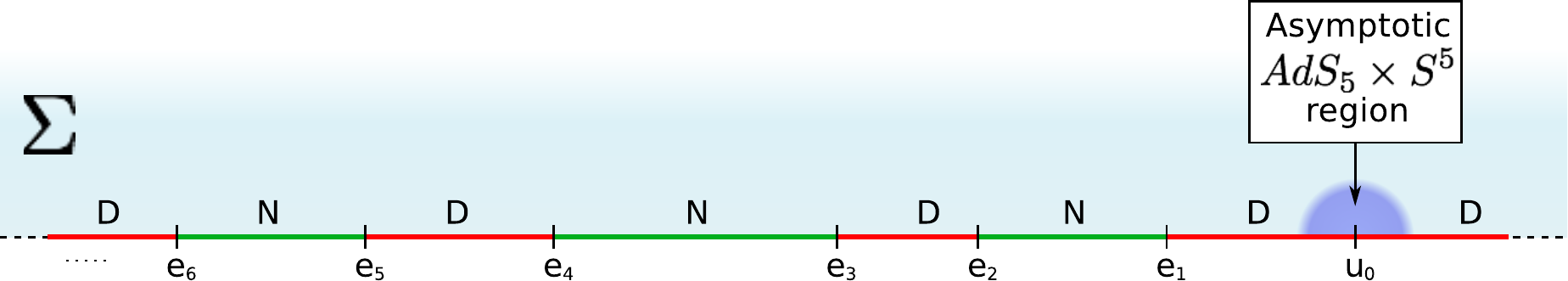}
\caption{The Riemann surface $\Sigma$ can be mapped to the upper-half plane. The harmonic function $h_1$ satisfies alternatively Neumann (N) and vanishing Dirichlet (D) boundary conditions along the real axis. In the neighborhood of the point $u_0$ the background is asymptotically $AdS_5 \times S^5$.}
\label{sigma}
\end{figure}

Let us begin with a description of the solutions found in \cite{D'Hoker:2007fq}.
In general these solutions are determined by two harmonic functions $h_1$ and $h_2$ defined on a two-dimensional Riemann surface $\Sigma$ which has disc topology. The harmonic function $h_2$ satisfies vanishing Dirichlet boundary conditions on the boundary of $\Sigma$, while $h_1$ satisfies alternating Neumann and vanishing Dirichlet boundary conditions (see Figure \ref{sigma})\footnote{We note that in general a solution is obtained for any choice of Riemann surface and any two harmonic functions.  In general the resulting geometry will contain singularities.  The specific choices presented here guarantee smooth solutions, but there may be other classes of smooth solutions or solutions with physical singularities.}.

We may choose $\Sigma$ to be the upper-half-plane.  We denote as $e_{1}$, $e_2$, etc. the points on the real axis where the boundary conditions of $h_1$ change.  In general, there is always an even number of such points. We denote this number by $2 g+ 2$. The functions $h_1$ and $h_2$ can be understood as hyper-elliptic functions on a Riemann surface without boundary of genus $g$, so we will loosely call the number $g$ the genus of the solution. Additionally, there is a singular point $u_0$ on the real axis, where the space-time is asymptotically $AdS_5 \times S^5$.  Using conformal transformations we can always choose the ordering:
\begin{align}
e_{2g+1} < ... < e_2 < e_1 < u_0
\end{align}
along with $e_{2g+2} = -\infty$ and the values $+\infty$ and $-\infty$ are identified.  The boundary conditions for the harmonic function $h_1$ can be written as:
\begin{align}
\label{boundaryconditions}
(e_{2g+1},e_{2g}),...,(e_{2i+1},e_{2i}),...,(e_1,u_0),(u_0,\infty): &\qquad {\rm vanishing \ Dirichlet} \nonumber \\
(e_{2g+2},e_{2g+1}),...,(e_{2j+2},e_{2j+1}),...,(e_{2},e_{1}): &\qquad {\rm Neumman} \ .
\end{align}
The harmonic functions obeying these boundary conditions satisfy the following equations:
\begin{align}
\label{harmfun}
\p_u h_1 \ du = - i \frac{P(u)}{(u-u_0)^2 s(u)} du \ ,
\qquad
\p_u h_2 \ du = i \frac{du}{(u-u_0)^2} \ ,
\end{align}
where $s(u)^2 = (u-e_1) \prod_{i=1}^{g} (u-e_{2i})(u-e_{2i+1})$ and $P(u)$ is a polynomial of degree $g+1$.  Note that $h_2$ can always be written in this form using conformal transformations.
The solutions depend on a total of $2g+2$ physical parameters%
\footnote{There are $2g+1$ parameters from the branch points $e_i$, one from $u_0$, $g+1$ from the zeros of the polynomial $P(u)$ plus one for its overall normalization, and two from the integration constant. Additionally there are $g+2$ constraints coming from the vanishing Dirichlet boundary conditions in the first line of $(\ref{boundaryconditions})$, and two for the shifts in $u$ which leave the form of $h_2$ invariant.}.
For $g=0$ the solution describes $AdS_5 \times S^5$ and the two parameters are the radius and the expectation value of the dilaton. Increasing the genus $g$ by one corresponds to adding two additional parameters for which we will give a detailed interpretation later.  A basis for the parameters is the point $u_0$, the overall coefficient of $h_1$ and the $2g+1$ branch points $e_i$ with the constraint $\sum_i e_i = 0$.  Alternatively, by using a scaling transformation, one may fix the value of $u_0$ and introduce on overall real coefficient in $h_2$.
\smallskip

For convenience, we introduce the following notations.  First we write the harmonic functions as
\begin{align}
h_1 = \A + \bar \A \qquad h_2 = \B + \bar \B \ ,
\end{align}
where $\A$ and $\B$ are holomorphic functions which are determined by the above equation up to an imaginary constant.  We introduce the dual harmonic functions $\tilde h_1$ and $\tilde h_2$ by
\begin{align}
\tilde h_1 = i (\A - \bar \A) \qquad \tilde h_2 = i (\B - \bar \B),
\end{align}
where the dual harmonic functions inherit the ambiguity in $\A$ and $\B$ and so are determined by $h_1$ and $h_2$ up to a constant.  This constant ambiguity corresponds to gauge transformations of the two-form potentials.  We also introduce the holomorphic quantity ${\cal C}$ defined by
\be \p_u {\cal C} = \A \p_u \B - \B \p_u \A, \ee
 whose constant ambiguity is related to gauge transformations of the four-form potential.  Finally we introduce the following combinations of $h_1$ and $h_2$
\begin{align}
W &= \p_u h_1 \p_{\bar u} h_2 + c.c. \cr
V &= \p_u h_1 \p_{\bar u} h_2 - c.c. \cr
N_1 &= 2 h_1 h_2 |\p_u h_1|^2 - h_1^2 W \cr
N_2 &= 2 h_1 h_2 |\p_u h_2|^2 - h_2^2 W
\end{align}
The full ten-dimensional geometry is given by a fibration of $AdS_2 \times S^2 \times S^4$ over $\Sigma$ with the metric in Einstein frame
\begin{align}\label{metricWilsonLine}
ds^2 = f_1^2 ds^2_{AdS_2} + f_2^2 ds^2_{S^2} + f_4^2 ds^2_{S^4} + 4 \rho^2 ds^2_{\Sigma} \ .
\end{align}
The warp factors are given by
\begin{align}
e^{4 \phi} = - \frac{N_2}{N_1} \ , \qquad \rho^8 = - \frac{W^2 N_1 N_2}{h_1^4 h_2^4} \cr
f_1^4 = - 4 e^{2 \phi} h_1^4 \frac{W}{N_1} \ , \qquad
f_2^4 = 4 e^{-2 \phi} h_2^4 \frac{W}{N_2} \ , \qquad
f_4^4 = 4 e^{-2 \phi} \frac{N_2}{W} \ .
\end{align}
The fluxes are given by
\begin{align}
H_3 = d b_1 \wedge \hat e^{01} \ , \qquad F_3 = d b_2 \wedge \hat e^{23} \ , \qquad
F_5 = -4 d j_1 \wedge \hat e^{0123} + 4 d j_2 \wedge \hat e^{4567} \ ,
\end{align}
where $\hat e^{01}$ is the unit volume form on $AdS_2$, $\hat e^{23}$ is the unit volume form on $S^2$, $\hat e^{4567}$ is the unit volume form on $S^4$ and
\begin{align}
\label{def:fluxes}
b_1 &= -2i \frac{h_1^2 h_2 V}{N_1} - 2 \tilde h_2 \ ,\cr
b_2 &= -2i \frac{h_1 h_2^2 V}{N_2} + 2 \tilde h_1 \ , \cr
j_2 &= i h_1 h_2 \frac{V}{W}  - \frac{3}{2} (\tilde h_1 h_2 - h_1 \tilde h_2) + 3 i ({\cal C} - \bar {\cal C})\ .
\end{align}
Note that $j_1$ is determined by the self-duality requirement of $F_5$.  All of the above formulas may be found in \cite{D'Hoker:2007fq} except for the expression for $j_2$ which was derived in \cite{Okuda:2008px}.
\smallskip

\begin{figure}[t]
\centering
\includegraphics[width=1\linewidth]{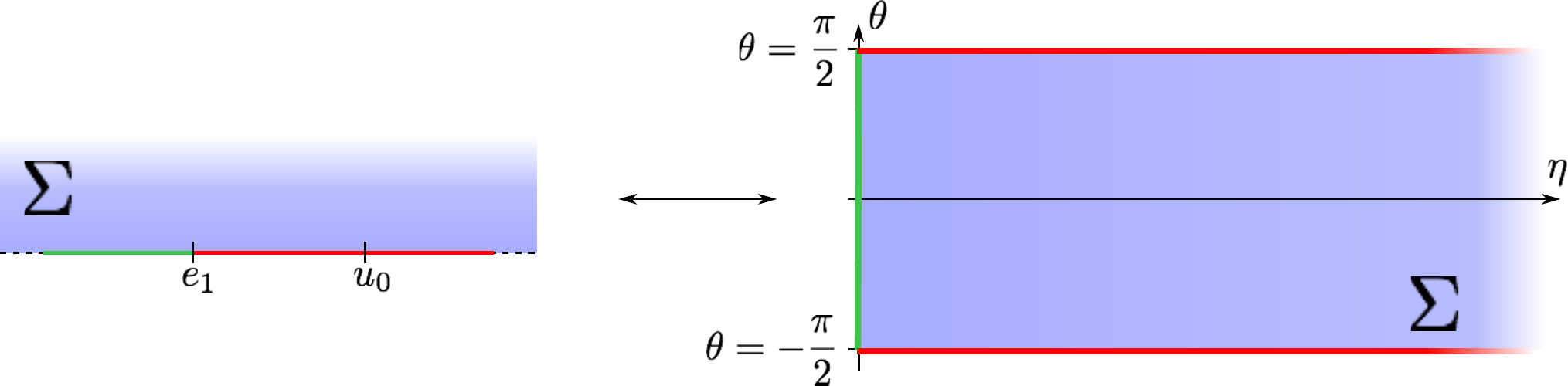}
\caption{The Riemann surface $\Sigma$ for the $g=0$ solution, which describes $AdS_5 \times S^5$.
\emph{Left:} The Riemann surface $\Sigma$ mapped to the upper-half plane.
\emph{Right:}  It is more convenient to map $\Sigma$ to a half-infinite rectangle. The vertical direction is parametrized by the polar angle of the five-sphere while the horizontal direction corresponds to a radial-like direction in $AdS_5$.}
\label{genus0}
\end{figure}

\paragraph{Genus-zero solution.}
For the case of $g=0$, we have simply $AdS_5 \times S^5$ and the harmonic functions are given by
\begin{align}
\label{hsAds5}
h_1 &= \frac{L^2}{4} e^{-\phi_0} \cosh(\eta+i \theta) + c.c. \cr
h_2 &= \frac{L^2}{4} e^{\phi_0} \sinh(\eta+i \theta) + c.c. \ ,
\end{align}
where $L$ is the radius 
and $e^{\phi_0}=\sqrt{g_s}$.
The metric reads:
\be ds^2 = L^2 \left(\cosh^2(\eta) \frac{dw^2 - dt^2}{w^2} + \sinh^2(\eta)d\Omega_{(2)}^2 + d\eta^2 + \cos^2 \theta d\Omega_{(4)}^2 + d\theta^2 \right)\ee
The coordinates $\eta$ and $\theta$ takes value in the range $0 \leq \eta \leq \infty$ and $-\pi/2 \leq \theta \leq \pi/2$.
In these coordinates, the Riemann surface $\Sigma$ is a semi-infinite rectangle parametrized by $\eta$ and $\theta$ (see Figure \ref{genus0}).
The map back to Poincare coordinates is given in (\ref{poincareToAdSxS}).

\begin{figure}[t]
\centering
\includegraphics[width=1\linewidth]{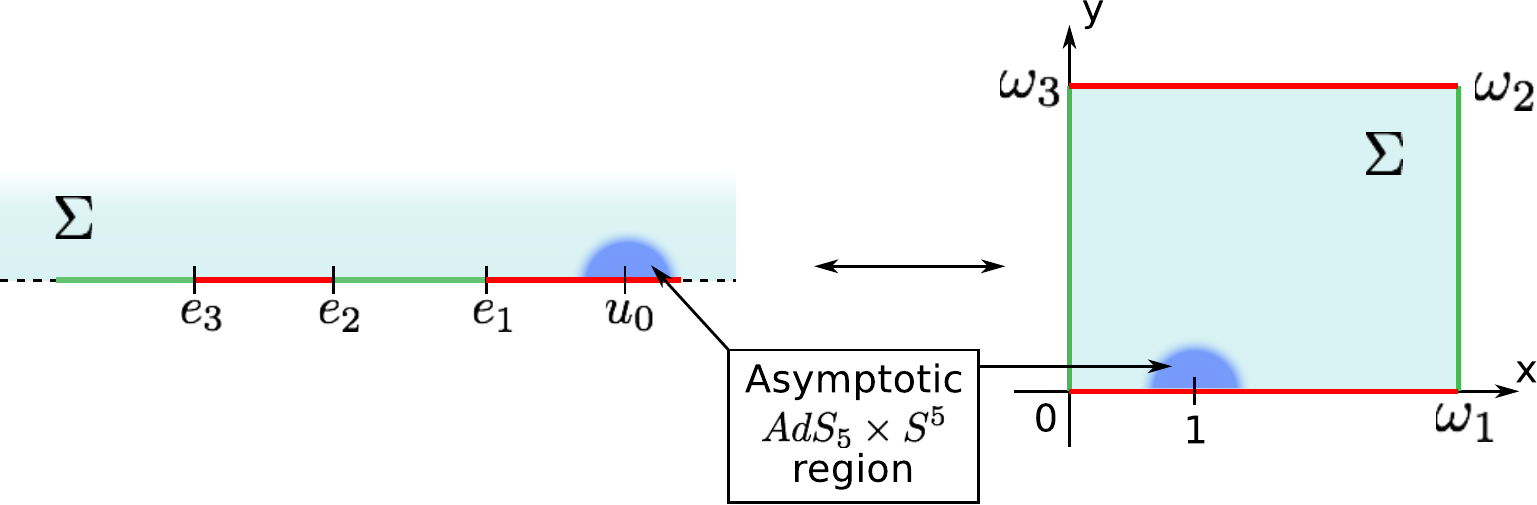}
\caption{\emph{Left:} The Riemann surface $\Sigma$ for the $g=1$ solution mapped to the upper-half plane.
\emph{Right:} The explicit solution takes a simpler form when we map $\Sigma$ to a rectangle in the upper-half plane delimited by $\omega_1$ and $\omega_3$  (the half-periods of the Weierstrass functions). The asymptotic $AdS_5 \times S^5$ region is mapped to the neighborhood of $1$.}
\label{genus1}
\end{figure}

\paragraph{Genus-one solution.}
For the case of $g=1$, the harmonic functions can be written explicitly in terms of Weierstrass functions.
The Weierstrass $\wp(z)$-function is defined by
\be (\wp'(z))^2 = 4 [\wp(z) - e_1][\wp(z) - e_2][\wp(z) - e_3],\ee
 together with the asymptotic condition $(\wp(z) - z^{-2}) |_{z=0} = 0$ which determines the integration constant.
The Weierstrass $\zeta$-function is defined by
\be \wp(z) = - \zeta'(z)\ee
 together with $(\zeta(z) - z^{-1})|_{z=0} = 0$.
The $e_i$'s satisfy the constraint\footnote{Note that the constraint on the $e_i$ can always be satisfied after a translation in the $u$-plane.} $e_1 + e_2 + e_3 = 0$.
Let us also introduce the $\omega_i$'s defined by $\wp(\omega_i) = e_i$ and satisfying $\omega_2 = \omega_1 + \omega_3$.
$\omega_1$ and $\omega_3$ are called the half-periods of the Weierstrass functions.
The rectangle delimited by the origin and the $\omega_i$'s forms a quarter of the fundamental domain for the Weierstrass functions (See Figure \ref{genus1}, \emph{right}).
For more details about these functions and their properties we refer to \cite{Bateman}.

The harmonic functions $h_1$ and $h_2$ take the form:
\begin{align}
\label{hsGen1}
h_1 &= \kappa_1  i \left( \zeta(x+iy - 1) + \zeta(x+iy + 1) - 2 \frac{\zeta(\omega_3)}{\omega_3} (x+iy) - c.c.  \right) \cr
h_2 &= \kappa_2 i \left( \zeta(x+iy-1) - \zeta(x+iy + 1) - c.c.  \right) \ .
\end{align}
The real coordinates $x$ and $y$ take respective values in the intervals $[0,\omega_1]$ and $[0,|\omega_3|]$ (see Figure \ref{genus1}).
The four parameters are the half-periods $\omega_1$ and $\omega_3$ which are respectively real and pure imaginary, and the two real coefficients $\kappa_1$ and $\kappa_2$. The parameters $\kappa_1$ and $\kappa_2$ are related to the asymptotic radius $L$ and dilaton $2\phi_0$ as\footnote{When comparing to \cite{D'Hoker:2007fq}, we have made an arbitrary scaling transformation in $z$ to set $w_0 = 1$ and introduced the arbitrary coefficient $\kappa_2$ in $h_2$.}
\begin{align}
\kappa_1 &= \frac{L^2}{8} e^{-\phi_0} \left( \wp(2) + \frac{\zeta(\omega_3)}{\omega_3} \right)^{-\frac{1}{2}} \ , \cr
\kappa_2 &= \frac{L^2}{8} e^{\phi_0} \left( \wp(2) + \frac{\zeta(\omega_3)}{\omega_3} \right)^{-\frac{1}{2}} \ .
\end{align}

%%%%%

\subsection{Computation of the charges}
\label{sec:charges}

Compared to $AdS_5 \times S^5$, the geometries with $g\ge 1$ contain several non-trivial 3-, 5- and 7-cycles.
The 3-cycles support non-trivial RR 3-form flux which corresponds to the presence of $D5$-branes.  The 5-cycles support non-trivial RR 5-form flux corresponding to the presence of additional $D3$-branes. Finally the Hodge dual of the 7-cycles support non-trivial NSNS 3-form flux corresponding to the presence of fundamental strings.
Thus a general solution contains $D5$- and $D3$-branes in addition to the expected fundamental strings.
In this section we carefully compute the various charges in order to work out the mapping with the probe brane discussion of section \ref{sec:probe}.
In general the genus $g$ solution contains $g$ additional 3- and 5-cycles so that raising the genus by one introduces another stack of $D3$- and $D5$-branes.

\begin{figure}[t]
\centering
\includegraphics[width=1\linewidth]{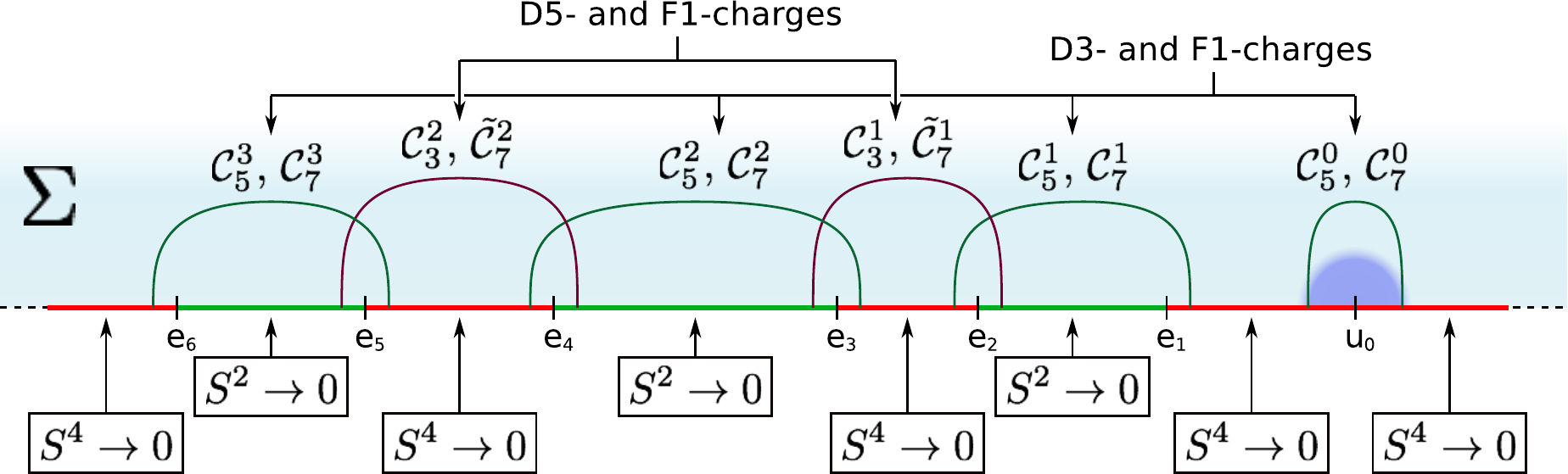}
\caption{Non-trivial 3-, 5- and 7-cycles are constructed as the fibration of various spheres on intervals in $\Sigma$. The endpoints of the intervals lie on the boundary of $\Sigma$ where either the four-sphere or the two-sphere degenerate. The cycles ${\cal C}_5^i$ (resp. ${\cal C}_7^i$) are built as the fibration of $S^4$ (resp. $S^4 \times S^2$) on intervals with endpoints on the degeneration locus of the $S^4$. The cycles ${\cal C}_3^i$ (resp. $\tilde {\cal C}_7^i$) are built as the fibration of $S^2$ (resp. $S^2 \times S^4$) on intervals with endpoints on the degeneration locus of the $S^2$.}
\label{sigmaCycles}
\end{figure}

First we introduce the following notations for the various cycles (see Figure \ref{sigmaCycles}):
\begin{itemize}
\item ${\cal C}_3^j$ is the three cycle formed by the fibration of $S^2$ over a line segment in $\Sigma$ with one endpoint in the interval $(e_{2j+2},e_{2j+1})$ and the other in $(e_{2j},e_{2j-1})$,
\item ${\cal C}_5^i$ is the five cycle formed by the fibration of $S^4$ over a line segment in $\Sigma$ with one endpoint in the interval $(e_{2i+1},e_{2i})$ and the other in $(e_{2i-1},e_{2i-2})$\footnote{We define $e_{2g+3} = -\infty$, $e_0 = e_{-1} = u_0$ and $e_{-2} = +\infty$.},
\item ${\cal C}_7^i$ is the seven cycle given by the warped product $S^2 \times {\cal C}_5^i$,
\item $\tilde {\cal C}_7^j$ is the seven cycle given by the warped product $S^4 \times {\cal C}_3^j$.
\end{itemize}
For the genus $g$ solution, the index $j$ runs from $g \geq j \geq 1$, while the index $i$ runs from $g+1 \geq i \geq 0$.
\smallskip

The definition of the charges in type IIB supergravity is subtle because of the Chern-Simons terms in the action  (see e.g. \cite{hep-th/0003037,Marolf:2000cb}).
We are interested in computing charges which are quantized and thus we need to compute the Page charge.  These are the charges which are local and quantized.  In general the Page charges are not gauge invariant under large gauge transformations.  In section \ref{sub:sugraYoung} we will link this ambiguity to the Hanany-Witten effect.  Our conventions are discussed in Appendix \ref{appconv} and the expressions for the charges are given in (\ref{app:chargedefs}), which we repeat here for convenience:
\begin{align}
\label{chargeform}
Q^{(j)}_{D5} &= \int_{{\cal C}^j_3} H_3 \cr
\tilde Q^{(j)}_{F1} &= \int_{\tilde {\cal C}^j_7} e^{-2 \phi} * H_{(3)} + C_{(4)} \wedge d C_{(2)} \cr
Q^{(i)}_{D3} &= \int_{{\cal C}^i_5} d C_{(4)} \cr
Q^{(i)}_{F1} &= \int_{{\cal C}^i_7} e^{-2 \phi} * H_{(3)} - C_{(2)} \wedge d C_{(4)} \ .
\end{align}
We remind the reader that $2 \phi = \Phi$ where $\Phi$ is the closed string dilaton field.  The reason for the different choice of Chern-Simons term in the definitions of the fundamental string charges $\tilde Q^{(j)}_{F1}$ and $Q^{(i)}_{F1}$ stems from the fact that $C_{(4)} \wedge d C_{(2)}$ is a well defined form on the $\tilde {\cal C}^j_7$ cycles, while $C_{(2)} \wedge d C_{(4)}$ is a well defined form on the ${\cal C}^i_7$ cycles.  The $i$-cycles are not all independent, as a result we have the following constraints on the charges
\begin{align}
\label{chargecons}
\sum_{i=0}^{g+1} Q^{(i)}_{D3} = 0, \qquad \qquad
\sum_{i=0}^{g+1} Q^{(i)}_{F1} = 0.
\end{align}
These constraints essentially state that the D3- and F1-charge $Q^{(0)}_{D3}$ and $Q^{(0)}_{F1}$ computed at infinity are the sum of the charges  $Q^{(i\neq 0)}_{D3}$ and $Q^{(i\neq 0)}_{F1}$ computed in the neighborhood of the sources.

To compute the charges we may always deform the integration contour so that it lies very close to the boundary of $\Sigma$ and we may work in a series expansion away from $\p \Sigma$.  This is useful as the boundary conditions for the harmonic functions limits the general form of such expansions.
Details are given in appendix \ref{app:compCharges}.  The reduced expressions for the charges are given by
\begin{align}
\label{explcharg}
Q^{(j)}_{D5} &= 2 i {\rm Vol}(S^2) \int_{e_{2j+1}}^{e_{2j}} d \A  + c.c. \cr
\tilde Q^{(j)}_{F1} &= - 12 i {\cal C}(\tilde e_j) \, {\rm Vol}(S^4) \, Q^{(j)}_{D5} + c.c. \cr
Q^{(i)}_{D3} &= 12 i \, {\rm Vol}(S^4) \, \int_{e_{2i}}^{e_{2i-1}} d {\cal C} + c.c. \qquad {\rm for} \; i \neq 0 \cr
Q^{(i)}_{F1} &= -2 i \A(\hat e_i) \, {\rm Vol}(S^2) \, Q^{(i)}_{D3} + c.c. \qquad {\rm for} \; i \neq 0
\end{align}
where $\hat e_i$ are arbitrary points in the intervals $(e_{2i},e_{2i-1})$ and $\tilde e_j$ are arbitrary points in the intervals $(e_{2j+1},e_{2j})$.\footnote{These formula can be understood as follows.  For the D5-brane charge, it turns out that the only non-vanishing contribution comes from the $\tilde h_1$ term in the definition of $b_2$ in (\ref{def:fluxes}).  This can be checked by showing that the remaining terms vanish when evaluated on $\p \Sigma$.  A similar argument works for the $D3$-brane charge so that the only non-trivial contribution comes from the ${\cal C}$ term in the definition of $j_2$ in (\ref{def:fluxes}).  For the fundamental charge, one finds that the only contribution comes from the Chern-Simons terms.  Additionally the gauge potentials $C_{(4)}$ and $C_{(2)}$ are constant on the intervals of their respective charges and we need only evaluate them at an arbitrary point in those intervals.}
The charges $Q^{(0)}_{D3}$ and $Q^{(0)}_{F1}$ can be computed using the charge conservation equations $(\ref{chargecons})$.  Additionally the D5-brane and D3-brane charges satisfy the inequalities
\begin{align}
Q^{(j)}_{(D5)} < 0 \qquad {\rm and} \qquad Q^{(i)}_{(D3)} > 0 \qquad {\rm for} \; i \neq 0 \ ,
\end{align}
which may be derived from the explicit expressions for the harmonic functions (\ref{harmfun}).
\smallskip

As discussed in Appendix \ref{appconv}, the charges are quantized in terms of the brane tension.
Introducing $N^{(i)}_{F1}$ as the number of fundamental strings, $N^{(i)}_{D3}$ as the number of D3-branes and $N^{(j)}_{D5}$ as the number of D5-branes, we have the following explicit formulas expressing the charges in terms of the number of branes
\begin{align}\label{Qs=Ns}
Q^{(0)}_{F1} &= -N^{(0)}_{F1} (4 \pi^2 \alpha^\prime)^3 \ , \qquad \qquad
Q^{(i)}_{F1} = N^{(i)}_{F1} (4 \pi^2 \alpha^\prime)^3 \ , \qquad {\rm for} \; i \neq 0\cr
Q^{(0)}_{D3} &= -N^{(0)}_{D3} (4 \pi^2 \alpha^\prime)^2 \ , \qquad \qquad
Q^{(i)}_{D3} = N^{(i)}_{D3} (4 \pi^2 \alpha^\prime)^2 \ , \qquad {\rm for} \; i \neq 0\cr
Q^{(j)}_{D5} &= - N^{(j)}_{D5} (4 \pi^2 \alpha^\prime) \ ,
\end{align}
where we have chosen the conventions so that the number of branes or strings is always positive in the canonical gauge discussed below.

The number of D3-branes measured at infinity is given by $N^{(0)}_{D3} = \sum_{i=1}^g N^{(i)}_{D3}$. It is also equal to the total number of D3-branes in the backreacting stack: $N^{(0)}_{D3} \equiv N$.
We can check that the asymptotic radius $L$ is related to the total number of D3-branes $N$ as (see Appendix \ref{App:asymptotics}):
\be L^4 = 4\pi N {\alpha'}^2 \ee
Note also that charge quantization implies that the genus of the solution $g$ cannot be greater than $N$.

The fundamental charges $Q^{(i)}_{F1}$ and $\tilde Q^{(j)}_{F1}$ are not invariant under gauge transformations of $C_{(2)}$ and $C_{(4)}$ respectively.  Indeed  we have:
\begin{align}\label{gaugeTransfoQ1s}
C_{(2)} \rightarrow C_{(2)} + \Delta_2 : \qquad Q^{(i)}_{F1} &\rightarrow Q^{(i)}_{F1} - \Delta_2 Q^{(i)}_{D3}
\end{align}
and:
\begin{align}\label{gaugeTransfoTQ1s}
C_{(4)} \rightarrow C_{(4)} + \Delta_4 : \qquad \tilde Q^{(j)}_{F1} &\rightarrow \tilde Q^{(j)}_{F1} + \Delta_4 Q^{(j)}_{D5}  \ .
\end{align}
This ambiguity is a manifestation of the Hanany-Witten effect, which claims that the number of fundamental strings is not an invariant quantity.  We will make this relation more precise in the next subsection.

Let us introduce some reasonable gauge choices for the charges $Q^{(i)}_{F1}$ (the story is similar for the charges $\tilde Q^{(j)}_{F1}$).
In order for the charge formula for the $Q^{(i)}_{F1}$ in (\ref{chargeform}) to be well defined, we must have $C_{(2)} = 0$ when $S^2$ shrinks to zero size.  This cannot be done globally and thus the charge formula for $Q^{(i)}_{F1}$ cannot cover the entire space without the use of gauge transformations.  We define a patch labeled by $k$ ($k \in \{1,...,g+1\}$) over which the charge formula is valid as the patch which covers the entire space except for the part of $\p \Sigma$ defined by $\bigcup_{i \neq k} (e_{2 i},e_{2 i-1})$, with the condition that $C_{(2)}$ vanishes on the interval $(e_{2 k},e_{2 k-1})$.  This is equivalent to the requirement $\im(\A(\hat e_{k})) = 0$ so that $Q^{(k)}_{F1} = 0$.  If we now wish to go to a new patch labeled by $l$ we must make a gauge transformation so that $Q^{(l)}_{F1} = 0$.  This corresponds to picking $\Delta_2 = Q^{(l)}_{F1}/Q^{(l)}_{D3}$ in \eqref{gaugeTransfoQ1s} so that the charges transform as
\begin{align}
\label{gaugetrans}
Q^{(i)}_{F1} &\rightarrow Q^{(i) \prime}_{F1} = Q^{(i)}_{F1} - \frac{Q^{(l)}_{F1}}{Q^{(l)}_{D3}} Q^{(i)}_{D3} \ .
\end{align}
In the remainder of the paper, unless otherwise mentioned, we shall choose the gauge with $k = g+1$ so that:
\be\label{canGauge} Q^{(g+1)}_{F1} = 0 \ee
This is equivalent to requiring:
\be \im(\A(\hat e_{g_+1})) = 0 \ee
We shall refer to this gauge as the canonical gauge.

We note that when we increase $g$ by one, the number of parameters increases only by two and so there must be a relation between the fundamental string charge and the D5- and D3-brane charges.  To make this relation manifest, we first consider the sum
\begin{align}
\sum_{j=n}^{g} Q^{(j)}_{D5} &= \sum_{i=n}^{g} 2 i {\rm Vol}(S^2) [\A(\hat e_i) - \A(\hat e_{i+1})] + c.c. \qquad \qquad n \in \{1,...,g\} \cr
&= 2 i {\rm Vol}(S^2) \A(\hat e_n) + c.c.
\end{align}
where in the second line, we have made use of the canonical gauge condition $\im(\A(\hat e_{g+1})) = 0$.  This allows us to solve for $\A(\hat e_n)$ in terms of the $Q^{(j)}_{D5}$ and express the number of fundamental strings as
\begin{align}\label{N1=N3N5}
N^{(i)}_{F1} &= N_{D3}^{(i)} \sum_{j=i}^{g} N^{(j)}_{D5} \qquad \qquad {\rm for} \, i \in \{1,...,g\}
\end{align}
along with $N^{(g+1)}_{F1} = 0$.  Note that this formula is manifestly consistent with charge quantization.  It also allows us to write the gauge transformation (\ref{gaugetrans}) in a way manifestly consistent with charge quantization as
\begin{align}
N^{(i)}_{F1} &\rightarrow N^{(i) \prime}_{F1} = N^{(i)}_{F1} - N^{(i)}_{D3} \sum_{j=l}^{g} N^{(j)}_{D5} \ .
\end{align}
Finally, the number of fundamental strings running to the boundary in the canonical gauge is given by
\begin{align}
N^{(0)}_{F1} = \sum_{i=1}^g N_{D3}^{(i)} \sum_{j=i}^{g} N^{(j)}_{D5} \ .
\end{align}
In the gauge with $Q^{(l)}_{F1} = 0$, the number of fundamental strings running to the boundary is related to the canonical gauge result by
\begin{align}
N^{(0) \prime}_{F1} = N^{(0)}_{F1} - N^{(0)}_{D3} \sum_{j=l}^{g} N^{(j)}_{D5} \ .
\end{align}
Note that they differ by an integer coefficient times $N^{(0)}_{D3}$ which is the rank of the gauge group in the dual CFT\footnote{This can be interpreted as introducing $\sum_{j=l}^{g} N^{(j)}_{D5}$ objects which are anti-symmetric combinations of $N^{(0)}_{D3}$ fundamental strings. From the gauge theory point of view they are colorless objects which do not change the representation of the Wilson loop.}.  When computing the charges $\tilde Q^{(j)}_{F1}$ a similar picture emerges.

%%%%%

\subsection{Associating Young tableaux to supergravity solutions}\label{sub:sugraYoung}

The supergravity solutions we discussed previously describe, in addition to fundamental strings ending on D3-branes, the gravity duals of half-BPS Wilson lines. These Wilson lines are conveniently labelled by irreducible representations of the gauge group $SU(N)$, or equivalently by Young tableaux.
In this paragraph we identify the Young tableau associated to each solution of \cite{D'Hoker:2007fq}. The picture we describe first appeared in \cite{Okuda:2008px}.  The explicit computation of the charges we performed in the previous subsection gives strong support in favor of this conjecture.

\begin{figure}[t]
\centering
\includegraphics[width=1\linewidth]{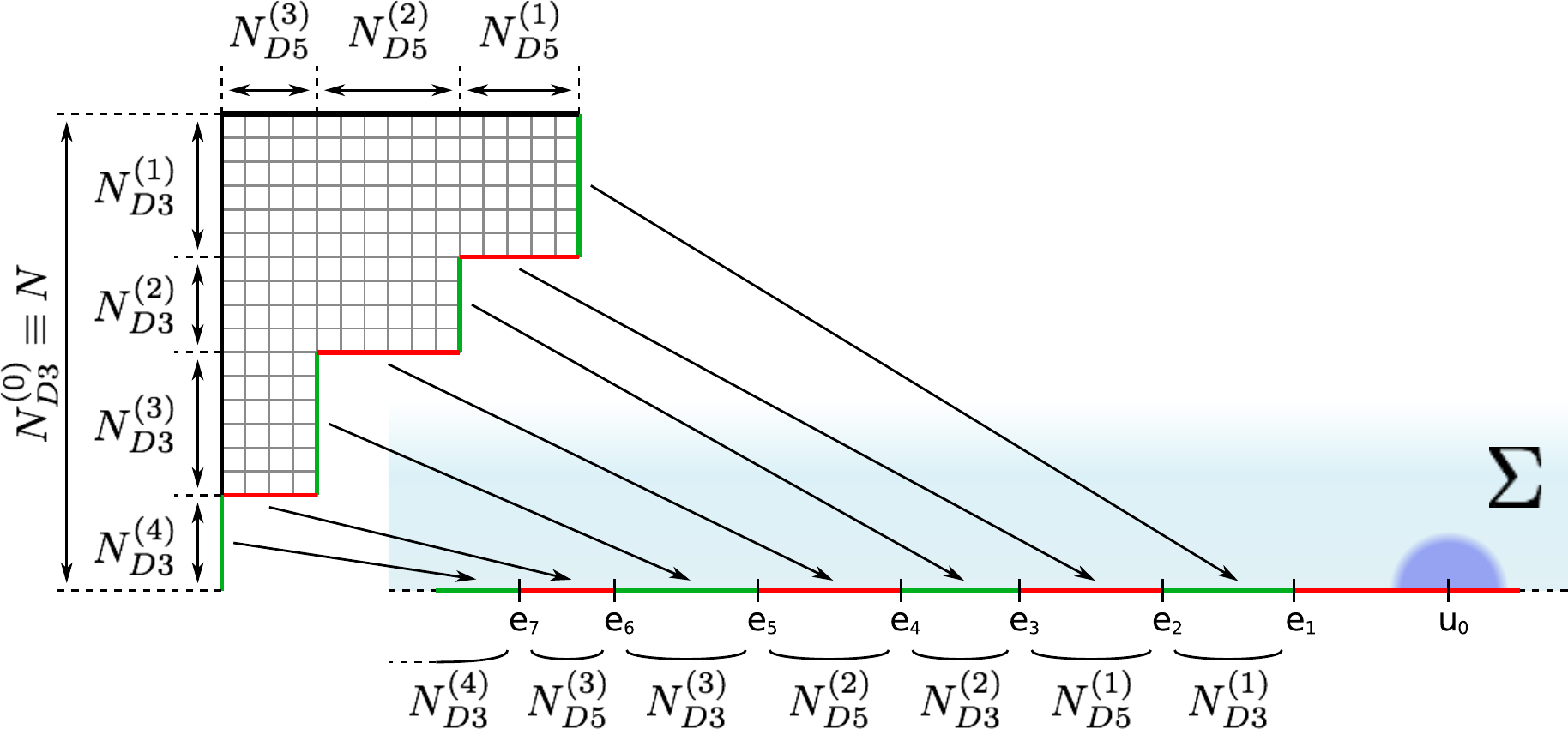}
\caption{The Young tableau associated to a given supergravity solution, in the case $g=3$.
There is a natural map between the boundary of the Young tableau and the boundary of the Riemann surface $\Sigma$.
The length of the segments on the boundary of the Young tableau are given by the numbers of D3- and D5-branes computed along the corresponding sections of the boundary of the Riemann surface $\Sigma$. }
\label{tableauSigma}
\end{figure}

\begin{figure}[t]
\centering
\includegraphics[width=1\linewidth]{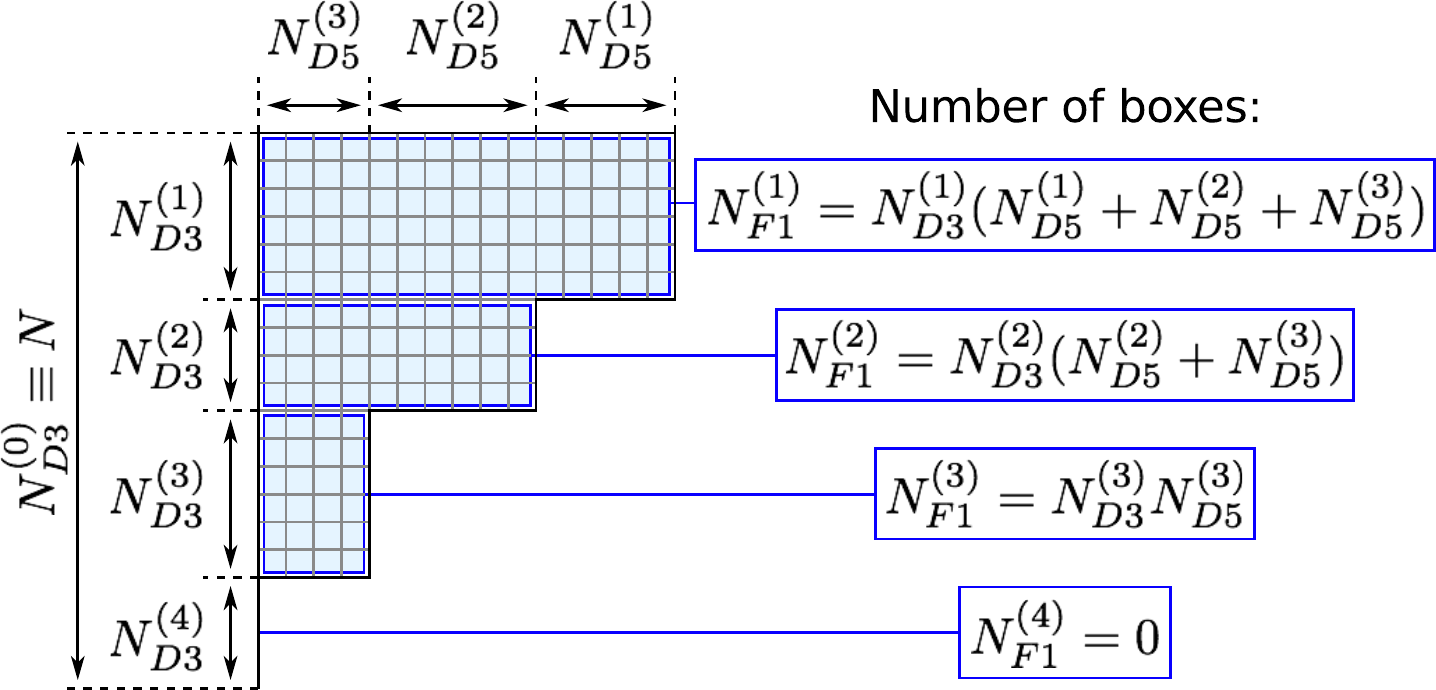}
\caption{The number of boxes in the Young tableau can be associated to numbers of fundamental strings. This imposes some constraints relating the numbers of D3-branes, D5-branes and fundamental strings and which are satisfied by the supergravity solutions. This picture holds in the canonical gauge \eqref{canGauge}.}
\label{tableauF1}
\end{figure}

Let us start from a generic Young tableau. We want to identify the supergravity solution dual to a Wilson line labelled by this tableau. To this end it is convenient to choose the $2g+2$ parameters defining the supergravity solution as the $g$ D5-charges $N^{(j)}_{D5}$, the $g+1$ D3-charges $N^{(i)}_{D3}$ and the asymptotic dilaton $2\phi_0$. For now we work in the canonical gauge.
The discussion of section \ref{sec:probe} suggests that each line of the Young tableau is associated to a D3-brane and each column to a D5-brane. This naturally leads to the following proposal.
We slice the Young tableau in horizontal rectangles (see Figure \ref{tableauF1}).
The number of such rectangles is the genus $g$.
The number of lines in the $i$-th horizontal rectangle (starting from the top) is $N_{D3}^{(i)}$.
The last D3-charge $N_{D3}^{(g+1)}$ is associated to an empty rectangle: when added to the total number of lines of the Young tableau, it provides the number of colors $N$ of the dual gauge theory (and thus the asymptotic curvature radius of the background).
We can also slice the tableau in $g$ vertical rectangles, and let $N^{(j)}_{D5}$ be the number of columns in the $j$-th vertical rectangle (starting from the right).
This gives precisely a mapping between the boundary of the Young tableau and the boundary of the Riemann surface $\Sigma$ (see Figure \ref{tableauSigma})

In section \ref{sec:probe} we also argue that each box is associated to a fundamental string.
Considering the horizontal slicing of the Young tableau, the number of boxes in the $i$-th rectangle has to be equal to $N_{F1}^{(i)}$.
For this picture to be consistent the following equation has to hold: $N^{(i)}_{F1} = N_{D3}^{(i)} \sum_{j=i}^{g} N^{(j)}_{D5}$ (see Figure \ref{tableauF1}). This is nothing but equation \eqref{N1=N3N5} that we derived from the supergravity solution.
Notice also that the canonical gauge choice \eqref{canGauge}, equivalent to demanding $N^{(g+1)}_{F1}=0$, is consistent with the fact that the $g+1$-th horizontal rectangle is empty.
Similarly the number of boxes in vertical rectangles is given by the charges $\tilde Q^{(j)}_{F1}$, and a similar discussion can be repeated for these charges.

\paragraph{Large gauge transformations and the Hanany-Witten effect.}
We saw previously that the number of fundamental strings $N_{F1}^{(i)}$ is not invariant under large gauge transformations. So one may worry that identifying fundamental strings with boxes of a Young tableau is not consistent.  However, we shall argue that the careful analysis of the large gauge transformations we did previously actually provides further evidence in favor of associating Young tableau with the supergravity solutions.

\begin{figure}[t]
\centering
\includegraphics[width=1\linewidth]{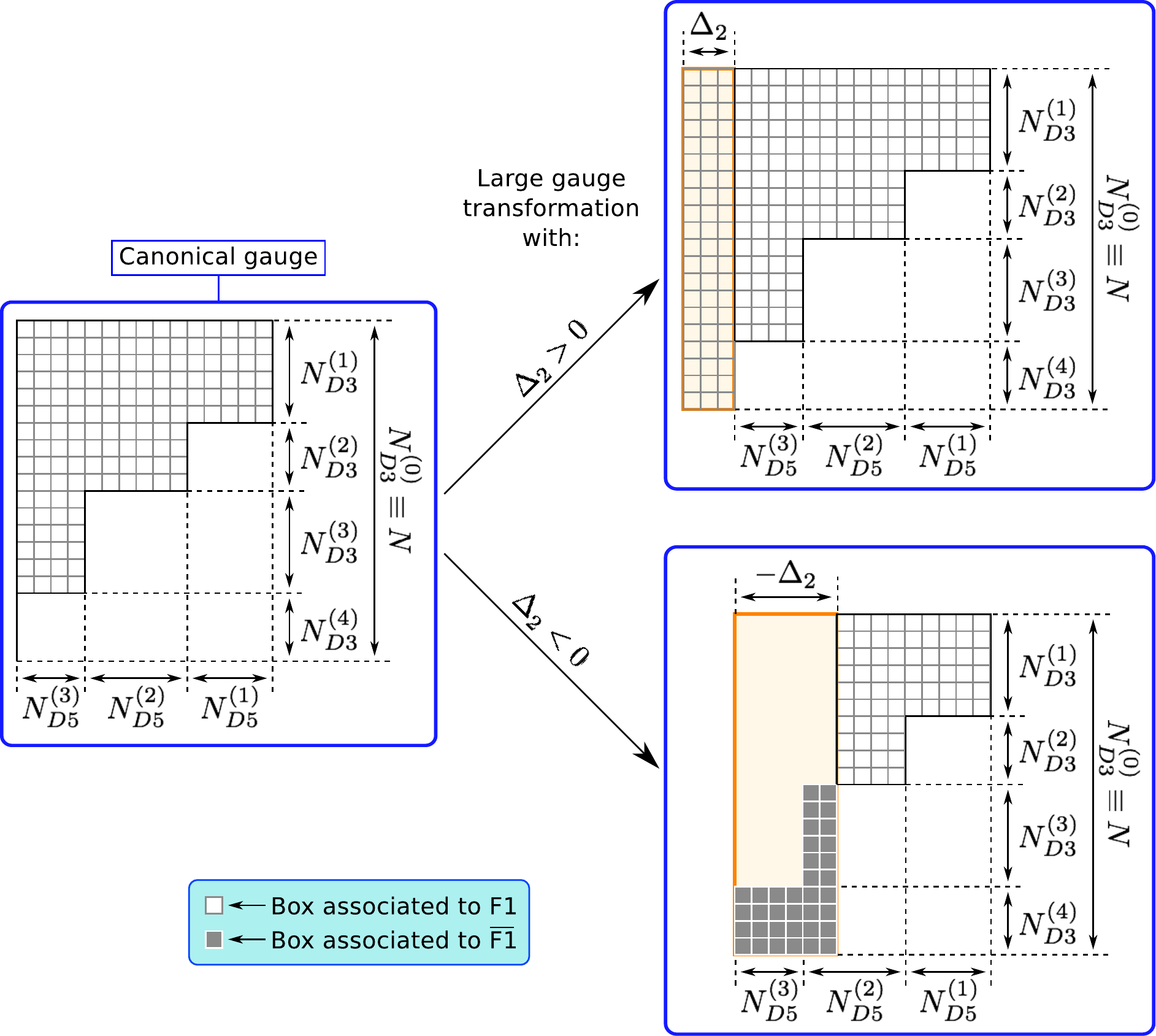}
\caption{Large gauge transformation of the type \eqref{gaugeTransfoQ1s} amounts to adding or removing columns of $N$ boxes to the Young tableau.
The boxes that have to be counted negatively are shaded.}
\label{tableauGauge}
\end{figure}

Under a large gauge transformation \eqref{gaugeTransfoQ1s} the number of fundamental strings $N^{(i)}_{F1}$ changes by $\Delta_2 N_{D3}^{(i)}$ units. Moreover the total number of fundamental strings changes by $\Delta_2 N_{D3}^{(0)}$ units.
Thus performing a large gauge transformation of parameter $\Delta_2$ amounts to adding $\Delta_2$ columns of $N$ boxes to the (extended) Young tableau, where $\Delta_2$ can be a positive or negative integer (see Figure \ref{tableauGauge}).
The canonical gauge choice \eqref{canGauge} can be understood as demanding the extended Young tableau to have exactly zero columns of $N$ boxes, or more simply put that we have a standard Young tableau.

In section \ref{sub:probeHW} we argued that the addition or subtraction of columns of $N$ boxes to the Young tableau was naturally related to the Hanany-Witten effect. More precisely, adding or removing $\Delta_2$ columns of $N$ boxes is equivalent to taking $\Delta_2$ D5-branes across the stack of D3-branes. The modification of the number of fundamental strings in this process via the Hanany-Witten effect is indeed equal to $N \Delta_2$.
Thus we conclude that the ambiguity in the number of fundamental strings in the supergravity solution is simply the manifestation of the Hanany-Witten effect.
Large gauge transformations of the two-form potential \eqref{gaugeTransfoQ1s} can be visualized as bringing D5-branes across the stack of D3-branes.

A parallel discussion can be done working with the fundamental charges $\tilde Q^{(j)}_{F1}$, that count the number of boxes in the vertical rectangles slicing the Young tableau. In that case large gauge transformations can be understood in the probe brane picture as taking some D3-branes from the stack and moving them across the D5-branes.

\paragraph{Reducible representations.}
We just argued that the smooth supergravity solutions found in \cite{D'Hoker:2007fq} are in one-to-one correspondence with the irreducible representations of $SU(N)$ that label half-BPS Wilson lines.  A natural question is then what are the holographic descriptions of Wilson lines in reducible representations?  We note that any reducible representation of $SU(N)$ can be decomposed into a set of irreducible representations. This perhaps suggests that the gravity dual of a Wilson line in a reducible representation should be understood as the quantum superposition of several supergravity solutions associated to irreducible representations.

This observation is related to the question of finding the supergravity solution describing a generic stack of fundamental strings ending on D3-branes. The stack of fundamental strings is generically associated to a multiple tensor product of fundamental representations, which is not an irreducible representation. We propose that there is no smooth supergravity solutions describing this stack. The state in the supergravity theory associated to this stack is rather the superposition of the supergravity solutions describing the irreducible representations that one finds in the decomposition of the tensor product of fundamentals.  It is possible that singular solutions exist, where the singularity encodes the fact that the underlying state is really reducible.  This is similar to the fuzzball proposal for black holes (see \cite{Mathur:2005zp,Skenderis:2008qn} and references therein).  If such singular solutions are found, its possible one may associate an entropy to them which somehow counts the number of irreducible representations encoded in the geometry.

%%%%%

\subsection{Charges for the genus one solution}

There exists a simple explicit expression for the harmonic functions defining the genus one solution, given by \eqref{hsGen1}. This allows us to derive explicit formulas for the charges that will be useful in the following sections. For the $g=1$ solution, the functions $\A$ and $\B$ are given by:
\begin{align}
\A(z) &= i  \kappa_1 \left( \zeta(x+iy-1) + \zeta(x+iy+1) - 2 \frac{\zeta(\omega_3)}{\omega_3} (x+iy) \right), \cr
\B(z) &= i \kappa_2 \left( \zeta(x+iy-1) - \zeta(x+iy+1) \right)
\end{align}
where we have chosen the canonical gauge for $\A$ while the constant ambiguity in $\B$ will not affect our results.
\smallskip

In this solution there is only one D5-charge to evaluate.
Formula \eqref{explcharg} leads to
\begin{align}\label{QD5g1}
Q^{(1)}_{D5} = 8 \pi i \left( \A(\omega_2) - \A(\omega_3) \right) + c.c.
= - \frac{16 i \pi^2 \kappa_1}{\omega_3}
\end{align}
where we have used $\zeta(\omega_2) = \zeta(\omega_1)+\zeta(\omega_3)$ and the Legendre's identity $\zeta(\omega_1)\omega_3 - \zeta(\omega_3)\omega_1 = i \frac{\pi}{2}$.

The computation of the D3-charges is slightly more involved.
The D3-charge is given by (\ref{explcharg}), which may be written as:
\be Q^{(i)}_{D3} = 32 i \pi^2 \int_{e_{2i}}^{e_{2i-1}} [ \partial (\A\B) - 2\B \partial \A] + c.c.
\qquad \qquad 2 \geq i \geq 1
\ee
First we note that the $\A\B$ is real when evaluated at any $\omega_i$ or at zero and thus the total derivative term does not contribute to the D3-brane charge. Let us introduce the function ${\cal F}$ defined as:
\begin{align}\label{defF}
{\cal F}(z) &\equiv - \int \B \partial \A
\end{align}
The explicit form of the function ${\cal F}(z)$ is computed in Appendix \ref{app:D3charge} and given in (\ref{f(I1)}).
In terms of ${\cal F}(z)$, the charges are given by
\begin{align}
Q_{D3}^{(0)} &= 64 i \pi^2 [{\cal F}(0) - {\cal F}(\omega_1)] + c.c. \cr
Q_{D3}^{(1)} &= 64 i \pi^2 [{\cal F}(\omega_1) - {\cal F}(\omega_2)] + c.c. \cr
Q_{D3}^{(2)} &= 64 i \pi^2 [{\cal F}(\omega_3) - {\cal F}(0)] + c.c. \ ,
\end{align}
where we have determined $Q_{D3}^{(0)}$ by the requirement $Q_{D3}^0 + Q_{D3}^1 + Q_{D3}^2 = 0$.  The explicit expressions in terms of $\omega_1$, $\omega_3$, $\kappa_1$ and $\kappa_2$ are
\begin{align}\label{QD3g1}
Q_{D3}^{(0)} &= 64 \pi^3 \kappa_1 \kappa_2 \bigg[ 4 \frac{\zeta(\omega_3)}{\omega_3} - 8 \wp(1) + \bigg( \frac{\wp''(1)}{\wp(1)} \bigg)^2 \bigg] \cr
Q_{D3}^{(1)} &= 128 \pi^2 \kappa_1 \kappa_2 i \bigg\{ \bigg[
4 \frac{\zeta(\omega_3)}{\omega_3} -8 \wp(1)
+ \frac{\wp''(1)^2}{\wp'(1)^2}  \bigg] \bigg(\omega_3 \zeta(1) - \zeta(\omega_3) \bigg) \cr
&  +  \bigg(\omega_3 \wp(1) + \zeta(\omega_3)\bigg) \frac{\wp''(1)}{\wp'(1)}  - \omega_3 \wp'(1) \bigg\}
\end{align}
while $Q_{D3}^{(2)}$ can be obtained using $Q_{D3}^{(0)} + Q_{D3}^{(1)} + Q_{D3}^{(2)} = 0$.

%%%%%%%%%%%%%%%%%%%%%%%%%%%%%%

\section{Probe brane limits}\label{sec:probeLim}

In this section we discuss the limits of the supergravity solutions in which one recovers the probe D5- and D3-brane descriptions discussed in section \ref{sec:probe}.
When the various D3-brane charges become small, except for one, the supergravity solution is associated to a Young tableau that is almost horizontal. According to the discussion of section \ref{sec:probe}, we should recover in this limit a probe D3-brane.  This probe D3-brane describes a BIon \cite{Callan:1997kz,Gauntlett:1999xz}: a D3-brane from the stack is pulled out into the bulk and forms a spike, because of the presence of the fundamental strings.
Alternatively, when the various D5-brane charges go to zero, the Young tableau is almost vertical and we should recover a probe D5-brane.  This brane results from the polarization of a stack of fundamental strings in the background fluxes \cite{Myers:1999ps,Emparan:1997rt,hep-th/0111156}.

We will show that these expectations are indeed realized. Thus we provide a \emph{bulk} derivation both of the D3-brane spike \cite{Callan:1997kz,Gauntlett:1999xz} and of the fundamental string polarization effect \cite{Myers:1999ps,Emparan:1997rt,hep-th/0111156}, that were previously studied using the DBI action.  We shall find remarkable agreement between the DBI and supergravity descriptions.

To study the probe limits, we use the genus one solution for which the explicit expressions of the harmonic functions $h_{1,2}$ are given in \eqref{hsGen1}. In this case the Young tableau is a rectangle.  We recall this solution is parametrized by four parameter which can be chosen as the half periods of the Weierstrass functions $\omega_1$ and $\omega_3$, as well as the asymptotic radius $L$ and dilaton $g_s$.  The half periods essentially determine the values of the charges and the small-charge limits are equivalent to degeneration limits of the rectangle of Figure \ref{genus1} (\emph{right}).

%%%%%%

\subsection{The probe D3-brane limit}

\begin{figure}[t]
\centering
\includegraphics[width=1\linewidth]{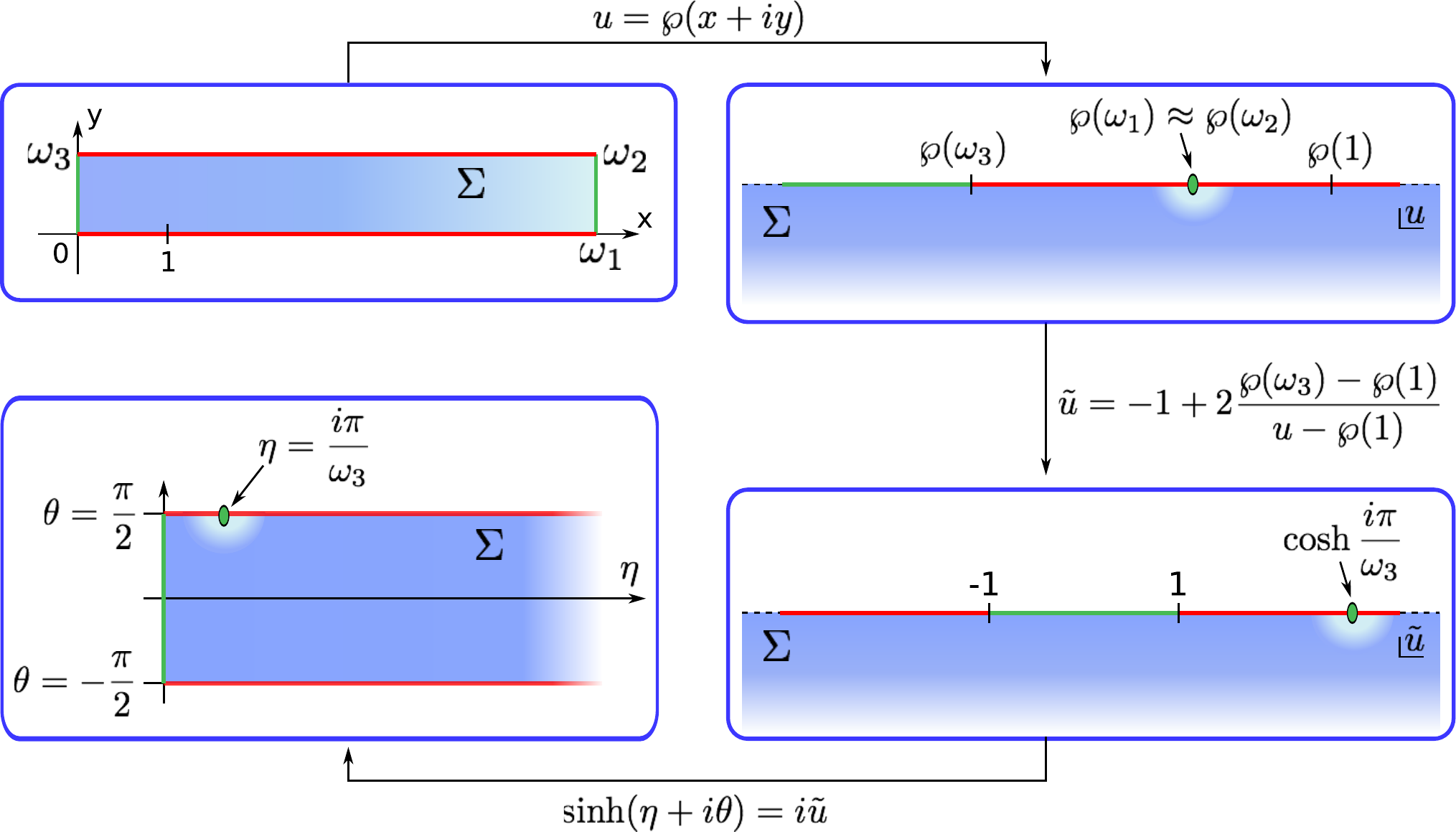}
\caption{Map from the rectangle associated to the genus-one solution (top-left) to the half-infinite rectangle associated to the genus-zero solution (bottom-left), in the limit $\omega_1 \to \infty$. The geometry differs from $AdS_5 \times S^5$ in the light-blue region. The green oval gives the position of the probe D3-branes.}
\label{dgenus1ProbeD3}
\end{figure}

Let us first consider the limit where the half period $\omega_1$ goes to infinity.
In terms of the Riemann surface $\Sigma$  of Figure \ref{genus1} (\emph{left}), the limit $\omega_1 \to \infty$  amounts to shrinking the segment $[e_2,e_1]$ down to zero size.
The amount of D3-charge carried by this segment goes to zero, while the segment $]-\infty,e_3]$ carries the $N$ units of D3-brane charge:
\begin{align}
Q_{D3}^{(1)} & \approx 0 \cr
Q_{D3}^{(2)} &\approx 4 \pi^3 L^4  = -Q_{D3}^{(0)}
\end{align}
The fact that the first D3-charge $Q_{D3}^{(1)}$ goes to zero implies that the Young tableau associated to this solution is almost a horizontal line.
In the limit $\omega_1 \to \infty$ the Weierstrass functions $\wp$ and $\zeta$ simplify to \cite{Bateman}:
\begin{align}\label{WeierstrassLim1a}
\wp(x) \approx -\frac{\pi^2}{12 \omega_3^2}\left(1+\frac{3}{\sinh^2\left(\frac{i\pi}{2\omega_3}x\right)}\right)\cr
\zeta(x) \approx \frac{\pi^2}{12 \omega_3^2}x + \frac{i\pi}{2 \omega_3}\coth\left(\frac{i\pi}{2\omega_3}x\right)
\end{align}
Moreover we have in this limit:
\be\label{WeierstrassLim1b} \wp(\omega_1) \approx \wp(\omega_2) \approx -\frac{1}{2} \wp(\omega_3) \ee
At the level of the supergravity solution the limit $\omega_1 \to \infty$ is smooth and the geometry reduces to $AdS_5 \times S^5$.
For large but finite $\omega_1$, we expect the geometry to be almost everywhere $AdS_5 \times S^5$, except for $u\approx e_{1,2}$ (or equivalently $x \approx \omega_1$). This portion of spacetime, where the geometry differs from $AdS_5 \times S^5$, is the neighborhood of the few D3-branes which are deformed by the presence of the fundamental strings.
We want to understand the location of this region inside $AdS_5 \times S^5$.
To this end we have to map the rectangle of Figure \ref{genus1} (\emph{right}), that gives a convenient parametrization of the genus one solution, to the half-infinite rectangle of Figure \ref{genus0} (\emph{right}), that gives a convenient parametrization of $AdS_5 \times S^5$ (the genus zero solution).
This mapping is described in Figure \ref{dgenus1ProbeD3}.
We demand that the segment $[0,\omega_3]$ of the rectangle that carries most of the D3-brane charge in the genus one solution be mapped to the left-border of the half-infinite rectangle that carries the D3-brane charge in the genus-zero solution.

First we map the rectangle parametrized by $x+iy$ to the lower-half plane parametrized by $u$ thanks to the Weierstrass $\wp$ function:
\be u = \wp(x+iy) \ee
Then we make a conformal transformation $u \to \tilde u$ that essentially rotates the boundary of the lower-half plane so that the asymptotic region lies now at infinity: $u=\wp(1) \to \tilde u = \infty$. We also demand that the segment of the boundary $]-\infty,\wp(\omega_3)]$ that carries most of the D3-brane charge is mapped to the segment $\tilde u \in [-1,1]$. The explicit form of the mapping is:
\be \tilde u = -1 + 2\frac{\wp(\omega_3)-\wp(1)}{u-\wp(1)} \ee
Finally we map the lower half plane to the half-infinite rectangle parametrized by $\eta$ and $\theta$:
\be \sinh(\eta + i \theta) = i \tilde u \ee
Using formulas \eqref{WeierstrassLim1a} and \eqref{WeierstrassLim1b} it is straightforward to check that along this chain of mapping, the genus-one harmonic functions \eqref{hsGen1} are mapped to the genus-zero harmonic functions \eqref{hsAds5}.
Additionally the segment $[\omega_1,\omega_2]$ is mapped to a point (in the limit $\omega_1 \to \infty$) on the boundary of the rectangle with coordinates $(\eta_{D3},\theta_{D3})$ given by:
\be\label{coordProbeD3} \eta_{D3} = \frac{i\pi}{\omega_3} \ , \ \theta_{D3}=\frac{\pi}{2} \ee
We identify this point with the locus of the probe D3-branes.
To compare this result with the DBI result \eqref{DBIProbeD3}, we notice that in the limit $\omega_1 \to \infty$ the D5-brane charge \eqref{QD5g1} simplifies to:
\begin{align}\label{D5chargeProbeD3}
Q_{D5}^{(1)} &= -4 e^{-\phi_0} L^2 \pi \sinh \left( \frac{i\pi}{\omega_3} \right) \ .
\end{align}
Note that $i/\omega_3$ can take any positive real value so that the D5-brane charge is unbounded as expected.
The charge quantization is easy to understand in this limit.  As usual the asymptotic radius $L$ is determined by the number of D3-branes in the asymptotic region, $N^{(0)}_{D3}$, while the number of D5-branes determines the half-period $\omega_3$ and the remaining number of D3-branes, $N^{(1)}_{D3}$ determines the half-period $\omega_1$.  Thus we see that the half-periods $\omega_1$ and $\omega_3$ are quantized.

Combining equations \eqref{coordProbeD3}, \eqref{D5chargeProbeD3} and \eqref{Qs=Ns}, we observe that the coordinate $\eta_{D3} $ is related to the number of D5-branes as:
\be\label{sugraProbeD3} \sinh \eta_{D3} = N_{D5} \pi \alpha' \frac{e^{\phi_0}}{L^2} \ee
Remember that the number of D5-branes $N_{D5}$ is nothing but the number of fundamental strings dissolved in each probe D3-brane, as is clear from the Young tableau picture. In particular for a single D3-brane $N_{D5} = N_{F1}$.
Remember also that the Einstein frame radius $L$ is given by $L^4 = 4\pi N {\alpha'}^2$, and $e^{\phi_0}=\sqrt{g_s}$.
Thus we find that formula \eqref{sugraProbeD3} is exactly equivalent to \eqref{DBIProbeD3}.
We conclude that our bulk analysis reproduces exactly the shape of the BIons previously derived using the DBI action \cite{Callan:1997kz,Gauntlett:1999xz,Drukker:2005kx}.  We note that this result is perhaps a bit surprising as the supergravity solution is strongly curved when the brane charges are small.  As a result one might expect that there are large $\alpha^\prime$ corrections and that these corrections are necessary in order for the supergravity and probe descriptions to match.  However, despite the strongly curved nature of the solutions, they reproduce exactly the DBI results.  This fact is most likely due to the large amount of supersymmetry the geometries possess and suggests that the geometries are not strongly corrected even when the D3-brane and D5-brane charges are small.

\paragraph{Beyond the probe limit.}
The supergravity solutions are valid beyond the probe limit.  Additionally, since the supergravity solutions reproduce correctly the DBI description of the probe branes, one may use the supergravity solutions to write, order by order, the corrections to the DBI result.
First we give the next order corrections to the Weierstrass function \eqref{WeierstrassLim1a}:
\begin{align}\label{WeierstrassLim1aa}
\wp(x) \approx &-\frac{\pi^2}{12 \omega_3^2}\left(1+\frac{3}{\sinh^2\left(\frac{i\pi}{2\omega_3}x\right)}\right) \cr
&
- \frac{\pi^2 \sin \left( \frac{\pi x}{\omega_3} \right)}{4 |\omega_3|^2 \sin^4 \left( \frac{\pi x}{2 \omega_3} \right) } e^{- \frac{2 \pi \omega_1}{|\omega_3|}} \bigg[6 \pi x - 8 \omega_3 \sin \left( \frac{\pi x}{\omega_3} \right) + \omega_3 \sin \left( \frac{2 \pi x}{\omega_3} \right) \bigg].
\end{align}
Using this expression, one can derive the first order corrections to the D3-brane charge formula:
\begin{align}
\label{correctedD3charges}
Q_{D3}^{(0)} &= -4 \pi^3 L^4 \ , \cr
Q_{D3}^{(1)} &= \frac{4 \pi^3 L^4}{|\omega_3|} e^{-2 \pi \left| \frac{\omega_1}{\omega_3} \right|} \bigg[ 6 \pi - 7 |\omega_3| \sinh \left( \frac{2 \pi}{|\omega_3|} \right) + 2 |\omega_3| \sinh \left( \frac{4 \pi}{|\omega_3|} \right) \bigg] \ , \cr
Q_{D3}^{(2)} &= 4 \pi^3 L^4 - \frac{4 \pi^3 L^4}{|\omega_3|} e^{-2 \pi \left| \frac{\omega_1}{\omega_3} \right|} \bigg[  6 \pi - 7 |\omega_3| \sinh \left( \frac{2 \pi}{|\omega_3|} \right) + 2 |\omega_3| \sinh \left( \frac{4 \pi}{|\omega_3|} \right) \bigg] .
\end{align}
Note the exponential suppression as compared to the probe D5-brane case (\ref{correctedD5charge}).
The probe-D3 profile gets a non-zero thickness when we take into account these corrections.
We leave the study of this effect for future work.

%%%%%

\subsection{The probe D5-brane limit}

\begin{figure}[t]
\centering
\includegraphics[width=1\linewidth]{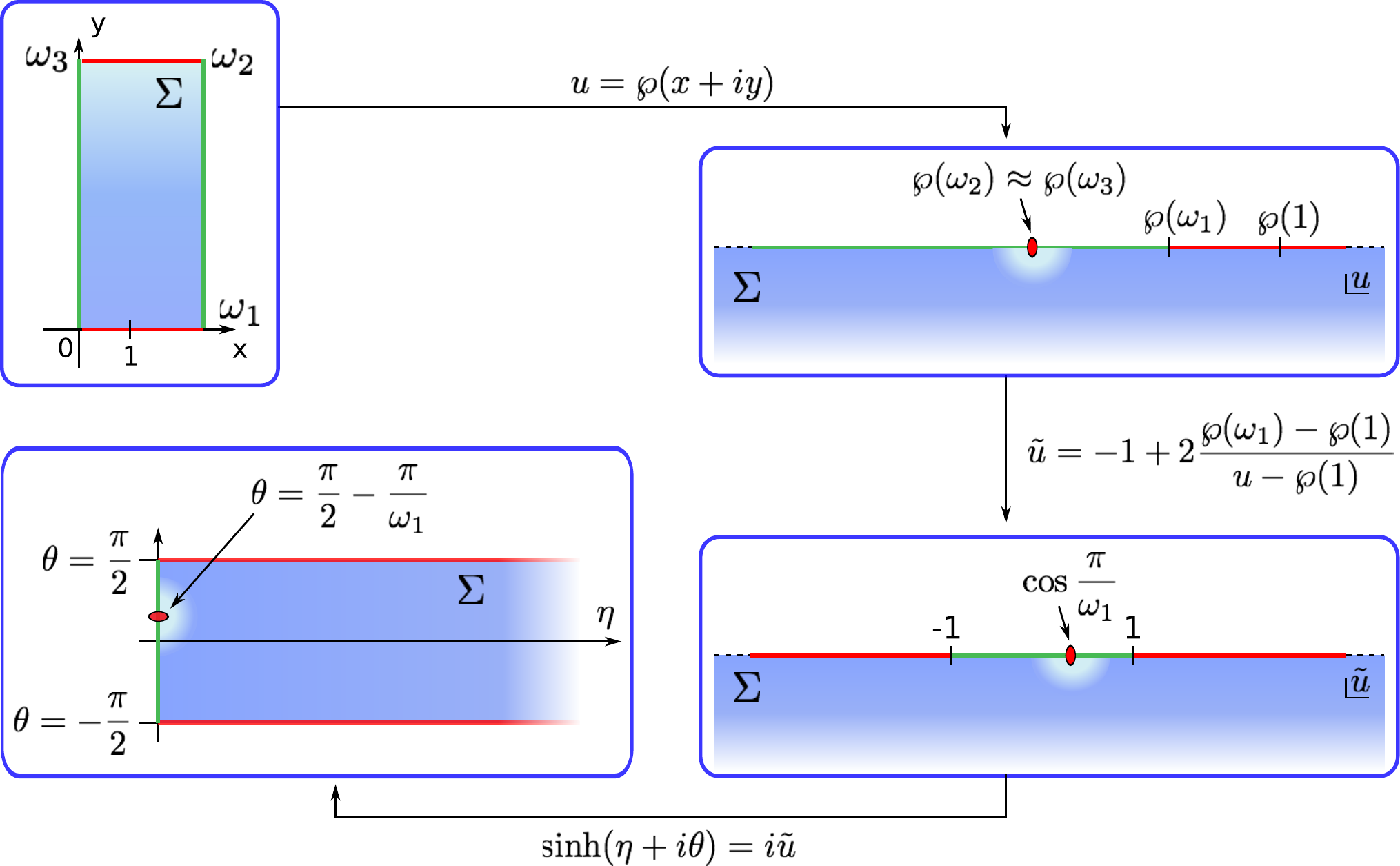}
\caption{Map from the rectangle associated to the genus-one solution (top-left) to the half-infinite rectangle associated to the genus-zero solution (bottom-left), in the limit $\omega_3 \to i\infty$. The geometry differs from $AdS_5 \times S^5$ in the light-blue region. The red oval gives the position of the probe D5-branes.}
\label{dgenus1ProbeD5}
\end{figure}

We now consider the limit $\omega_3 \to i \infty$ with $\omega_1$ held fixed. In this limit the D5-brane charge $Q_{D5}^{(1)}$ goes to zero, while the D3-brane charges $Q_{D3}^{(1)}$ and $Q_{D3}^{(2)}$ remain finite.
The corresponding Young tableau is almost vertical.
In the limit $\omega_3 \to i \infty$ the Weierstrass functions $\wp$ and $\zeta$ simplify to \cite{Bateman}:
\begin{align}\label{WeierstrassLim2a}
\wp(x) \approx \frac{\pi^2}{12 \omega_1^2}\left(-1+\frac{3}{\sin^2\left(\frac{\pi}{2\omega_1}x\right)}\right)\cr
\zeta(x) \approx \frac{\pi^2}{12 \omega_1^2}x + \frac{\pi}{2 \omega_1}\cot\left(\frac{\pi}{2\omega_1}x\right)
\end{align}
and we have in this limit:
\be\label{WeierstrassLim2b} \frac{1}{2}\wp(\omega_1) \approx -\wp(\omega_2) \approx -\wp(\omega_3) \ee
In terms of the Riemann surface $\Sigma$ of Figure \ref{genus1} (\emph{left}), the limit $\omega_3 \to i \infty$  amounts to shrinking the segment $[e_3,e_2]$ down to zero size.
In this limit the geometry reduces smoothly to $AdS_5 \times S^5$ \cite{D'Hoker:2007fq}.
For large but finite $|\omega_3|$, we expect the geometry to be  $AdS_5 \times S^5$ everywhere except for $u\approx e_{2,3}$ (or equivalently $y \approx |\omega_3|$), which is the region where the probe D5-branes are located.
To identify the shape of this region we follow the same strategy as in the probe D3-brane case:
we map the rectangle of Figure \ref{genus1} (\emph{right}) to the half-infinite rectangle of Figure \ref{genus0} (\emph{right}).
This mapping is described in Figure \ref{dgenus1ProbeD5}.
This time, the segments $[0,\omega_3]$ and $[\omega_2,\omega_1]$ both carry a non-negligible amount of D3-brane charge, while the segment $[\omega_3,\omega_2]$ carries almost no D5-brane charge. Consequently we demand that the combination of segments $[0,\omega_3] \bigcup [\omega_3,\omega_2] \bigcup [\omega_2,\omega_1]$ be mapped to the left-border of the half-infinite rectangle that carries the D3-brane charge in the genus-zero solution.

First we map the rectangle parametrized by $x+iy$ to the lower-half plane parametrized by $u$ thanks to the Weierstrass $\wp$ function:
\be u = \wp(x+iy) \ee
Then we make a conformal transformation $u \to \tilde u$ so that the asymptotic region lies now at infinity: $u=\wp(1) \to \tilde u = \infty$. We also demand that the (punctured) segment  $]-\infty,\wp(\omega_1)]$ that carries the D3-brane charge is mapped to the segment $\tilde u \in [-1,1]$. The explicit form of the mapping is:
\be \tilde u = -1 + 2\frac{\wp(\omega_1)-\wp(1)}{u-\wp(1)} \ee
Finally we map the lower half plane to the half-infinite rectangle parametrized by $\eta$ and $\theta$:
\be \sinh(\eta + i \theta) = i \tilde u \ee
Using formulas \eqref{WeierstrassLim2a}, \eqref{WeierstrassLim2b} we see that the genus-one harmonic functions \eqref{hsGen1} are mapped to the genus-zero harmonic functions \eqref{hsAds5}.
Additionally the segment $[\omega_3,\omega_2]$ that carries the D5-brane charge is mapped to a point on the boundary of the rectangle with coordinates $(\eta_{D5},\theta_{D5})$ given by:
\be\label{coordProbeD5} \eta_{D5} = 0 \ , \ \theta_{D5}=\frac{\pi}{2} - \frac{\pi}{\omega_1} \ee
We identify this point with the locus of the probe D5-branes.

To compare this result with the DBI result \eqref{thetaD5}, we compute the value of the D3-brane charges in the limit $\omega_3 \to i \infty$. Using the explicit expression for these charges given in \eqref{QD3g1} and formulas \eqref{WeierstrassLim2a},\eqref{WeierstrassLim2b} (see also appendix \ref{app:D3charge}), we obtain:
\begin{align}\label{3chargeProbeD5}
Q_{D3}^{(0)} &= -4 \pi^3 L^4 \cr
Q_{D3}^{(1)} &= 4 \pi^3 L^4 \bigg(\frac{1}{\omega_1} -\frac{1}{2\pi}\sin \left( \frac{2 \pi}{\omega_1} \right) \bigg) \cr
Q_{D3}^{(2)} &= 4 \pi^3 L^4 \bigg(1 - \frac{1}{\omega_1} + \frac{1}{2\pi}\sin \left( \frac{2 \pi}{\omega_1} \right) \bigg) \ .
\end{align}
Note that the period $\omega_1$ is bounded from below by $1$.  As a result the D3-brane charges $Q_{D3}^{(1)}$ and $Q_{D3}^{(2)}$ are bounded by $|Q_{D3}^{(0)}|$ as expected.
Combining now equations \eqref{coordProbeD5} and \eqref{3chargeProbeD5}, we observe that the D3-brane charges are related to the angle $\theta_{D5}$ as:
\begin{align}\label{sugraThetaD5b}
\theta_{D5} + \sin \theta_{D5} \, \cos \theta_{D5} = \frac{\pi}{2} + \pi \frac{Q_{D3}^{(1)}}{Q_{D3}^{(0)}}
\end{align}
In terms of the number of D3-branes, this gives:
\begin{align}\label{sugraThetaD5}
\theta_{D5} + \sin \theta_{D5} \, \cos \theta_{D5} = \frac{\pi}{2} - \pi \frac{N_{D3}^{(1)}}{N}
\end{align}
Remember that $N_{D3}^{(1)}$ is the number of fundamental strings dissolved in each D5-brane. In the limit where the number of D5-branes is equal to one, we have $N_{D3}^{(1)}=N_{F1}^{(1)}$. We conclude that formula \eqref{sugraThetaD5} matches exactly with the DBI result \eqref{thetaD5}.

It is interesting to discuss the meaning of the large gauge transformation in this context. First let us consider the case where $Q_{D3}^{(1)}$ is small and $|Q_{D3}^{(0)}| \approx |Q_{D3}^{(2)}|$. We see from \eqref{sugraThetaD5b} that the angle $\theta$ is approximatively equal to $\pi/2$, meaning that the  probe D5-brane is sitting close to the North pole of the five-sphere. This situation is well understood in the canonical gauge
%added:
where $N_{F1}^{(1)}$ is small and $N_{F1}^{(2)}$ is equal to zero: 
we have a small number of fundamental strings sitting on the North pole of the sphere, and they polarize accordingly into a small D5-brane.
Let us now consider the case where $Q_{D3}^{(2)}$ is small while $|Q_{D3}^{(1)}| \approx |Q_{D3}^{(0)}|$. Now the angle $\theta$ is approximatively equal to $-\pi/2$, meaning that the probe D5-brane is sitting close to the South pole of the five-sphere. This configuration is best understood in a gauge such that $N_{F1}^{(1)}$ is equal to zero and $N_{F1}^{(2)}$ is a small (negative) number: we have a few anti-strings sitting on the South-pole of the sphere, that polarize into a small D5-brane.

Similar to the probe D3-brane limit, one may infer the corrections to the probe approximation.  With this view in mind, we give the first order correction to the D5-brane charge:
\begin{align}
\label{correctedD5charge}
Q_{D5}^{(1)} &= - 4 \pi e^{-\phi_0} L^2 \frac{\omega_1}{|\omega_3|} \sin \left( \frac{\pi}{\omega_1} \right) \ .
\end{align}

%%%%%%%%%%%%%%%%%%%%%%%%%%%%%%%%%%%%%%%

\section{Gravitational interactions between open strings on backreacting D3-branes}\label{sec:potential}

In this section we study the gravitational interaction between open strings ending on backreacting D3-branes.
Open strings also exchange other modes of the closed strings, in particular the NSNS 2-form and the dilaton.
We focus on the graviton since the other forces are presumably irrelevant in phenomenological set-ups.
The implications of this computation for phenomenology are discussed in \cite{BE}

%%%%%

\subsection{Generalities}

\begin{figure}[t]
\centering
\includegraphics[width=0.7\linewidth]{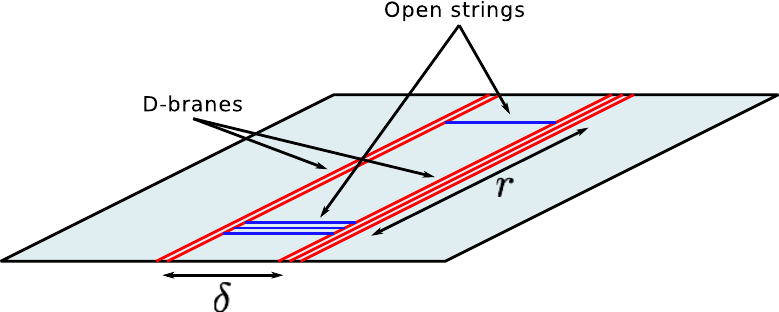}
\caption{We study the gravitational attraction between a stack of open strings and a single string, separated by a distance $r$. The strings are stretched between parallel stacks of D3-branes.}
\label{gravityProbe}
\end{figure}

\paragraph{Probe brane approximation.}
First we compute the gravitational potential between open strings ending on probe D3-branes. The result is rather trivial but the purpose is to illustrate the generic strategy we will adopt when the D-brane backreaction is taken into account.
We consider a stack of $N_{F1}$ fundamental strings stretched between two parallel stacks of (probe) D3-branes in ten-dimensional Minkowski spacetime (see Figure \ref{gravityProbe}). The distance between the D-brane stacks is denoted by $\delta$, so that the inertial mass of each string is equal to $\delta/2\pi\alpha'$.
We consider another fundamental string stretching between the two stacks. The distance between the single string and the stack of strings is denoted by $r$. We want to compute the gravitational attraction that the single string feels due to the stack of strings.
To this end we evaluate the Nambu-Goto action for the single string in the gravitational field created by the stack of strings.
At large distance, that is for $r \gg \delta$, the spacetime metric resulting from the gravitational backreaction of the stack of fundamental strings is well approximated by the 10-dimensional Schwarzschild metric:
\be ds^2 = -\left(1-2G_N^{(10)}\frac{N_{F_1}\delta}{2\pi \alpha'}\frac{1}{r^7}\right) dt^2 + \left(1-2G_N^{(10)} \frac{N_{F_1}\delta}{2\pi \alpha'}\frac{1}{r^7}\right)^{-1} dr^2 + r^2 d\Omega_8^2 \ee
where $G_N^{(10)}$ is the ten-dimensional Newton's constant.
Next we evaluate the Nambu-Goto action for the single string. We obtain:
\begin{align}
S_{NG} &= -\frac{1}{2\pi \alpha'} \int d\tau d \sigma \sqrt{-\det(g_{ab} \p_i x^a \p_j x^b)}\cr
& = -\int d\tau \left( \frac{\delta}{2\pi\alpha'} + E_{probe}(r) \right)
\end{align}
where $E_{probe}(r)$ is the gravitational potential. In the large $r$ limit the potential reads:
\be\label{Eprobe} E_{probe}(r) = -G_N^{(10)}\frac{N_{F_1}\delta}{2\pi \alpha'}\frac{\delta}{2\pi \alpha'}\frac{1}{r^7} \ee
Note that $\frac{N_{F_1}\delta}{2\pi \alpha'}$ and $\frac{\delta}{2\pi \alpha'}$ are respectively the inertial masses of the stack of strings and of the single string. So we simply re-derived Newton's law in ten dimensions. As expected, strings on probe D-branes behave like point-particles at large distances. We will see that this is not the case when the D-brane backreaction is taken into account.

\begin{figure}[t]
\centering
\includegraphics[width=1\linewidth]{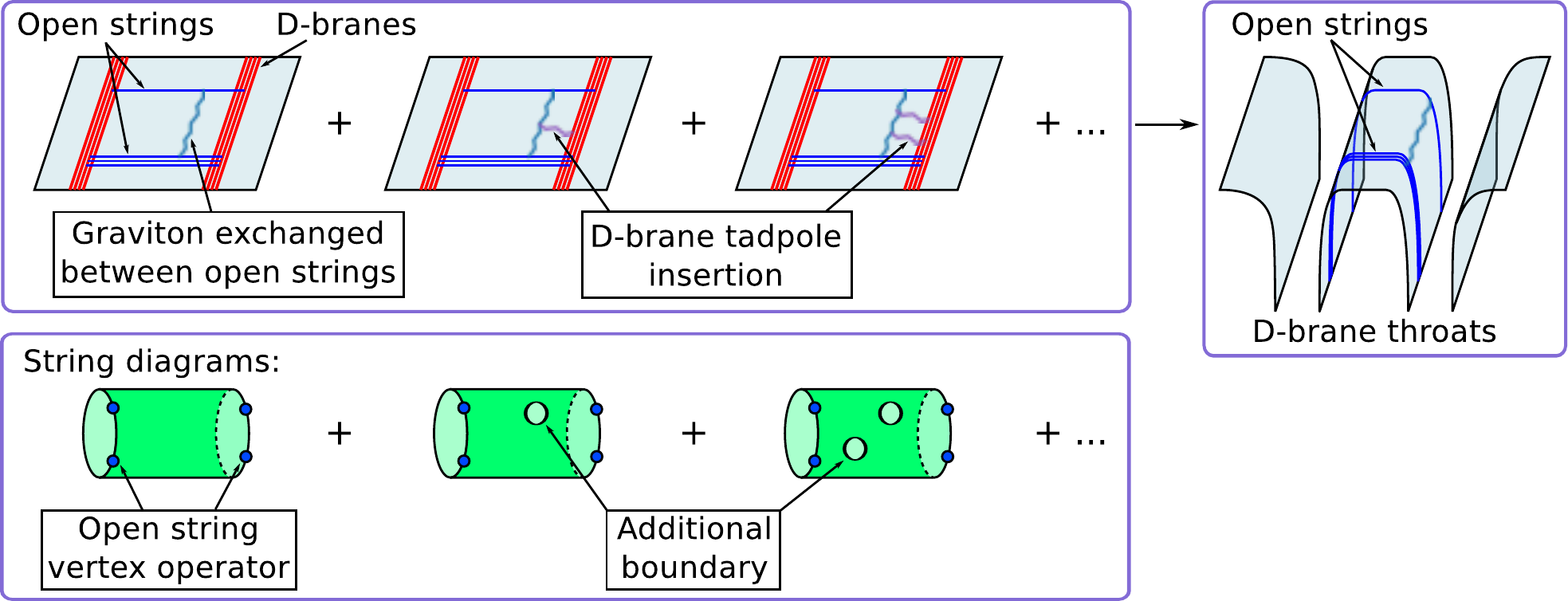}
\caption{\emph{Top:} The graviton exchange between open strings receives corrections from D-brane tadpole insertions.
\emph{Bottom:} At the level of the worldsheet the tadpole insertions correspond to the additions of extra boundaries on the worldsheet.
\emph{Right:} To take into account the full D-brane backreaction, we compute the graviton exchange in the backreacted D-brane background.}
\label{gravityTadpoles}
\end{figure}

\paragraph{The relevance of the D-brane backreaction.}
The D-brane backreaction takes the form of non-vanishing tadpoles for the closed-string modes and in particular for the graviton.
D-brane tadpole insertions affect the graviton exchange between open strings (see Figure \ref{gravityTadpoles}).

On the open string worldsheet, a D-brane tadpole insertion corresponds to the addition of an extra boundary to the worldsheet, with the proper D-brane boundary conditions.
Adding a boundary to the worldsheet changes the genus by one unit, and thus the string diagrams receive additional factors of $g_s$. Taking into account the Chan-Paton indices adds another factor of $N$.
Thus a tadpole insertion comes with a factor of $g_s N$.
This might lead one to believe that the D-brane backreaction is negligible as soon as $g_s N$ is small.
However we would like to point out a possible loophole in this argument. Let us consider the D3-brane metric in 10D Minkowski spacetime:
\be\label{metricD3} ds^2 = H(u)^{-\frac{1}{2}} dx_\mu dx^\mu + H(u)^{\frac{1}{2}}\left(du^2 + u^2 d\Omega_{(5)}^2\right) \ee
where $H(u) = 1+L^4/u^4$, and $L$ is given by $L^4 = 4\pi g_s N {\alpha'}^2$. The coordinate $u$ is  the radial coordinate\footnote{One should not confuse this radial coordinate with the complex coordinate we used previously to parametrize the upper-half plane.} away from the D3-branes, that are sitting at $u=0$. From this metric it is clear that the backreaction of the D3-branes is negligible as soon as $L^4/u^4 \ll 1$ and $H(u) \approx 1$, namely  for $g_s N \ll u^4/{\alpha'}^2$. Even if $g_s N$ is large, the backreaction is always negligible far enough from the D3-branes (that is for large $u$). Conversely, even if $g_s N$ is small, the backreaction is always important very close to the D3-branes (that is for small $u$).
The right parameter to estimate the relevance of the D3-brane backreaction is really  $g_s N {\alpha'}^2/ u^4$, rather than $g_s N$.
There always exists segments of the open strings which are very close to the D-brane where the backreaction is never expected to be small. Thus it is possible that the D-brane backreaction has a noticeable affect on the open string physics, even at small $g_s N$.
Notice also that $u < \sqrt{\alpha'}$ does not mean that the distance away from the D3-branes is smaller than the fundamental string scale. Actually the proper distance between any point of the bulk with $u\neq 0$ and the D3-branes at $u=0$ is infinite, because of the infinite redshift at the horizon.
Of course we should not forget that for small $g_s N$ the geometry \eqref{metricD3} is strongly curved and thus $\alpha'$ corrections should not be neglected. Unfortunately the full string theory in the background \eqref{metricD3}  (supported by RR fluxes) is still poorly understood, so we will not attempt to discuss such $\alpha'$ corrections.

In the following we will assume that $g_s N$ is large so that the supergravity description of the bulk gravity is reliable. To take into account the full backreaction of the D3-branes, we compute the gravitational potential between the open strings in the background of the D3-branes \eqref{metricD3} (see Figure \ref{gravityTadpoles}, \emph{right}).
Such elongated open strings have been considered previously in the literature (see e.g. \cite{hep-th/9602043,Rey:1998ik,Maldacena:1998im,hep-th/0204196,hep-th/0604123}), but a lot of works remains to be done to fully understand the physics of these strings.

\begin{figure}[t]
\centering
\includegraphics[width=1\linewidth]{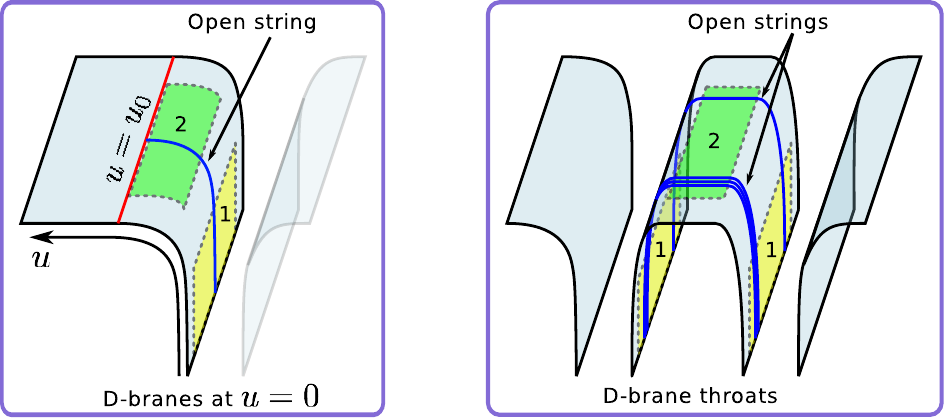}
\caption{\emph{Left:} A string stretched between a probe brane at $u=u_0$ and a backreacting stack of D-branes at $u=0$. The section of the string in region 2 carries most of the inertial mass of the string. The section of the string in region 1 is infinitely long, but carries almost no mass.
\emph{Right:} The configuration of Figure \ref{gravityProbe}, once the D-brane backreaction is taken into account.
When computing the gravitational potential, we find that the sections of the strings contained in region 2 essentially provide the usual Newtonian potential. Nevertheless, we show that the sections of the strings contained in regions 1 provide a non-negligible contribution.}
\label{inertialMass}
\end{figure}

\paragraph{The open string inertial mass.}
Let us consider a single open string in the D3-brane background \eqref{metricD3} (see Figure \ref{inertialMass},\emph{ left}). For simplicity we assume that this string is straight: it is localized in all spacelike directions expect for the radial direction $u$. It extends from $u=u_0$ up to $u=0$. Demanding that the string extends up to the horizon at $u=0$ is equivalent to demanding that one endpoint of the open string is attached to the backreacting D3-branes. We can think of the other endpoint as being attached to a probe D3-brane located at $u=u_0$.
The inertial mass of this string can be easily evaluated by integrating the (redshifted) string tension along the string. This is equivalent to evaluating the Nambu-Goto action for the string. We find:
\be S_{NG} = -\frac{1}{2\pi \alpha'} \left(\int dt\right)  m \ee
with the inertial mass $m$ given by:
\be m = u_0 \alpha'.\ee
This is in agreement with the probe brane picture.
Notice however that the proper length of the string is now infinite:
\be \int_{u=0}^{u_0} ds =  \int_{u=0}^{u_0} \left( 1 + \frac{L^4}{u^4}\right)^{\frac{1}{4}} = \infty .\ee
The inertial mass is finite because the string tension is infinitely redshifted near the horizon.
This implies that the section of the string very close to the horizon (region 1 in Figure \ref{inertialMass}), despite its infinite length, carries a very small part of the total inertial mass of the string.
Most of the inertial mass of the string is localized far away from the horizon (region 2 in Figure \ref{inertialMass}), in the region where the redshift is smallest.

Our purpose is to compute the gravitational potential between one stack of fundamental strings and one single string.
Let us consider for instance the configuration of Figure \ref{inertialMass}, \emph{right}.
The sections of the strings contained in region 2 provide the usual Newtonian contribution to the gravitational potential. Indeed they carry most of the inertial mass of the strings. In ten-dimensional Minkowski spacetime, this contribution is essentially the potential \eqref{Eprobe} that we computed in the probe brane approximation.
In the following we will show that the sections of the strings that are located closer to the horizon in region 1 provide a non-negligible contribution to the gravitational potential.

\begin{figure}[t]
\centering
\includegraphics[width=1\linewidth]{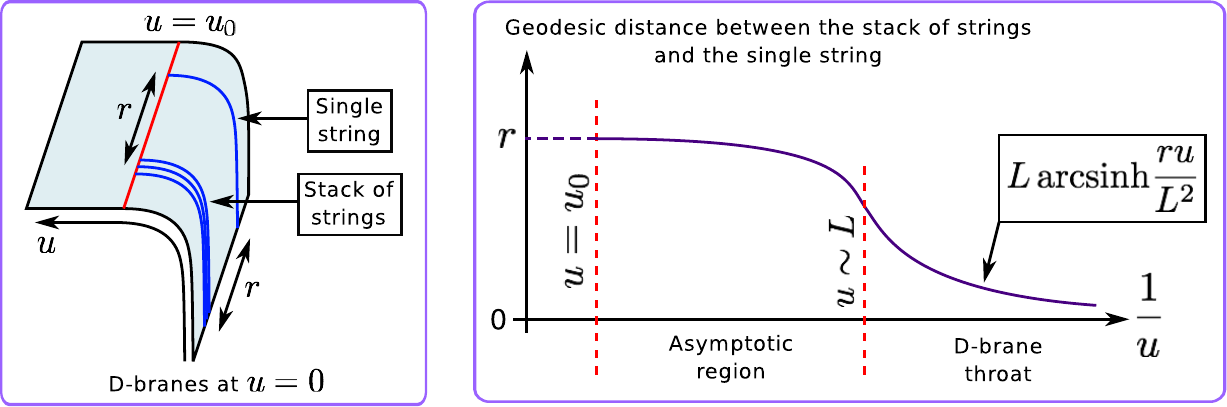}
\caption{\emph{Left:} The coordinate distance $r$ between the stack of strings and the single string is constant.
\emph{Right:} On the other hand the geodesic distance between the stack of strings and the single string goes to zero near the horizon of the D3-branes.}
\label{geodesicDistance}
\end{figure}

\paragraph{The distance between the strings.}
We consider now a stack of straight strings located at $r=0$ in the background \eqref{metricD3}, and a single straight string located at constant $r$ (see Figure \ref{geodesicDistance}, \emph{left}).
We assume that the distance $r$ between the stack and the single string is much larger than all the other length scales in the problem.
The coordinate distance $r$ is the distance that an observer living on the D-branes would measure.
More generally, it is natural to identify the four-dimensional Minkowski metric $dx_\mu dx^\mu$ that appears in \eqref{metricD3} with the metric that an observer on the branes would use to measure distances.
In the asymptotic flat region, the coordinate distance $r$ coincides with the geodesic distance between the stack of strings and the single string.
However, as we get closer to the D3-brane horizon, the geodesic distance between the stack of strings and the single string goes to zero (see Figure \ref{geodesicDistance}, \emph{right}).
Let us evaluate this geodesic distance more precisely.
As we look at the geometry close to the D3-brane horizon, it is convenient to introduce the coordinate $z$ related to $u$ as: $z=L^2/u$.
Then the geometry \eqref{metricD3} reduces to $AdS_5 \times S^5$ in the usual Poincare coordinates.
The geodesic distance between a point of the single string with coordinates $(r,z)$ and the stack at $r=0$ is equal to (see appendix \ref{app:paraAdS}):
\be\label{geoDistStrings} L\, \mathrm{arcsinh}\frac{r}{z} \approx_{z\to \infty} L\frac{r}{z} \ee

Formula \eqref{geoDistStrings} implies that the geodesic distance between the stack of strings and the single string goes to zero close to the horizon,  even if $r$ is very large.
This suggests that the sections of the strings closer to the horizon provide a larger contribution to the gravitational potential than may have been expected.
At the same time the strings' tension is also redshifted close to the horizon. To really understand the contribution to the gravitational potential coming from the near-horizon region, a detailed computation has to be performed.

\paragraph{Zoom on the near-horizon region.}
To compute the contribution  to the gravitational potential coming from the section of the strings closest to the horizon, we work in the near-horizon geometry of the D3-branes.
In the following we consider strings in Poincare $AdS_5 \times S^5$, that extend up to the Poincare horizon where the D3-branes are sitting.
We consider straight strings only for simplicity, that are extended only along the $z$ direction in the Poincare coordinates \eqref{PoincareAdS}.
For a straight string that extends from the Poincare horizon (at $z=\infty$) up to a point $z=z_0$, the inertial mass is equal to:
\be m = \frac{1}{2 \pi \alpha'} \frac{\sqrt{g_s}L^2}{z_0} . \ee
If the string extends up to the AdS boundary (i.e. $z_0=0$), the inertial mass is infinite.

Since we are going to perform a computation in the near-horizon of the D3-branes, our results hold independently of the details of the asymptotic geometry. The contribution to the gravitational potential that we will obtain is universal.
Additionally, we expect our conclusions to apply to oscillating strings as well. Indeed the wavelength of any oscillation is infinitely redshifted near the horizon, and all strings appear straight in this region.
This is also valid for oscillating strings extended in the bulk with both endpoints on the backreacting D-branes.

\begin{figure}[t]
\centering
\includegraphics[width=0.9\linewidth]{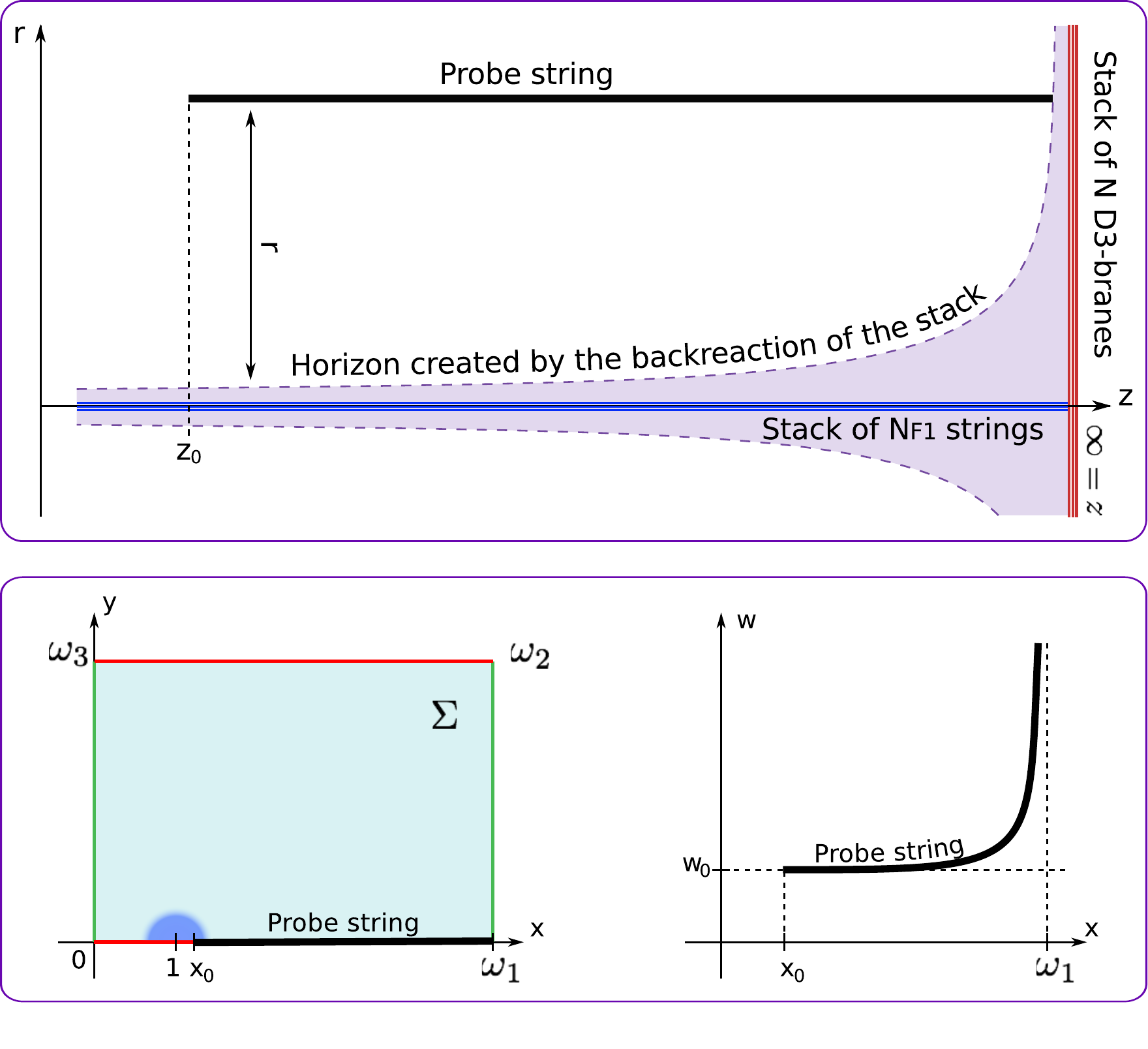}
\caption{\emph{Top}: We consider a stack of strings and a probe string in the near-horizon of the D3-branes. To compute the gravitational force felts by the single string, we have to backreact the stack of strings.
\emph{Bottom:} We embed the single string in the supergravity background describing the backreacting stack of strings.}
\label{choicePath}
\end{figure}

\paragraph{The choice of path.}
We consider a stack of straight strings in $AdS_5 \times S^5$  located at $r=0$ and extended along the $z$ direction up to the horizon, in the Poincare coordinates. Additionally a probe string is located at constant $r$ and extends up to the horizon (see Figure \ref{choicePath}, \emph{top}).
All strings are sitting at the same point of the five-sphere. This is the case for instance if all the strings are stretched between two stacks of D3-branes as in Figure \ref{gravityProbe}, since the position on the $S^5$ in the near-horizon can be understood as the angle along which the strings extend away from the backreacting stack in the asymptotic region.

To evaluate the gravitational potential between the stack of fundamental strings and the single string, we evaluate the Nambu-Goto action for the probe string in the background resulting from the backreaction of the stack of open strings.
In the near-horizon of the D3-branes, this background is given by the supergravity solution of \cite{D'Hoker:2007fq} that we discussed in the previous sections.
For simplicity we focus on the genus-one solution corresponding to a stack of fundamental strings in a representation labeled by a rectangular Young tableau.
Remember that we can compute two independent D3-brane charges $Q_{D3}^{(1)}$ and $Q_{D3}^{(2)}$ in this solution.
We assume $Q_{D3}^{(1)} \ll Q_{D3}^{(2)}$.
Consequently we can without ambiguity associate the segment $]-\infty, e_3]$ on the boundary of the Riemann surface $\Sigma$ (see Figure \ref{genus1}, \emph{left}) with the stack of D3-branes (see section \ref{sec:probeLim}). Indeed this segment carries most of the D3-brane charge.
The segment $[e_2, e_1]$ is associated to the D3-brane spikes that are formed because of the presence of the stack of fundamental strings.

We now have to embed the probe string into the geometry \eqref{metricWilsonLine}.
A priori there is no rule that fixes this embedding. This ambiguity essentially results from the diffeomorphism invariance of gravitational theory.
More generally, we have to keep in mind that the gravitational potential that we want to compute is not a proper observable since it is not invariant under diffeomorphisms.
It is well-known that there are no local observables in a theory of gravity.
In order to define a gravitational potential, one has to make a comparison of the deformed geometry with the undeformed case, in a canonical choice of coordinates defined by a choice of observer.  Our strategy will be to specify a set of mild conditions that the embedding of the probe string in the geometry \eqref{metricWilsonLine} has to satisfy.  These conditions will be enough to fix completely the gravitational potential, up to a numerical coefficient.

We assume that the probe string lies on the boundary of the Riemann surface $\Sigma$, on the segment $[1, \omega_1]$ of the fundamental domain of the Weierstrass functions (see Figure \ref{choicePath}, \emph{bottom left}).
The four-sphere factor of the metric  \eqref{metricWilsonLine} vanishes on this segment, so this choice amounts to demanding that the backreacting stack of fundamental strings and the probe string sit at the same point of the five-sphere in $AdS_5 \times S^5$.
The string extends along the $x$ direction up to the point $x=\omega_1$.
It is localized in all other spacelike directions except for the $AdS_2$ radial direction $w$.
At the point $x=\omega_1$ the string reaches the horizon at $w=\infty$  (see Figure \ref{choicePath}, \emph{bottom right}).
This amounts to demanding that the open string is attached to the backreacting stack of D3-branes.
The coordinates of the other endpoint of the string are denoted by $(x_0, w_0)$.
We assume that this endpoint is lying in the asymptotic region where the geometry is $AdS_5\times S^5$.
We can think of this endpoint as being attached to a probe D3-brane.
In the asymptotic $AdS_5\times S^5$ region, we can use the change of variables described in appendix \ref{app:paraAdS} to write the coordinates of the endpoint $(x_0, w_0)$ in terms of the Poincare coordinates $(z_0,r)$:
\be\label{w0(r)} w_0 = \sqrt{r^2 + z_0^2} \ee
\be\label{x0(r)} x_0 = \frac{z_0}{r} \left(1+\sqrt{1+\frac{r^2}{z_0^2}}\right) \ee
Notice that for a given $z_0$ (that is, for a fixed mass of the probe string), $x_0$ goes to 1 as $r$ goes to infinity. Consequently the assumption that the endpoint lies in the asymptotic $AdS_5\times S^5$ region is always justified when $r$ is large enough.

The profile of the string is given by a function $w(x)$, with $w(x_0)=w_0$ and $w(\omega_1)=\infty$.
We demand that the function $w(x)$ is strictly increasing, which means that the probe string does not turn back.
Additionally, an important condition is that the probe string profile reduces to the straight string profile of Figure \ref{choicePath} (\emph{top}) in the limit where the number of fundamental strings in the stack goes to zero. In that limit the backreaction of the stack fades away and the spacetime reduces to $AdS_5\times S^5$. This condition can be written as:
\be\label{straightLim} \lim_{N_{F1}^{(1)}\to 0} w(x) = r \frac{x^2+1}{2x} \ee
The right-hand side of the previous equation is the profile of a straight string in the parametrization \eqref{paramAdSWeierstrass} of $AdS_5$ (see appendix \ref{app:paraAdS}).

A possible choice of path that satisfies all these conditions is the geodesic emanating from the boundary and reaching the horizon.
The analytic expression for these geodesics is given in appendix \ref{app:geodesic}.
There is another natural choice for the path.
We know that parallel strings ending on D3-branes form a BPS configuration.
So there exists a path in the geometry \eqref{metricWilsonLine} such that a string following this path does not break any supersymmetry.
We can also chose to put the probe string along this path\footnote{A string following this path with the right orientation would not feel any attraction from the stack of strings, as the gravitational attraction would be canceled by other interactions. We are mostly interested in a string with the opposite orientation so that the forces do not cancel. More generally, we only consider gravity since the other bulk interactions are presumably irrelevant in a phenomenological context.}.  It is important to note that the path of minimal action does not satisfy the last condition \eqref{straightLim} that we spelled out.  This is because the minimal action configuration corresponds to the case when the strings join and so do not extend up to the horizon.  In fact, this path is relevant for the holographic computation of the gauge potential \cite{Rey:1998ik,Maldacena:1998im}, as we discuss further in section (\ref{sec:holgcomp}).

%%%%%

\subsection{The $1/r$ behavior of the potential}\label{sub:1/r}

On general grounds, we expect the Nambu-Goto action for the probe string to take the following form:
\be S_{NG} = \int d\tau \int d\sigma  \mathcal{L}_{NG} = - \left( m_0 + E(r) \right) \int d\tau \ee
where $\mathcal{L}_{NG}$ is the Nambu-Goto Lagrangian for the probe string in the geometry \eqref{metricWilsonLine}, $m_0$ is the inertial mass of the probe string, and $E(r)$ is the potential energy. The potential energy is expected to be negative since gravity is attractive.
In this section we show that the potential $E(r)$ goes like $1/r$ at large $r$.

\paragraph{Dimensional analysis.}
Let us begin with elementary dimensional analysis.
We use worldsheet diffeomorphism invariance to identify the worldsheet coordinates $(\tau,\sigma)$ with the spacetime coordinates $(t,x)$.
The Nambu-Goto action reads:
\begin{align}\label{SNG} S_{NG} = &   -\frac{\sqrt{g_s}}{2\pi \alpha'} \int dt \left[ \int_{x_0}^{\omega_1} dx \sqrt{ g_{tt} \left(g_{xx} + \left( \frac{\p w(x)}{\p x}\right)^2 g_{ww}\right)} \right]
\end{align}
The Nambu-Goto action is defined with the string-frame metric, which explains the additional overall factor of $\sqrt{g_s}$.
Let us consider the integral between square brackets in the previous expression.
This integral has the dimensions of a length.
We have three independent length scales in this problem: $L$, $r$ and $z_0$ that are respectively related to the curvature radius of the geometry, the distance between the strings and the inverse mass of the probe string\footnote{The string scale $\alpha'$ is not a relevant length scale in this problem.}.
First note that the radius $L$ appears in the metric \eqref{metricWilsonLine} as an overall factor only\footnote{We take the four parameters for the genus one solution as the radius $L$, the string coupling $g_s$ and the half-periods $\omega_1$ and $\omega_3$.}.
So we can factor out the radius dependence of the previous integral.
This provides an overall factor of $L^2$, and what remains is independent of $L$.
When $r \to \infty$, the Nambu-Goto action goes to a constant proportional to the inertial mass of the probe string $\sim 1/z_0$.
We deduce that we have the following expansion at large $r$:
\be\label{largerExpforSNG} \int_{x_0}^{\omega_1} dx \sqrt{ g_{tt} \left(g_{xx} + \left( \frac{\p w(x)}{\p x}\right)^2 g_{ww}\right)}
= L^2 \left( \frac{\#}{z_0} + \frac{\#}{r} + \frac{\# z_0}{r^2} + ... \right) \ee
where the $\#$'s are dimensionless coefficients that depend on the various charges in the problem.

The first term in the expansion \eqref{largerExpforSNG} gives the inertial mass of the string.
The subleading terms give the potential $E(r)$.
This suggests that the leading contribution to the potential energy goes like $1/r$ at large $r$.
Before we prove that the coefficient of this $1/r$ term is non-zero, let us make another remark about the expansion \eqref{largerExpforSNG}.
We see that when $z_0$ goes to zero, namely when the probe string goes up to the boundary, the potential has to go exactly like $1/r$.
This is in agreement with the four-dimensional scale invariance of the problem.
In the case $z_0=0$  the inertial mass of the string diverges and the Nambu-Goto action has to be regularized (for instance by subtraction of the inertial mass).
For a finite $z_0$, the probe string has a finite mass, and the scaling symmetry along the AdS radial direction is broken (or equivalently the four-dimensional scale invariance is broken). Then subleading corrections to potential energy are allowed.

\paragraph{The asymptotic contribution.}
In this paragraph we show that the coefficient of the $1/r$ term in the potential is non-zero.
More precisely we evaluate the Nambu-Goto action on the section of the probe string contained in the asymptotic region where the geometry \eqref{metricWilsonLine} reduces to a small perturbation of $AdS_5\times S^5$.
We will show that this section of the string provides a contribution to the potential that goes like $1/r$, with a non-vanishing negative coefficient.
We work with the asymptotic form of the metric \eqref{metricWilsonLine}.
In appendix \ref{App:asymptotics} it is shown that the leading correction to the $AdS_5$ metric is given by:
\be\label{asympAdSMetric} ds^2 = L^2 \frac{dz^2 + \left(1-h \frac{z^2}{r^2}\right) dx_\mu dx^\mu}{z^2} \ee
where $h$ is a positive coefficient that can be read off from equation \eqref{asympMetric}.
The asymptotic expansion is organized in powers of $z/r$ which is essentially the inverse geodesic distance away from the backreacting stack (see \eqref{geoDistStrings}).
The asymptotic form of the metric \eqref{asympAdSMetric} is reliable for $z/r$ small enough.
Next we compute the Nambu-Goto action for the section of the string lying in the region of spacetime where the geometry is well-approximated by the metric \eqref{asympAdSMetric}.
For definiteness we work in the region defined by $z < \alpha r$, where $\alpha$ is a number that is small enough.

First let consider a string lying at constant $r$  in the geometry \eqref{asympAdSMetric}. An open string is really expected to extend from $z=z_0$ up to $z=\infty$, but we focus on the section of the string for which $z < \alpha r$ and the metric \eqref{asympAdSMetric} is reliable.
For $r$ large enough, we have $z_0 < \alpha r $ and there is always a part of the string in the asymptotic region.
We use worldsheet diffeomorphism invariance to identify the worldsheet coordinates $(\tau,\sigma)$ with the spacetime coordinates $(t,z)$.
We evaluate the contribution to the Nambu-Goto action of the section of the string in the asymptotic region:
\begin{align}\label{compEAsympt} S_{NG} \supset &   -\frac{\sqrt{g_s}}{2\pi \alpha'} \int dt  \int_{z_0}^{\alpha r} dz \sqrt{ g_{tt} g_{zz}} \cr
 &  = -\frac{\sqrt{g_s}}{2\pi \alpha'} \int dt \int_{z_0}^{\alpha r} dz \sqrt{ \frac{L^4}{z^4}\left( 1-h \frac{z^2}{r^2} \right)}\cr
& = -\frac{\sqrt{g_s}}{2\pi \alpha'} \int dt \left(\frac{L^2}{z_0} - \frac{L^2}{\alpha r} - \frac{h\alpha }{2}\frac{L^2}{r} + \mathcal{O}\left(\frac{1}{r^2}\right)\right)
\end{align}
The first term in the previous result is the inertial mass of the string. The second part is the potential.
We see that the potential goes like $1/r$ at large $r$. The potential is negative, as expected since gravity is attractive.

Let us now consider a more general case:  we allow for a non trivial profile $r(z)$ for the string.
The Nambu-Goto action now reads:
\begin{align} &   -\frac{\sqrt{g_s}}{2\pi \alpha'} \int dt \int_{z_0}^{\alpha r} dz \sqrt{ g_{tt} \left(g_{zz} +\left( \frac{\p r(z)}{\p z}\right)^2 g_{rr}\right)}
 \end{align}
 We demand that close to the boundary $r(z)$ goes to a constant $r$, so in the asymptotic region we can expand $\p_z r(z)$ as a series in $z/r$:
\be \frac{\p r}{\p z} = \# \frac{z}{r} + \# \frac{z^2}{r^2} + ... \ee
where the $\#$'s stand for arbitrary coefficients.
Then it is straightforward to generalize the previous computation. We find that once again the $1/r$ term in the potential comes with a coefficient that is non-vanishing and negative.

We just evaluated the contribution to the potential of the section of the string contained in the region $z \in [z_0, \alpha r]$.
We saw that this contribution goes like $1/r$ at large $r$.
Let us now briefly argue that the contribution coming from the other part of the string contained in the near-horizon region $z > \alpha r $ cannot cancel the leading $1/r$ term in the potential.
When $r$ goes to infinity, the amount of inertial mass contained in the near-horizon region goes to zero.
Consequently the Nambu-Goto action evaluated in the near-horizon region $z > \alpha r $ goes like $\mathcal{O}(1/r)$.
Moreover the Nambu-Goto action is manifestly negative.
We deduce that the contribution of the near-horizon region to the potential may increase the magnitude of the $1/r$ contribution from the asymptotic region, but may never cancel it.

Actually the computation \eqref{compEAsympt} suggests that the contribution to the gravitational potential coming from the near-horizon region is  dominant with respect to the contribution from the asymptotic region. Indeed the potential obtained in \eqref{compEAsympt} diverges when the coefficient $\alpha$ gets large.
This simply follows from the fact that the perturbation of the $AdS_5$ metric, that is responsible for the non-zero potential, gets larger close to the horizon.
We deduce that an asymptotic computation similar to \eqref{compEAsympt} cannot lead to a good estimate of the coefficient of the $1/r$ term in the potential.
We will compute this coefficient using the full non-linear supergravity solution in section \ref{sub:coefE}

Finally let us make a short digression and compute the gravitational potential felt by a string that does not extend up to the horizon.
Let us consider a string that extends from $z=z_0$ up to $z=z_1$. If the distance $r$ is big enough, this string is lying entirely in the asymptotic region where the metric \eqref{asympAdSMetric} is reliable. Generalizing the computation \eqref{compEAsympt} we find that the gravitational potential felt by this string goes like $1/r^2$. This can be explained thanks to an elementary Newtonian analysis. The backreacting stack of fundamental strings polarize into a complicated configuration of D3- and D5-branes. The objects with highest dimensionality are the D5-branes, that give rise to the potential with the slowest decay at large distance. The gravitational potential created by D5-branes in ten dimensions indeed goes like $1/r^2$. This point is further discussed in appendix \ref{App:asymptotics} below equation \eqref{asympMetricB}.
We conclude that the $1/r$ behavior of the gravitational potential is only valid for open strings that extend up to the horizon created by the D-branes. For objects of finite size, the decay of the gravitational potential is faster.

\paragraph{The case of the geodesic path.}
In order to make the previous arguments more concrete, we now study an  example.
We consider a probe string following a spacelike geodesic that extends up to the horizon\footnote{
The proper length  of a spacelike geodesic that reaches the horizon is infinite.
The value of the Nambu-Goto action on such a section of geodesic is finite and provides an interesting measure of the distance away from the horizon.
}.
These geodesics are described in appendix \ref{app:geodesic}.
Let us explicitly show that the gravitational potential felt by this string indeed goes like $1/r$ at large $r$.

We parametrize the worldsheet of the string with the coordinates $(\tau, \sigma)$. We choose $\tau = t$ and $\sigma=x$. The Nambu-Goto action reads:
\begin{align} S_{NG} & = -\frac{\sqrt{g_s}}{2\pi \alpha'} \int dt \int_{x_0}^{\omega_1} dx \sqrt{-g_{tt}\left( g_{xx} + g_{ww} (\p_x w)^2 \right) } \cr
& = -\frac{\sqrt{g_s}}{2\pi \alpha'} \int dt \int_{x_0}^{\omega_1} dx  \frac{f_1(x)}{w(x) \sqrt{V(x)}}
\end{align}
The function $V(x)$ is defined in \eqref{defV} and the geodesic path $w(x)$ is given in \eqref{w(x)}.
First let us investigate the $r$-dependence of the Nambu-Goto Lagrangian $\mathcal{L}_{NG}$:
\be\label{LNGgeod} \mathcal{L}_{NG} = \frac{f_1(x)}{\sqrt{V(x)}}\frac{1}{w(x)} \ee
The first factor $ f_1(x)/\sqrt{-V(x)}$  depends only on $x$, and is manifestly independent of $r$.
The $r$-dependence of the Nambu-Goto Lagrangian \eqref{LNGgeod} is hidden in the factor $1/w(x)$. More precisely $r$ is related to the coordinates of the starting point of the probe string $(x_0,w_0)$  through equations \eqref{w0(r)}, \eqref{x0(r)}.
We assume $r \gg z_0$, which means that the distance between the strings is greater than the inverse masse of the probe string. In this limit \eqref{w0(r)} and \eqref{x0(r)} reduce to:
\be\label{w0(r)Lim} w_0 = r + \mathcal{O}\left(\frac{1}{r}\right) \ee
\be\label{x0(r)Lim} x_0 = 1+ \mathcal{O}\left(\frac{1}{r}\right)  \ee
Using the explicit expression \eqref{w(x)}  we rewrite $1/w(x)$ as:
\begin{align}\label{w(x)b} \frac{1}{w(x)}
& = \frac{1}{w_0} \exp \left( \int_{1}^{x_0}  \frac{C}{f_1^2} \frac{1}{\sqrt{V}}  \right) \exp \left( -\int_{1}^x  \frac{C}{f_1^2} \frac{1}{\sqrt{V}}  \right) \end{align}
The first factor in the previous expression becomes at large $r$:
\be \frac{1}{w_0} = \frac{1}{r} +  \mathcal{O}\left(\frac{1}{r^3}\right) \ee
Using \eqref{x0(r)Lim}, we find that the second factor in \eqref{w(x)b} simplifies in the large $r$ limit:
\be \exp \left( \int_{1}^{x_0}  \frac{C}{f_1^2} \frac{1}{\sqrt{V}}  \right)  = 1+ \mathcal{O}\left(\frac{1}{r}\right) \ee
Finally the third factor in \eqref{w(x)b} does not depend on $r$.
Thus at large $r$ the Nambu-Goto Lagrangian \eqref{LNGgeod} behaves like:
\be\label{LNGgeod(r)} \mathcal{L}_{NG} = \frac{1}{r}\frac{f_1(x)}{\sqrt{V(x)}}\exp \left( -\int_{1}^x  \frac{C}{f_1^2} \frac{1}{\sqrt{V}}  \right)  +\mathcal{O}\left(\frac{1}{r^2}\right)  \ee
where the $r$-dependence is now explicit.

Now we study the $r$-dependence of the potential energy.
The potential energy can be written as:
\be\label{E=L-m} E(r) = \frac{\sqrt{g_s}}{2\pi \alpha'} \int _{x_0}^{\omega_1} dx \mathcal{L}_{NG} - m_0 \ee
The inertial mass of the probe string $m_0$ is given by the value of the Nambu-Goto action for a probe string in $AdS_5 \times S^5$:
\be\label{m=intL} m_0 = \frac{\sqrt{g_s}}{2\pi \alpha'}\int_{x_0}^\infty dx  \mathcal{L}_{NG}^{(0)} \ee
where $\mathcal{L}_{NG}^{(0)}$ is the Nambu-Goto Lagrangian in $AdS_5 \times S^5$.
Here it is convenient to use the coordinate system defined in \eqref{paramAdSWeierstrass} to parametrize $AdS_5 \times S^5$.
This coordinate system arises in the small-charge limit of the genus one solution ($\omega_1\to \infty$, $\omega_3 \to i\infty$).
In this coordinate system the $AdS_5 \times S^5$ Nambu-Goto Lagrangian reads:
\be\label{LNGAdS} \mathcal{L}_{NG}^{(0)} = \frac{1}{r} \frac{2(x^2+1)}{(x^2-1)^2} \ee
The reason why the integral of this Lagrangian density  \eqref{m=intL} is independent of $r$ is that the starting point of the integration domain $x_0$ depends on $r$ (see \eqref{x0(r)}). The $r$-dependence cancels out.
Thus we can rewrite \eqref{E=L-m} as:
\be E(r) =  \frac{\sqrt{g_s}}{2\pi \alpha'} \left(\int_{x_0}^{\omega_1} dx \mathcal{L}_{NG} - \int_{x_0}^{\infty} dx \mathcal{L}_{NG}^{(0)} \right) \ee
Our goal is to show that  the potential energy $E(r)$ goes like $1/r$ at large $r$.
To this end we  cut the integration domains $[x_0,\omega_1]$ and $[x_0,\infty[$ into two pieces: $x<x_1$ and $x>x_1$.
We chose $x_1 \in\ ]1,\omega_1]$ so that $x_1$ lies in the asymptotic region: the metric for $x \in ]1,x_1]$ is well-approximated by the asymptotic expansion \eqref{asympAdSMetric}.
We take $r$ large enough so that $1>x_0>x_1$.

First let us evaluate the contribution to the potential coming from the region $x>x_1$. We find:
\be\label{geodEpart1} \int_{x_1}^{\omega_1} dx \mathcal{L}_{NG} - \int_{x_1}^{\infty} dx \mathcal{L}_{NG}^{(0)} \propto \frac{1}{r} + \mathcal{O}\left(\frac{1}{r^2}\right)
\ee
The $r$-dependence is easy to extract in this case, since the boundaries of the integration domains are manifestly independent of $r$. From equations \eqref{LNGgeod(r)} and \eqref{LNGAdS}, we see that the $r$-dependence at large $r$ reduces to an overall factor of $1/r$.
Numerics show that $\mathcal{L}_{NG} < \mathcal{L}_{NG}^{(0)}$ in this region. In particular $\mathcal{L}_{NG}$ vanishes at $x=\omega_1$ while $ \mathcal{L}_{NG}^{(0)}$ is strictly positive there. We deduce that the $1/r$ term on the right-hand side of \eqref{geodEpart1} comes with a non-vanishing negative coefficient.

Next we evaluate the contribution to the potential coming from the asymptotic region $x_0>x>x_1$.
Using the asymptotic form of the AdS metric \eqref{asympAdSMetric} we obtain (see \eqref{compEAsympt}):
\begin{align}\label{geodEpart2a}  \int_{x_0}^{x_1} & dx ( \mathcal{L}_{NG}- \mathcal{L}_{NG}^{(0)}) \approx
- \int_{z_0}^{z_1} dz \frac{L^2}{r^2} \frac{h}{2}
 = -\frac{h}{2} \frac{L^2(z_1-z_0)}{r^2}
\end{align}
The endpoint of the integration domain $z_1$ is related to $x_1$ as (see \eqref{z(x,w)}):
\be z_{1} = w(x_1) \frac{x_{1}^2-1}{x_{1}^1+1}\ee
Moreover we have $w(x_1) = r + \mathcal{O}(1)$ (see \eqref{w(x)b}).
We deduce:
\be\label{geodEpart2} \int_{x_0}^{x_1}dx \mathcal{L}_{NG} - \int_{x_0}^{x_1} dx \mathcal{L}_{NG}^{(0)} \propto \frac{1}{r} + \mathcal{O}\left(\frac{1}{r^2}\right)
\ee
It is clear from \eqref{geodEpart2a} that the coefficient of the $1/r$ term is once again non-vanishing and negative.
Combining equations \eqref{geodEpart1} and \eqref{geodEpart2}, we have shown that for a string following a spacelike geodesic, the potential  goes like $1/r$.

%%%%%

\subsection{The dependence of the gravitational potential on $g_s$ and $N$}\label{sub:coefE}

Previously we showed that the gravitational potential goes like $1/r$ at large $r$. Now we would like to get more information about the coefficient of this $1/r$ term.
More precisely we would like to determine how the string coupling $g_s$ and the number of D3-branes $N$ enter in this coefficient.

The genus one solution has four parameters that can be chosen as the asymptotic radius $L$, the asymptotic dilaton $g_s$, and the periods of the Weierstrass functions $\omega_1$ and $\omega_3$.
Consequently the gravitational potential also depends on these four parameters.
Let us consider the expression \eqref{SNG} for the Nambu-Goto action.
We want to understand the $g_s$ and $L$ dependence of this expression.
There is a manifest factor of $\sqrt{g_s}$ in front of the action.
The metric that appears in the Lagrangian depends on $L$ only through an overall factor of $L^2$, and is independent of $g_s$.
We deduce that the potential energy can be written as:
\be\label{Etildew1w3} E(r) = \frac{1}{r} \frac{\sqrt{g_s}L^2}{\alpha'} \tilde{E}(\omega_1,\omega_3) + \mathcal{O}\left(\frac{1}{r}\right) \ee
where $\tilde{E}(\omega_1,\omega_3)$ is a dimensionless function that depends only on the periods $\omega_1$ and $\omega_3$ and is independent of $g_s$ and $L$.

Now  we would like to write the potential energy in terms of the number of branes $N_{D3}^{(1)}$ and $N_{D5}^{(1)}$, instead of the periods $\omega_1$ and $\omega_3$.
Using the explicit expressions for the charges \eqref{QD3g1} and \eqref{QD5g1}, we can write :
\be N_{D3}^{(1)} = \frac{L^4}{{\alpha'}^2} \tilde{N}_{D3}(\omega_1,\omega_3) \ee
\be N_{D5}^{(1)} = \frac{L^2}{\sqrt{g_s}{\alpha'}} \tilde{N}_{D5}(\omega_1,\omega_3) \ee
where $\tilde{N}_{D3}(\omega_1,\omega_3)$ and $\tilde{N}_{D5}(\omega_1,\omega_3)$ are independent of $g_s$ and $N$.
So we can build two dimensionless quantities that are independent of $g_s$ and $N$: these are $N_{D3}^{(1)} \frac{{\alpha'}^2}{L^4} $ and $N_{D5}^{(1)}\frac{\sqrt{g_s}{\alpha'}} {L^2}$.
Now we can use these two quantities instead of the periods $\omega_1$ and $\omega_3$ to parametrize the genus one solution.
On general grounds the function $\tilde{E}(\omega_1,\omega_3)$ can be expanded as a series in these two parameters:
\be\label{E(N3,N5)} \tilde{E}(\omega_1,\omega_3) = \sum_{m=0}^\infty \sum_{n=0}^\infty \# \left( N_{D3}^{(1)} \frac{{\alpha'}^2}{L^4}  \right)^m  \left( N_{D5}^{(1)}\frac{\sqrt{g_s}{\alpha'}} {L^2}\right)^n \ee
where the $\#$'s are some numerical coefficients.
We only include positive powers since we do not expect the potential to diverge when either the D3- or the D5-charge goes to zero.
We assume that $g_s$ is small, that $L^2/\alpha' = \sqrt{4\pi N}$ is large and that the charges are not too large so that the parameters $N_{D3}^{(1)} \frac{{\alpha'}^2}{L^4} $ and $N_{D5}^{(1)}\frac{\sqrt{g_s}{\alpha'}} {L^2}$ are small with respect to one.
This is consistent with our initial assumption that $N_{D3}^{(1)}\ll N$.
Thus the expansion \eqref{E(N3,N5)} is dominated by the terms with small powers $m$ and $n$.

Now we use one assumption we made about the string profile.
We demanded that when the number of fundamental strings in the backreacting stack $N_{F1}^{(1)}$ goes to zero and the geometry reduces to $AdS_5 \times S^5$, the string profile reduces to the straight string in $AdS_5 \times S^5$ (see Figure \ref{choicePath} (\emph{top}) and equation \eqref{straightLim}).
In this limit the gravitational potential goes to zero.
When either the number of D3-branes, $N_{D3}^{(1)}$, or the number of D5-branes, $N_{D5}^{(1)}$, goes to zero, the number of fundamental strings $N_{F1}^{(1)} = N_{D3}^{(1)}N_{D5}^{(1)}$  goes to zero as well.
This implies that all the terms with $m=0$ or $n=0$ in the expansion \eqref{E(N3,N5)} must come with a vanishing coefficient.
Consequently the leading term in the expansion \eqref{E(N3,N5)} is proportional to the number of fundamental strings:
\be\label{tE(F1)} \tilde{E}(\omega_1,\omega_3) =  \#  N_{F1}^{(1)} \frac{{\alpha'}^3\sqrt{g_s}}{L^6} + \mathrm{subleading} \ee
We deduce for the gravitational potential energy:
\be\label{E(gs,L)} E(r) = \# \frac{N_{F1}^{(1)} }{r} \frac{g_s {\alpha'}^2}{L^4}  + \mathrm{subleading} \ee
where $\#$ is a numerical constant that depends on the precise choice of string profile and can be evaluated numerically.
Since in Einstein frame, the asymptotic radius $L$ is related to the total number of D3-branes $N$ as $L^4=4\pi N$, we can write the potential energy in terms of $g_s$ and $N$ as:
\be\label{E(gs,N)} E(r) = \# \frac{N_{F1}^{(1)} }{r} \frac{g_s}{N}  + \mathrm{subleading} \ee
From equation \eqref{E(gs,N)} it is clear that at large $N$, the potential goes to zero. This can be understood as follows. In the large $N$ limit, spacetime becomes flat. In this limit our set-up reduces to infinite straight strings in 10-dimensional Minkowski spacetime. We know from standard Newtonian gravity that the gravitational potential between such objects goes like $1/r^6$. Thus the vanishing of the $1/r$ term in the potential is expected.
This is to be contrasted with the dual gauge theory interactions for which the potential typically grows with the number of colors $N$.

For phenomenological applications it is also convenient to re-write the energy as:
\be\label{Ebraneworld} E(r) = \# \frac{{\alpha'}^4 g_s^2}{(\sqrt{g_s}L^2)^3} \frac{N_{F1}^{(1)} }{\alpha'}  \frac{\sqrt{g_sN}}{r}  + \mathrm{subleading} \ee
where $g_s^{1/4}L$ is the asymptotic string frame radius and  $\frac{{\alpha'}^4 g_s^2}{(\sqrt{g_s}L^2)^3} $ can be understood as the four-dimensional Newton's constant on the braneworld \cite{hep-th/9906064,hep-th/9906182}.
We observe that the gravitational potential \eqref{Ebraneworld} comes with an additional factor of $\sqrt{g_sN}$ with respect to the usual four-dimensional Newtonian potential.
In terms of the D-brane worldvolume theory with natural coupling $g_s N$, this is suggestive of a strong-coupling effect.
From the bulk perspective this strong coupling effect is not surprising: the leading contribution to the potential comes from the sections of the strings very close to the D-branes where the backreaction is large.

The subleading terms in the expressions \eqref{E(gs,N)} and \eqref{Ebraneworld} for the potential include terms that come with a higher (negative) power of $r$. They also include terms that go like $1/r$ but that come with higher powers of the charges $N_{D3}^{(1)}$ and $N_{D5}^{(1)}$. These terms are suppressed by additional powers of $g_s$ and/or $1/N$.

\paragraph{Comparison with the linear computation}
In \cite{Benichou:2010sj} the gravitational potential between open strings was computed using linearized gravity.
This linear approximation is enough to understand the $1/r$ behavior of the potential, as we explained in section \ref{sub:1/r}.
However  the linear treatment is not precise enough to provide the correct coefficient for the potential. In particular the $g_s$ and $N$ dependence cannot be computed reliably.
The reason is that the sections of the strings closest to the horizon contribute more to the potential, and the linear approximation breaks down in this region.

There is another point that is missed in the linear treatment of the backreaction of the stack of fundamental strings.
This stack polarizes because of the presence of background fluxes. As a consequence, the horizon surrounding the stack of strings is blown up. This effectively lowers the coefficient of the gravitational potential

%%%%%

\subsection{Open strings versus point particles}\label{sub:stringsVsPoints}
Let us make two rather elementary but  important observations about the gravitational potential energy \eqref{E(gs,N)}:
\begin{itemize}
\item It goes like $1/r$ despite the fact that gravity propagates in 10 non-compact dimensions.
\item It is independent of the inertial masses of the strings.
\end{itemize}
It turns out that these two surprising features can be understood thanks to a simple observation: in our computation the open strings behave like infinitely extended objects and not like point particles.

To illustrate this claim, let us study a very simple set-up. We consider an extended string of size $l$ and mass $m$ in D-dimensional Minkowski spacetime. We assume $D>4$. We compute the gravitational potential created by this object.
At large distances $r\gg l$, the object behaves like a point-like mass and the potential goes like $m/r^{D-3}$. But at small distances $r\ll l$, the object appears infinitely extended and the potential goes like $m/(l r^{D-4})$. In this case the potential does not depend on the mass of the object but rather on its tension $m/l$. Moreover the decay of the potential is slower.

In our case the D-brane backreaction stretches the open strings up to an infinite length.
Consequently the open strings interact like infinitely extended objects: the gravitational potential depends on their tension rather than their inertial masses, and the decay of the potential energy is slower than what is expected for point-like objects.

There is still one contribution to the gravitational potential that is proportional to the inertial masses of the strings.
Let us consider the string configuration of Figure \ref{inertialMass}, \emph{right}.
Previously we evaluated the contribution to the gravitational potential that comes from region 1 in Figure \ref{inertialMass}, close to the D-branes' horizon.
There is also a contribution to the gravitational potential that comes from  the flat asymptotic region (region 2  in Figure \ref{inertialMass}). Since most of the inertial mass of the strings is located in this region, this contribution is proportional to the masses.
The length of the sections of the strings contained in this region is finite. Consequently these sections of the strings  behave like point-like masses at large distance.
The contribution to the potential coming from this region goes like $m^2/r^7$, where $m$ is the inertial masses of the strings.
It is clear that this term is subdominant at large $r$. Even if region 1 in Figure \ref{inertialMass} contains a negligible portion of the inertial masses of the strings, the decay of the potential is slower in this region because the strings behave like infinitely extended objects. Consequently the contribution from this region dominates at large $r$.

In a phenomenological context, one should consider D3-branes in a four-dimensional compactification of string theory.
In that case the gravitational potential contains the usual Newtonian term that goes like $m^2/r$ \cite{Benichou:2010sj}.
The throat-contribution to the potential that we computed previously gives a stringy correction to the Newtonian potential.
This correction implies a violation of the equivalence principle on the D3-branes. These phenomenological aspects are further discussed in \cite{BE}.

Remember that because of the infinite redshift at the horizon, the geodesic distance between the single string and the stack of strings goes to zero near the D-brane horizon.
 Thus the leading term in the potential energy  \eqref{E(gs,N)} comes from a small-distance effect in the bulk, even if the coordinate distance $r$ is very large.
 This is reminiscent of an IR/UV duality.

In our  computation, it is crucial that we consider extended strings that stretch up to the horizon.
The potential energy between point-like particles in the bulk would also be affected by the D-branes' backreaction, but there would be no dramatic enhancement of the gravitational force as the one we observe here.  The surprising behavior of the potential energy \eqref{E(gs,N)} is the result of a stringy effect.

%%%%%

\subsection{Comparison with the holographic computation of the gauge potential}\label{sec:holgcomp}

Using the AdS/CFT correspondence \cite{hep-th/9711200,hep-th/9802109,hep-th/9802150}, we can also extract from the supergravity solutions of \cite{D'Hoker:2007fq} a potential energy in the dual gauge theory.
Let us now discuss this computation.
In the dual theory the stack of open strings describes a set of massive quarks, whereas the single string corresponds to a single massive anti-quark. To compute the gauge potential between these objects, the AdS/CFT dictionary tells us to evaluate the string theory path integral with given open-string asymptotic conditions close to the AdS boundary \cite{Rey:1998ik,Maldacena:1998im}.
In this computation it is crucial that we consider open strings that extend up to the boundary.

\begin{figure}[t]
\centering
\includegraphics[width=1\linewidth]{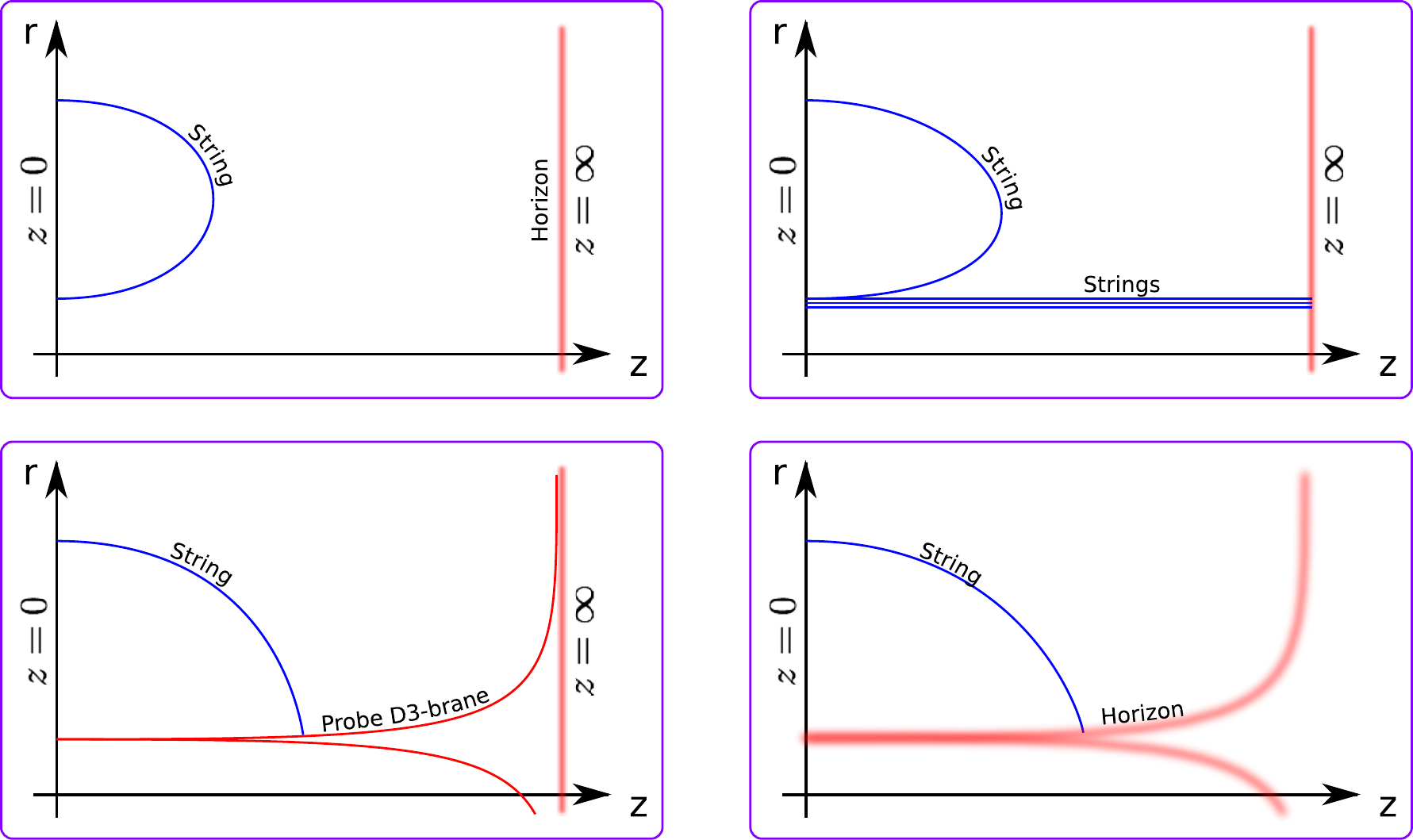}
\caption{Various string configurations relevant for the holographic computation of the gauge potential between: a quark and an anti-quark (\emph{top left}), several quarks and an anti-quark (\emph{top right}), many symmetrized quarks and an anti-quark (\emph{bottom left}), and lots of quarks and an anti-quark (\emph{bottom right}).}
\label{dGaugePotential}
\end{figure}

When the stack consists of a single string, it is well known that the path integral is dominated by a saddle point corresponding to a string hanging from two points of the boundary \cite{Rey:1998ik,Maldacena:1998im} (see Figure \ref{dGaugePotential}, \emph{top left}).
When the stack consists of several strings, then only one of the strings in the stack can recombine with the single string to form a hanging string. Charge conservation imposes that all the other strings in the stack extend up to the horizon  (see Figure \ref{dGaugePotential}, \emph{top right}).
Next we consider the case where the number of strings in the stack $N_{F_1}$ is of order $N$.
We assume for simplicity that these strings are symmetrized, so that the stack of open strings can be represented by a probe D3-brane of the type described in section \ref{subprobed3}.
Then the path integral is dominated by a saddle point that can be described as follows: the single string extends from the boundary and ends up on the probe D3-brane (see Figure \ref{dGaugePotential}, \emph{bottom left}). Moreover the string follows the path that minimizes its action.
This picture is reminiscent of the holographic computation of three-point functions that involve two heavy and one light operators \cite{arXiv:1008.1059}.
Finally when the number of strings in the stack is even bigger, say of order $N^2$, then we have no choice but to describe the stack of strings in terms of the supergravity solutions of \cite{D'Hoker:2007fq}.
The dominant saddle point can now be described as follows: the single string extends from the boundary up to the horizon, following the path that minimizes its action (see Figure \ref{dGaugePotential}, \emph{bottom right}).

Let us now stress the differences between the holographic computation of the gauge potential, and the computation of the gravitational potential.
The prescription for the gauge potential is to compute the full string theory path integral with given asymptotic conditions at the boundary.
This prescription ensures that what we are computing is a proper observable of the gravitational theory.
Consequently it corresponds to an observable of the dual gauge theory, namely the gauge  potential.
On the other hand the gravitational potential is not a proper observable of the gravitational theory (even though it is an important and useful quantity to compute): there are no local observables in diffeomorphism invariant theories.
Thus there is no observable in the dual gauge theory related to the gravitational potential.

At the technical level, the main difference between the two computations comes from the choice of profile for the single string.
In the gauge theory computation, we have to pick the unique profile that minimizes the action.
In the gravitational potential computation, we want the string to be as straight as possible.
In particular we chose a profile that reduces to the straight-string profile in the limit where the geometry reduces to $AdS_5 \times S^5$.
The minimal action profile does not satisfy this condition, as should be clear from Figure \ref{dGaugePotential}: the minimal action profile goes to the hanging-string profile when the number of strings in the stack goes to zero (there is a discontinuity at zero).
From the point of view of the holographic Wilson line computation, the string profile that we use to compute the gravitational potential is an off-shell field configuration and thus gives an irrelevant contribution to the path integral.  From the point of view of the gravitational potential computation, extremizing the action would make no sense as it would correspond to letting the string fall at the minimum of its gravitational potential, holding an arbitrary point of the string fixed, before measuring the gravitational force.

In the gravitational potential computation we insist on having all strings stretching down to the Poincare horizon in the $AdS_5 \times S^5$ limit, since this is really where the backreacting D3-branes are sitting, and we want to study open strings ending on backreacting D3-branes.
In the computation of the gravitational potential, whether or not the open strings extend up to the AdS boundary is irrelevant, since the leading contribution to the potential energy comes from the section of the strings that are very close to the horizon.

In the holographic gauge potential computation, the open strings stretch up to the boundary. Consequently the Nambu-Goto action evaluated on the minimal path action diverges. One natural way to regularize this divergence is to subtract the value of the Nambu-Goto action evaluated on a ``straight" path (for instance a spacelike geodesic),  generalizing the prescription of \cite{Rey:1998ik,Maldacena:1998im}. This regularization scheme gives the gauge potential schematically as:

\begin{align} E_{gauge} \left(\int d \tau\right) &= S_{NG}^{\mathrm{(minimal\ action)}} - S_{NG}^{\mathrm{(straight\ path)}} \cr
&= S_{NG}^{\mathrm{(minimal\ action)}} - (\mathrm{Inertial\ mass} + E_{gravity})\left(\int d \tau\right)
\end{align}

\paragraph{Computation of the gauge potential.}
In this paragraph we generalize the arguments of section \ref{sub:1/r} and \ref{sub:coefE} to compute the gauge potential between quarks and anti-quarks in the dual gauge theory.
We consider the Nambu-Goto action evaluated on the minimal action path.
The argument of section \ref{sub:1/r} applies with no need for modifications, and we find that the potential goes like $1/r$.
Actually since we now consider strings that go up to the boundary, there is no subleading corrections to the $1/r$ term in the potential, as explained below equation \eqref{largerExpforSNG}.
This is in agreement with four-dimensional conformal invariance.
Let us now compute the $g_s$ and $N$ dependence of the gauge potential, following the strategy of section \ref{sub:coefE}.
We can still write the gauge potential energy as in \eqref{Etildew1w3}:
\be E_{gauge}(r) = \frac{1}{r} \frac{\sqrt{g_s}L^2}{\alpha'} \tilde{E}_{gauge}(\omega_1,\omega_3) \ee
and the function of the periods $\tilde{E}_{gauge}(\omega_1,\omega_3)$ can be expanded as in \eqref{E(N3,N5)}:
\be\label{Eg(N3,N5)}\tilde{E}_{gauge}(\omega_1,\omega_3) = \sum_{m=0}^\infty \sum_{n=0}^\infty \# \left( N_{D3}^{(1)} \frac{{\alpha'}^2}{L^4}  \right)^m  \left( N_{D5}^{(1)}\frac{\sqrt{g_s}{\alpha'}} {L^2}\right)^n \ee
Now comes the main difference: when the number of  strings in the stack goes to zero, the minimal-action path does not go to the straight string path, but rather to the hanging string path. Thus the potential does not vanish in this limit (there is a discontinuity at zero).
This implies that the term with $m=n=0$ in the expansion \eqref{Eg(N3,N5)}  comes with a non-zero coefficient:
\be \tilde{E}_{gauge}(\omega_1,\omega_3) = \mathrm{constant} + \mathrm{subleading} \ee
We deduce that the gauge theory potential goes like:
\be E_{gauge}(r) =  \frac{1}{r} \left( \# \sqrt{g_s N}+ \mathrm{subleading}\right )\ee
where $\#$ is a numerical factor.
The $\sqrt{g_s N}$ behavior is in agreement with the results of \cite{Rey:1998ik,Maldacena:1998im}.
The subleading corrections contain terms that depend on the D3-brane and D5-brane charges. These terms are corrected by powers of $g_s$ and/or $1/N$.  It would be interesting to work out the full modifications to the gauge theory potential when the number of D3-branes and D5-branes in not assumed to be small.

%%%%%%%%%%%%%%%%%%%%%%%%%%%%%%%%%%%%%%%

\section{Discussion}\label{sec:discussion}

\paragraph{Summary of the results.}
Let us briefly summarize the results presented in this paper.
In the first part we studied in detail the supergravity solutions found in \cite{D'Hoker:2007fq} that describe open strings ending on backreacting D3-branes. In particular we carefully computed the various charges. This lead us to a one-to-one mapping between the supergravity solutions and the Young tableaux labeling irreducible representations of $SU(N)$.
We also understood the large gauge transformations of the RR-potentials, which change the number of fundamental strings in the supergravity geometries, as the manifestation of the Hanany-Witten effect.
Then we studied small-charge limits of the supergravity solutions, and obtained the profiles for probe D3- and D5-branes in $AdS_5\times S^5$, which agreed precisely with a DBI analysis. This provides a supergravity derivation of some non-trivial features of open string theory: the creation of D3-brane spikes from fundamental strings ending on D3-branes, and the polarization of stacks of open strings into D5-branes in the presence of 5-form flux.

In the second part, we discussed the interactions between open strings ending on backreacting D3-branes.
In particular, we computed the gravitational potential between a stack of open strings and a single string.
We found that this potential goes like $1/r$ even though gravity propagates in ten non-compact dimensions and that it violates the equivalence principle. We explained these surprising features by the observation that the open strings behave like infinitely extended objects near the horizon of the D3-branes.  We also contrasted this computation with the holographic computation of the dual gauge theory potential.

\paragraph{Consequences for stringy brane-world scenarios.}
In most type II phenomenological models, the Standard Model is realized on a configuration of D-branes organized in a four-dimensional compactification of string theory.
Our results suggest that gravity in these models is richer than expected, when the D-brane backreaction is properly taken into account.
In particular it is not correct to work with the naive Kaluza-Klein reduction of the 10-dimensional gravity on the compactification manifold, treating the D-branes as probes.
These aspects are further discussed in \cite{BE}, where we argue that an observer living on such a brane-world would detect a violation of the equivalence principle at very low temperature.

The open strings we considered are straight strings, so we can ask whether the gravitational potential we computed is also valid for oscillating strings.
We expect our results to hold for oscillating strings as well, since the wavelength of any oscillation would be redshifted to infinity near the D-brane horizon. All the strings appear straight close to the horizon, in agreement with the fact that the open strings satisfy Neumann boundary conditions along the D-brane worldvolume.
In particular our results are relevant for the gravitational interactions between matter particles of the standard model, that are realized as open strings stretched between different stacks of branes.
An interesting question concerns the gravitational interactions for the massless open string modes, for instance the photon. Such massless open strings may not extend in the bulk at all, thus it is not clear that our computations are relevant in this case. The full string theory is necessary to rigorously study the gravitational behavior of massless particles in braneworld scenarios.

So far, our computation of the gravitational potential applies to isolated stacks of D3-branes.
It would be interesting to extend this computation first to the case of Dp-branes, and then to the case of intersecting brane configurations.
This would bring us closer to realistic models, and would lead to a refinement of the experimental predictions.
We expect that the Dp-brane backreaction will always induce an enhancement of the gravitational potential.
Indeed it seems to be a generic fact that extremal Dp-brane backreaction stretches the open strings up to an infinite proper length.

In this work we used a supergravity description of the bulk gravity. For this approach to be reliable, the bulk curvature has to be small enough. In particular we need $g_s N$ to be larger than one, so it is legitimate to wonder whether our conclusions still hold when $g_s N$ becomes small.
The supergravity solutions we work with preserves 16 supercharges, so one might hope that $\alpha'$ corrections can be safely neglected. Additionally we showed that the gravitational potential goes like $1/r$, instead of the $1/r^7$ expected for instance in ten-dimensions. It would be surprising that this $1/r$ contribution to the potential would suddenly disappear at some value of the couplings.  Thus we expect our computation of the gravitational potential to be qualitatively reliable even at smaller coupling.

\paragraph{About the effective field theory description of open string theory.}
We argued that the surprising behavior of the gravitational potential can be explained by the fact that the open strings behave like infinitely long objects near the D-brane horizon. An interesting question is whether we can reproduce this potential using an effective field theory description of the open string theory.

From a ten-dimensional bulk perspective, it is not clear that this is possible.
A natural starting point would be to couple the four-dimensional DBI action to the ten-dimensional supergravity. However this is notoriously difficult when the D-brane backreaction is taken into account. The main issue is that the throat created by Dp-branes of codimension greater than 2 is infinitely long. If we define the DBI action on the backreacting D-brane worldvolume, the couplings to the bulk degrees of freedom are redshifted down to zero. Some progress has been realized recently to overcome this difficulty (see e.g. \cite{arXiv:0705.3212,arXiv:0812.3820,arXiv:0912.3039}).
However even if this problem were solved, it is not clear that such an approach would capture the corrections to the gravitational potential that we compute. Indeed  as we stressed in section \ref{sub:stringsVsPoints}, the fact that we consider strings and not point-like particles is crucial for our computation.

Another open question is to find a four-dimensional effective field theory description of the gravitational interactions between elementary particles in a stringy brane-world scenario.  For an observer living on the branes, the open strings look like point particles, so such an effective field theory should exist.  Our results show that this field theory would necessarily deviate from Einstein gravity. In particular the equivalence principle should be violated at low temperatures \cite{BE}.

\paragraph{Reducible representations.}
We have shown that the supergravity solutions of \cite{D'Hoker:2007fq} describe complicated configurations of D3-branes, D5-branes and fundamental strings. It is striking that these solutions are non-singular.
We have proposed an explanation for this fact: these supergravity solutions are associated to Wilson lines in irreducible representations of $SU(N)$. In other words, we can associate these supergravity solutions to pure quantum states.  On the other hand we have shown that none of the solutions of \cite{D'Hoker:2007fq} describe Wilson lines in reducible representations.  This implies that irreducible Wilson lines might be described by singular solutions.  If such singular solutions are found, it is possible that one may associate an entropy to them which somehow counts the number of irreducible representations encoded in the geometry.  This is similar to the fuzzball proposal for black holes (see \cite{Mathur:2005zp,Skenderis:2008qn} and references therein).

%%%%%%%%%%%%%%%%%%

\section*{Acknowledgments}
We would like to thank Costas Bachas, Cliff Burgess, Ben Craps, Frederik Denef, Hai Lin, Rodolfo Russo, Nathan Seiberg and Jan Troost for useful discussions and correspondence.
This research is supported in part by the Belgian Federal Science Policy Office through the Interuniversity Attraction Pole Programme IAP VI/11-P.
R.B. is a Postdoctoral researcher of FWO-Vlaanderen, and is supported in part through project G011410N.
J.E. is supported by the FWO - Vlaanderen, Project No. G.0235.05.

%%%%%%%%%%%%%%%%%%%

\appendix
\section{Supergravity conventions}
\label{appconv}
We briefly summarize here our conventions.  The type IIB bosonic fields are the metric $g_{MN}$, the dilaton $\Phi = 2 \phi$, the axion $\chi$, the RR 2-form $C_{(2)}$ with field strength $F_{(3)} = d C_{(2)}$, the NSNS 2-form $B_{(2)}$ with field strength $H_{(3)} = d B_{(2)}$ and the RR 4-form $C_{(4)}$ with field strength $F_{(5)} = d C_{(4)}$.
The equations derive from the action
\begin{align}
\label{actIIB}
S_{IIB} &=
{1 \over 2 \kappa_{(10)}^2} \int d^{10}x \sqrt{g} \bigg \{
R - {4 \over 2} \p_M \phi \p^M \phi
- \half e^{4 \phi} \p_M \chi \p ^M \chi
- { 1 \over 2} e^{-2 \phi} |H_{(3)}|^2
\cr & \qquad
- {1 \over 2} e^{2 \phi} |F_{(3)} - \chi H_{(3)}|^2
- {1  \over 4} | \hat F_{(5)}|^2 \bigg \}
- {1 \over 4 \kappa _{10}^2} \int d^{10}x \
C_{(4)} \wedge H_{(3)} \wedge F_{(3)}
\end{align}
in the following sense. The field equations are derived by first requiring that $S$
be extremal under  arbitrary variations of the fields  $g_{MN}$, $\phi$,
$B_{(2)}$, $C_{(2)}$ and $C_{(4)}$;  and second by imposing the self-duality condition on $\hat F_{(5)}$ as a supplementary equation.
The coupling constant $\kappa_{(10)}^2$ is given by:
\be 2 \kappa_{(10)}^2 = (2\pi)^7 {\alpha'}^4 \ee
Next we consider the definition and quantization of charges.  We define the charges as follows
\begin{align}
\label{app:chargedefs}
Q_{D5} &= \int_{V_{(4)}} d F_{(3)} = \int_{S_{(3)}} F_{(3)} \ , \cr
Q_{D3} &= \int_{V_{(6)}} d F_{(5)} = \int_{S_{(5)}} F_{(5)} \ , \cr
Q_{F1} &= \int_{V_{(8)}} d (e^{-2 \phi} * H_{(3)}) + d C_{(4)} \wedge d C_{(2)} = \int_{S_{(7)}} (e^{-2 \phi} * H_{(3)}) + C_{(4)} \wedge d C_{(2)} \ , \cr
&\phantom{= \int_{V_{(8)}} d (e^{-2 \phi} * H_{(3)}) + d C_{(4)} \wedge d C_{(2)} } = \int_{S_{(7)}} (e^{-2 \phi} * H_{(3)}) - C_{(2)} \wedge d C_{(4)} \ ,
\end{align}
where the $V_{(p+1)}$ are volumes containing the p-branes while the $S_{(p)} = \p V_{(p+1)}$ are surfaces which enclose the p-branes.  The charges are quantized in terms of the brane tension $T_{(p)}$ as
\begin{align}
Q_{Dp} = 2 \kappa_{(10)}^2 T_{(p)} N_{Dp}  \ ,
\end{align}
where $N_{Dp}$ is the number of Dp-branes and the brane tension is given by
\begin{align}
T_{(p)}^2 = \frac{\pi}{\kappa_{(10)}^2} (4 \pi^2 \alpha^\prime)^{3-p} \ .
\end{align}
In particular it satisfies the quantization constraint
$2 \kappa_{(10)}^2 T_{(p)} T_{(p-6)} = 2 \pi$
and the T-duality constraint
$2 \pi (\alpha^\prime)^\half T_{(p+1)} = T_{(p)}$.
For example, the expression for $T_{(p)}$ can be determined by comparing the supergravity and string theory calculations for the exchange of closed strings between a pair of p-branes as in \cite{hep-th/9510017}.

%%%%%%%%%%%%%%%%%%%%%%%%%%%%%%%

\section{Parametrizations of $AdS_5$}\label{app:paraAdS}

\paragraph{Poincare coordinates.}
The $AdS_5$ metric takes the form:
\be\label{PoincareAdS} ds^2 = L^2 \frac{dz^2 - dt^2 + dr^2 + r^2 d\Omega_{(2)}^2}{z^2} \ee
The AdS boundary is located at $z=0$, while the Poincare horizon is lying at $z=\infty$.

The spacelike geodesics in the $(r,z)$ plane  are half-circles with center on the AdS boundary at $z=0$.
In the limit where the radius of the circle goes to infinity, we obtain a geodesic that reaches the horizon at $z=\infty$.
This geodesic follows a line $r=cst$.
A probe string placed along such a straight geodesic extremizes the Nambu-Goto action.

Let us compute the geodesic distance between the point of coordinates $(z,r)$ and the line $r=0$. We find that the geodesic that minimizes the length is the half circle with center at the origin $(z=0,r=0)$ and with radius $\sqrt{z^2 + r^2}$.
The proper length of the section of this geodesic that connects the point $(z,r)$ to the point $(\sqrt{z^2 + r^2},0)$ is given by:
\be\label{geoDistStack} L\, \mathrm{arcsinh}\left(\frac{r}{z}\right) \ee

\paragraph{$AdS_2 \times S^2$ parametrization.}
Let us introduce the coodinates $\eta$ and $w$ such that:
\be\label{poincareToAdSxS} z = \frac{w}{\cosh(\eta)} \qquad ; \qquad r = w \tanh(\eta) \ee
The $AdS_5$ metric becomes:
\be\label{SU11SU2AdS} ds^2 = L^2 \left(\cosh^2(\eta) \frac{dw^2 - dt^2}{w^2} + \sinh^2(\eta)d\Omega_{(2)}^2 + d\eta^2 \right) \ee
The boundary lies at:
\begin{align}
& (z=0, r>0)  \Leftrightarrow (w>0, \eta=\infty) \cr
& (z=0, r=0) \Leftrightarrow (w=0, \eta \ge 0)
\end{align}
The Poincar\'e horizon lies at:
\be (z=\infty, r\ge 0) \Leftrightarrow (w \to \infty, \eta \to 0)\ \mathrm{with}\ w \eta \sim r \ee
The section of a straight geodesic defined by $r=cst$, $z>z_0$ in the Poincare coordinates reads in the $(w,\eta)$ coordinates:
\be w\tanh(\eta) = r \qquad ; \qquad 
\sinh(\eta)<\frac{r}{z_0}\ee

\paragraph{Small-charge limit of the genus one geometry.}
The Riemann surface $\Sigma$ for the genus one solution is conveniently chosen as the rectangle of Figure \ref{genus1} parametrized by the coordinates $(x,y)$.
In the limit where the charges of the genus one solution go to zero (that is for $\omega_1 \to \infty$, $\omega_3 \to i \infty$), the geometry degenerates to $AdS_5 \times S^5$.
The parametrization that naturally appears in this limit is:
\be (x+iy)^2 = \frac{i \sinh(\eta + i \theta) - 1}{i \sinh(\eta + i \theta) + 1} \ee
where $\theta$ is the polar angle of the 5-sphere and $\eta$ is the radial coordinates in the $AdS_2\times S^2$ parametrization of $AdS_5$ \eqref{SU11SU2AdS}.
Let us set the 5-sphere polar angle to $\theta=\pi/2$.
This amounts to taking $x>1$ and $y=0$. We obtain:
\be\label{x(eta)} x^2 = \frac{\cosh(\eta)+1}{\cosh(\eta)-1} \quad \Leftrightarrow \quad  \cosh(\eta) = \frac{x^2+1}{x^2-1}\ee
The $AdS_5$ metric then becomes:
\be\label{paramAdSWeierstrass} \frac{ds^2}{L^2} = \left( \frac{x^2+1}{x^2-1}\right)^2  \frac{dw^2 - dt^2}{w^2} + \frac{4 x^2}{(x^2-1)^2}d\Omega_{(2)}^2  + \frac{4 dx^2}{(x^2-1)^2} \ee
The Poincare coordinate $z$ can be written in terms of $x$ and $w$ as:
\be\label{z(x,w)} z = w \frac{x^2-1}{x^2+1}\ee
The boundary lies at:
\begin{align}
& (z=0, r>0)  \Leftrightarrow (w>0, x=1) \cr
& (z=0, r=0) \Leftrightarrow (w=0, x\ge 1)
\end{align}
The Poincar\'e horizon lies at:
\be (z=\infty, r\ge 0) \Leftrightarrow (w \to \infty, x \to \infty)\ \mathrm{with}\ w/x \sim r/2 \ee
The section of a straight geodesic defined by $r=cst$, $z>z_0$ in the Poincare coordinates becomes in this parametrization:
\be \frac{2x}{x^2+1} = \frac{r}{w} \qquad ; \qquad
x>x_{0} = \frac{z_0}{r}\left(1+\sqrt{1+\frac{r^2}{z_0^2}}\right)\ee

%%%%%%%%%%%%%%%%%%%%%%%%%%%

\section{Computation of the charges: technical details}\label{app:compCharges}

In section \ref{sec:charges} we described the computation of the various charges in the supergravity solutions.
In this appendix we gives some more details about this computation.
To compute the charges it is convenient choose our integration contours to lie along the boundary of $\Sigma$.
Consequently, we first examine the expansions of the fluxes near the boundary of $\Sigma$.
We take $\Sigma$ as the upper-half plane (see Figure \ref{sigma}), and the boundary of $\Sigma$ is the real line.
We parametrize the upper-half plane with $u=\nu + i \epsilon$, where $\nu$ and $\epsilon$ are real.
On the boundary there are two types of regions to consider, those where $h_1$ satisfies the Dirichlet boundary conditions and those where it satisfies Neumann boundary conditions.  We may expand $\A$ and $\B$ in series away from the boundary as
\begin{align}
\A(\nu + i \epsilon) &= i \, a + \A(\nu) + \A'(\nu) \epsilon + ... , \nonumber\\
\B(\nu + i \epsilon) &= i \, b + \B(\nu) + \B'(\nu) \epsilon + ... ,
\end{align}
where $a$ and $b$ are real and provide the arbitrary imaginary constants in the definition of $\A$ and $\B$.
Demanding either vanishing Dirichlet or Neumann conditions for $\A$ and vanishing Dirichlet conditions for $\B$, we have
\begin{align}
{\rm Dirichlet}: \qquad \A(\nu) = - \bar \A(\nu) \qquad \qquad \B(\nu) = - \bar \B(\nu) \nonumber\\
{\rm Neumann}: \qquad \A(\nu) = \bar \A(\nu) \qquad \qquad \B(\nu) = - \bar \B(\nu)
\end{align}
We find the following behavior for the fluxes along Dirichlet segments:
\begin{align}
&dj_2 = 3 i [b \, \A'(\nu) - a \, \B'(\nu)] d\epsilon + {\cal O}(\epsilon) \ , \nonumber\\
&e^{-2 \phi} * H_{(3)} = {\cal O}(\epsilon^4) \ .
\end{align}
From these expressions, we conclude that $Q_{D3}^{(i)}$ and $Q_{F1}^{(i)}$ do not depend on the choice of endpoints for the integrals and furthermore that the non-vanishing contribution to these charges comes from integrating the fluxes along the Neumann segments of the boundary.  We also note that the only contribution to $Q_{D3}^{(i)}$ comes from the ${\cal C}$ term, and as a result may write
\begin{align}
\label{D3charge}
Q_{D3}^{(i)} = 3 i \, {\rm Vol}(S^4) \, \int_{e_{2i}}^{e_{2i-1}} d {\cal C} + c.c. \qquad {\rm for} \; i \neq 0
\end{align}
For the Neumann segments, we find:
\begin{align}
&b_2 = 4 a + {\cal O}(\epsilon^2)\ , \nonumber\\
&e^{-2 \phi} * H_{(3)} = {\cal O}(\epsilon^2) \ .
\end{align}
We see that $b_2$ is constant along Neumann segments and moreover the contribution comes from the constant term in $\A$, so that we may write
\begin{align}
Q^{(i)}_{F1} = -8 i \A(\hat e_i) \, {\rm Vol}(S^2) \, Q^{(i)}_{D3} + c.c. \qquad {\rm for} \; i \neq 0
\end{align}
where $\hat e_i$ are arbitrary points in the intervals $(e_{2i},e_{2i-1})$.
The $Q^{(j)}_{D5}$ are associated to the Dirichlet segment. They can also be computed to give
\begin{align}
\label{D5charge}
Q^{(j)}_{D5} &= 2 i {\rm Vol}(S^2) \int_{e_{2j+1}}^{e_{2j}} d \A  + c.c. \nonumber\\
&= 2 i {\rm Vol}(S^2) [\A(\hat e_j) - \A(\hat e_{j+1})] \ .
\end{align}
Finally, we can also obtain an explicit expression for $\tilde Q_{F1}^{(j)}$ by noting that the only contribution to $j_2$ along the Dirichlet segments comes from the ${\cal C}$ term, which is constant. As a result, we may write $\tilde Q_{F1}^{(j)}$ as
\begin{align}
\tilde Q_{F1}^{(j)} = - 12 i {\cal C}(\hat e_j) \, {\rm Vol}(S^4) \, Q^{(j)}_{D5} + c.c. \ ,
\end{align}
where $\hat e_j$ are arbitrary points in the intervals $(e_{2j+1},e_{2j})$.

%%%%%%%%%%%%%%%%%%%%%%%%%%%%%%%%%%%

\section{D3-brane charges for the genus one solution}\label{app:D3charge}

In this appendix we compute explicitly the function ${\cal F}(z)$ defined in \eqref{defF}.
This leads to the explicit values of the D3-charges for the genus one solution given in \eqref{QD3g1}.
To compute ${\cal F}(z)$, we first write $\B$ and the derivative of $\A$ as
\begin{align}
\partial \A &= -i \kappa_1 \left( 2\frac{\zeta(\omega_3)}{\omega_3} + 2 \wp(1) + \frac{\wp''(1)}{(\wp(z) - \wp(1))} + \frac{\wp'(1)^2}{(\wp(z) - \wp(1))^2}
\right) \cr
\B &= i \kappa_2 \frac{\wp'(1)}{\wp(z) - \wp(1)} \ ,
\end{align}
where we have made use of the identities $\wp(u+v) = \wp(v) - \frac{1}{2} \p_v [(\wp'(u) - \wp'(v))/(\wp(u)-\wp(v))]$ and $\zeta(u-v) - \zeta(u+v)= - 2\zeta(v) + \wp'(v)/(\wp(z) - \wp(z))$.
The function ${\cal F}(z)$ can then be written as
\begin{align}\label{f(I1)} {\cal F}(z) &\equiv - \int \B \partial \A \\
&= -\frac{\kappa_1 \kappa_2}{2} \wp'(1) \bigg[ 4 \left( \frac{\zeta(\omega_3)}{\omega_3} + \wp(1) \right) I_1(z,v)
+ \frac{\wp''(1)}{\wp'(1)} \frac{\p I_1(z,v)}{\p v}
+ \frac{\p^2 I_1(z,v)}{\p v^2} \bigg] \bigg|_{v=1} \nonumber
\end{align}
where we have introduced the integral
\be I_1(z,v) = \int \frac{1}{(\wp(z) - \wp(v))} \ee
and made use of a recursion relation obtained by differentiating the above integral with respect to $v$.
The integral can be written in terms of the Weierstrass $\sigma$-function
\footnote{The Weierstrass $\sigma$-function is defined in terms of the Weierstrass $\zeta$-function as $\zeta(z) = \sigma'(z)/\sigma(z)$.} up to an irrelevant constant as
\begin{align}\label{I1uv}
& I_1(z,v) = \frac{\log \sigma(z-v) - \log \sigma(z+v) + 2 z \zeta(v)}{\wp'(v)} \ .
\end{align}
Formulas \eqref{f(I1)} and \eqref{I1uv} provide an explicit expression for the function ${\cal F}(z)$.

To obtain the D3-charges \eqref{QD3g1}, we will be interested in evaluating the quantities \eqref{I1uv} at the branch points $z = \omega_i$ with $i = 1,2,3$ and $z = 0$.
First we find\footnote{To compute this we have made use of the identity $\zeta(z+ 2 \omega_i) = - \omega(z) \exp[2 \zeta(\omega_i) (z+\omega_i)]$.}
\begin{align}
\log \frac{\sigma(\omega_i-v)}{\sigma(\omega_i+v)}  = - 2 \zeta(\omega_i) v \ .
\end{align}
Thus we have
\begin{align}
I_1(\omega_i,v) = 2 \frac{\omega_i \zeta(v) - v \zeta(\omega_i)}{\wp'(v)} \ .
\end{align}
Finally, we obtain the following values for the integral
\begin{align}
I_1(\omega_3,v) &= 2 \frac{\omega_3 \zeta(v) - v \zeta(\omega_3)}{\wp'(v)} \cr
I_1(\omega_2,v) &= I_1(\omega_3,v) + {\rm real \, part} \cr
I_1(\omega_1,v) &= {\rm real \, part} \cr
I_1(0,v) &= \frac{i \pi }{\wp'(v)}
\end{align}
where in the last line, we have used that $\sigma$ is an odd function.
Explicitly the D3-brane charges $Q_{D3}^{(0)}$ and $Q_{D3}^{(1)}$ read:
\begin{align}
Q_{D3}^{(0)} &= 64 \pi^3 \kappa_1 \kappa_2 \bigg[ 4 \frac{\zeta(\omega_3)}{\omega_3} - 8 \wp(1) + \bigg( \frac{\wp''(1)}{\wp(1)} \bigg)^2 \bigg] \cr
Q_{D3}^{(1)} &= 128 \pi^2 \kappa_1 \kappa_2 i \bigg\{ \bigg[
4 \frac{\zeta(\omega_3)}{\omega_3} -8 \wp(1)
+ \frac{\wp''(1)^2}{\wp'(1)^2}  \bigg] \bigg(\omega_3 \zeta(1) - \zeta(\omega_3) \bigg) \cr
&  +  \bigg(\omega_3 \wp(1) + \zeta(\omega_3)\bigg) \frac{\wp''(1)}{\wp'(1)}  - \omega_3 \wp'(1) \bigg\}
\end{align}
while $Q_{D3}^{(2)}$ can be obtained using $Q_{D3}^{(0)} + Q_{D3}^{(1)} + Q_{D3}^{(2)} = 0$.

\paragraph{Small-charge limits.}
In the limit $\omega_1 \rightarrow \infty$ we find that ${\cal F}(\omega_3)$ and ${\cal F}(\omega_2)$ vanish, while
\begin{align}
{\cal F}(0) = i \frac{L^4 \pi}{32} \ ,
\end{align}
along with
\begin{align}
e^{\phi_0} \kappa_1 = e^{-\phi_0} \kappa_2 =  \frac{L^2}{4\pi} \frac{\omega_3}{i} \sinh \left( \frac{\pi i}{\omega_3} \right) \ .
\end{align}
In the limit $\omega_3 \rightarrow i \infty$ we find that
\begin{align}
{\cal F}(\omega_3) &= {\cal F}(\omega_2) = \frac{i \pi^2}{4 \omega_1^3 \sin^2 \left( \frac{\pi}{\omega_1} \right)} \bigg[ 2 \pi - \omega_1 \sin \left( \frac{2\pi}{\omega_1} \right) \bigg] \cr
{\cal F}(0) &= \frac{i \pi^3}{\omega_1^2 \sin^2 \left( \frac{\pi}{\omega_1} \right)} \ ,
\end{align}
along with
%\begin{align}
%\wp(2) + \frac{\zeta(\omega_3)}{\omega_3} = \frac{\pi^2}{4 \omega_1^2 \sin^2 \left( \frac{\pi}{\omega_1} \right)} \ .
%\end{align}
\begin{align}
e^{\phi_0} \kappa_1 = e^{-\phi_0} \kappa_2 =  \frac{L^2}{4\pi} \omega_1 \sin \left( \frac{\pi}{\omega_1} \right) \ .
\end{align}

%%%%%%%%%%%%%%%%%%%%%%%

\section{Asymptotic expansion}\label{App:asymptotics}

In this appendix we study the expansion of the fields near the point $u_0$ that corresponds to the asymptotic $AdS_5 \times S^5$ region.
Expanding the harmonic functions around $u = u_0$, we obtain
\begin{align}\label{asymptoticABb}
\A = \frac{i}{(u-u_0)} [a_0 - \sum_{n=2}^{\infty} a_n (u-u_0)^n ] + i a_1 \ ,
\qquad
\B = - \frac{i}{(u-u_0)} [\tilde b_0 - \sum_{n=2}^{\infty} \tilde b_n (u-u_0)^n ] + i \tilde b_1 \ ,
\end{align}
where the $a_i$ and $\tilde b_i$ are real constants.  Here we allow $\B$ to take a more general form than in (\ref{harmfun}), which can be recovered by setting $\tilde b_0 = 1$ and $\tilde b_n = 0 $ for $n > 0$.  Note the absence of a possible logarithm term in the above, which is necessary in order for the geometry to be asymptotically $AdS_5 \times S^5$.  Examining the expansion of $\p_u h_1$ given in (\ref{harmfun}), this leads to the requirement
\begin{align}
\frac{P'(u_0)}{P(u_0)} = \frac{s'(u_0)}{s(u_0)} \ .
\end{align}
Writing $z = u_0 + \varepsilon e^{i \vartheta}$, and substituting the above expansions into the general solution, we obtain
\begin{align}
e^{4 \phi} \sim \frac{\tilde b_0^2}{a_0^2} \frac{\tilde b_0 a_2 - a_0 \tilde b_2}{\tilde b_0 a_2 + a_0 \tilde b_2} + {\cal O}(\varepsilon^4) \ , \qquad
f_4^8 \sim 1024 (\tilde b_0 a_2 - a_0 \tilde b_2)^2 \sin^8(\vartheta) + {\cal O}(\varepsilon) \ , \nonumber\\
f_1^8 \sim f_2^8 \sim \frac{64 a_0^4 \tilde b_0^4}{(\tilde b_0 a_2 - a_0 \tilde b_2)^2} \frac{1}{\varepsilon^8} \bigg( 1 + {\cal O}(\varepsilon) \bigg) \ , \qquad
\rho^8 \sim \frac{4 (\tilde b_0 a_2 - a_0 \tilde b_2)^2}{\varepsilon^8} \bigg( 1 + {\cal O}(\varepsilon) \bigg) \ .
\end{align}
Making the substitution $\varepsilon=\lambda e^{-\eta}$ with $\lambda^2 = 2 |a_0 \tilde b_0| / |\tilde b_0 a_2 - a_0 \tilde b_2|$ and defining $L^4 = |32 (\tilde b_0 a_2 - a_0 \tilde b_2)|$, one recognizes the asymptotic $AdS_5 \times S^5$ metric:
\begin{align}
ds^2 = L^2 \left( \frac{e^{2 \eta}}{4} ds^2_{AdS_2} + \frac{e^{2 \eta}}{4} ds^2_{S^2} + d\eta^2 + d\vartheta^2 + \sin^2(\vartheta) ds^2_{S^4} \right) \ .
\end{align}
Expanding the potentials, we find as $\varepsilon \rightarrow 0$,
\begin{align}
&d j_2 \sim - 32 \sin^4(\vartheta) [(\tilde b_0 a_2 - a_0 \tilde b_2) d \vartheta + (\tilde b_0 a_3 - a_0 \tilde b_3) \sin(\vartheta) d\varepsilon] + {\cal O}(\varepsilon) \ , \nonumber \\
&b_2  \sim -4 \frac{\tilde b_0 a_1 a_2 + \tilde b_0 a_0 a_3 - \tilde b_2 a_0 a_1 - \tilde b_3 a_0^2}{\tilde b_0 a_2 - a_0 \tilde b_2} + {\cal O}(\varepsilon) \ , \nonumber\\
&\frac{e^{-2\phi} f_2^2 f_4^4}{f_1^2} *_2 d b_1 \sim \# \sin^5(\vartheta) d\varepsilon + {\cal O}(\varepsilon) \ .
\end{align}
The first term determines the D3-brane charge, while the second and third terms enter the computation of the F1-charge.  Note that the $d\varepsilon$ components of the above formula all vanish at $\vartheta = 0, \pi$ so that the charges do not depend on the endpoints chosen for the integration curve.  Secondly we note that the third line does not contribute to the F1-charge.  The charges are given by
\begin{align}
\label{boundcharge}
Q_{D3}^0 &= 12 \pi (a_2 \tilde b_0 - a_0 \tilde b_2) {\rm Vol(S^4)} = {\rm Vol(S^5)} L^4 \nonumber\\
Q_{F1}^0 &= -48 \pi [a_1(\tilde b_0 a_2 - \tilde b_2 a_0) + a_0(\tilde b_0 a_3 - \tilde b_3 a_0) ] {\rm Vol(S^4)} {\rm Vol(S^2)} \cr
&= - 128 [a_0(\tilde b_0 a_3 - \tilde b_3 a_0) ] {\rm Vol(S^5)} {\rm Vol(S^2)} - 4 a_1 {\rm Vol(S^2)} Q_{D3}
\end{align}
Note that the F1-charge computed above is not gauge invariant, in particular it is not invariant under shifts of $a_1$ by a constant.  This stems from the fact that the Page charge is not gauge invariant.

\paragraph{The linear perturbation of the $AdS_5$ metric.}
Next we compute the leading correction to the asymptotic $AdS_5$ metric.
For simplicity we focus here on the boundary of the Riemann surface where the four-sphere degenerates.
This means that we are sitting on a pole of the five-sphere.
Mapping the Riemann surface $\Sigma$ to the upper-half plane as in Figure \ref{sigma}, we work on the real axis in the neighborhood of the point $u_0$.
We write the coordinate $u$ parametrizing the upper half plane as $u=\nu + i \epsilon$ with $\nu$ and $\epsilon$ real.
We focus on the five-dimensional $AdS$-part of the metric:
\be ds^2 = f_1^2(u) \frac{dw^2-dt^2}{w^2} + f_2^2(u) d\Omega_2^2 + 4 \rho^2(u) d\nu^2 + ... \ee
We can reparametrize the Riemann surface $\Sigma$ so that the asymptotic expansion \eqref{asymptoticABb} takes the form:
\begin{align}\label{asymptoticAB}
\A = \frac{i}{(u-u_0)} [a_0 - \sum_{n=2}^{\infty} a_n (u-u_0)^n ] + i a_1 \ ,
\qquad
\B = - \frac{ib}{(u-u_0)}+ i \tilde b_1 \ ,
\end{align}
The asymptotic expansion of the functions defining the metric becomes:
\begin{align}
& f_1^2(u) = 2 \sqrt{2} a_0 \sqrt{\frac{b}{a_2}}\frac{1}{(u-u_0)^2} - 4 \sqrt{2} a_0 a_3 \sqrt{\frac{b}{a_2^3}}\frac{1}{u-u_0} + \frac{\sqrt{2b}(5a_2^3 + 12 a_0 a_3^2 - 10 a_0 a_2 a_4)}{a_2^{\frac{5}{2}}} + \mathcal{O}(u-u_0) \cr
& f_2^2(u) = 2 \sqrt{2} a_0 \sqrt{\frac{b}{a_2}}\frac{1}{(u-u_0)^2} - 4 \sqrt{2} a_0 a_3 \sqrt{\frac{b}{a_2^3}}\frac{1}{u-u_0} + \frac{\sqrt{2b}(a_2^3 + 12 a_0 a_3^2 - 10 a_0 a_2 a_4)}{a_2^{\frac{5}{2}}} + \mathcal{O}(u-u_0) \cr
& \rho^2 = \sqrt{2 a_2 b} \frac{1}{(u-u_0)^2} + 2 \sqrt{2} a_3 \sqrt{\frac{b}{a_2}} \frac{1}{u-u_0} - \frac{\sqrt{b}(a_2^3+4a_0 a_3^2 - 10 a_0 a_2 a_4)}{\sqrt{2} a_0 a_2^{\frac{3}{2}}} + \mathcal{O}(u-u_0)
\end{align}
We want to change the coordinates from $(\nu,w)$ to $(z,r)$ so that the metric takes the canonical form (see e.g. \cite{hep-th/0002091}):
\be\label{canonicalPerturbedAdS5} ds^2 = L^2 \frac{dz^2 + (\eta_{\mu \lambda} + h_{\mu \lambda})dx^\mu dx^\lambda }{z^2} + ... \ee
We work perturbatively in $z/r$, which is essentially the inverse of the geodesic distance away from the stack of strings in the Poincare coordinates (see \eqref{geoDistStack}). We assume that $\nu-u_0$ is of order $z/r$. More precisely, we set:
\begin{align}
& \nu-u_0 = A_1 \frac{z}{r} + A_2 \left(\frac{z}{r}\right)^2 + A_3 \left(\frac{z}{r}\right)^3+...\cr
& w = r\left( B_0 + B_1\frac{z}{r}+ B_2 \left(\frac{z}{r}\right)^2 + B_3 \left(\frac{z}{r}\right)^3 + ... \right)
\end{align}
The coefficients $A_i$, $B_i$ are fixed by demanding the metric to take the form \eqref{canonicalPerturbedAdS5}.
We find:
\begin{align}
& A_1 = \sqrt{\frac{a_0}{2 a_2}} \quad ; \quad A_2 = -\frac{a_0 a_3}{2 a_2^2} \quad ; \quad A_3 =  \sqrt{\frac{a_0}{2 a_2}}\frac{18 a_0 a_3^2 - 10 a_0 a_2 a_4 - 3 a_2^3}{16 a_2^3} \cr
& B_0 = 1 \quad ; \quad B_1 = 0 \quad ; \quad B_2 = \frac{1}{2} \quad ; \quad B_3 = 0
\end{align}
At this order in the asymptotic expansion, the metric takes the form:
\be\label{asympMetricB} ds^2 =  L^2\frac{dz^2 + \left(1+ \frac{5(a_2^3+2a_0a_3^2 - 2a_0a_2a_4)}{8a_2^3}\frac{z^2}{r^2}\right)\eta_{\mu \nu} dx^\mu dx^\nu }{z^2} + ... \ee
with $L^2 = \sqrt{32 |a_2 b|}$.

\paragraph{Explaining the $z^2/r^2$ perturbation.}
We found that the leading correction to the asymptotic metric goes like $z^2/r^2 \sim 1/l^2$, where $l$ is the geodesic distance away from the stack of strings. Let us try to understand why the geodesic distance appears in this way.

Generically for an object extended in $p$ spacelike dimensions in a D-dimensional spacetime, we expect from Newtonian gravity a backreaction that decays like $1/l^{D-p-3}$, where $l$ is the geodesic distance away from the object. Here we assumed that $l$ is much smaller than the typical size of the object. We also assumed that the curvature radius of spacetime is much bigger than $l$ so that a flat space approximation is reliable.
For a stack of long strings in ten dimensions, we would expect a linear perturbation of the metric that goes like $1/l^6$.
However we saw in section \ref{sec:probe} that the stack of strings polarizes in a complicated configuration of D3- and D5-branes.
The objects with the highest dimensionality are the D5-branes,
%added:
so they create the backreaction with the slowest decay. 
 Newtonian gravity tells us that the backreaction of D5-branes in 10-dimensions goes like $1/l^2$.
This is precisely what equation \eqref{asympMetricB} gives us.

Notice that the D5's wrap an $S^4$ inside the $S^5$.
We can perform a dimensional reduction along this $S^4$.
Then the polarized stack of strings in ten dimensions essentially becomes  a non-polarized stack of strings in a 6-dimensional spacetime $AdS_5 \times S^1/\mathbb{Z}_2$.
The gravitational backreaction of such a stack of strings was computed using linearized gravity in \cite{Benichou:2010sj}.

\paragraph{The $g_s$ and $N$ dependence of the metric perturbation.}
Now we want to find the $g_s$ and $N$ dependence of the leading correction to the $AdS_5$ metric.
For simplicity we focus on the genus-one solution that is parametrized by asymptotic radius $L$ and dilaton $g_s$, and the periods $\omega_1$ and $\omega_3$.
We follow a strategy similar to the one used in section \ref{sub:coefE} to obtain the $g_s$ and $N$ dependence of the gravitational potential.
The coefficient of the perturbation in the metric \eqref{asympMetricB} is: $ 5(a_2^3+2a_0a_3^2 - 2a_0a_2a_4)/8a_2^3$.
Comparing the expansions \eqref{asymptoticAB} with the explicit form of the harmonic functions \eqref{hsGen1}, we deduce that this coefficient is independent of the asymptotic radius $L$ and dilaton $g_s$. Thus this coefficient depends only on the periods  $\omega_1$ and $\omega_3$. Alternatively we can express it in terms of the D3- and D5- numbers as:
\begin{align}\label{pert(NDp)}  \frac{5(a_2^3+2a_0a_3^2 - 2a_0a_2a_4)}{8a_2^3} & =  \sum_{m=0}^\infty \sum_{n=0}^\infty \# \left( N_{D3}^{(1)} \frac{{\alpha'}^2}{L^4}  \right)^m  \left( N_{D5}^{(1)}\frac{\sqrt{g_s}{\alpha'}} {L^2}\right)^n
\end{align}
where the $\#$'s are numerical coefficients.
In the limit where the number of fundamental strings $N_{F1}^{(1)}= N_{D3}^{(1)} N_{D5}^{(1)} $ goes to zero, the metric reduces to $AdS_5 \times S^5$ and the perturbation vanishes. Thus all terms with $m$ or $n$ equal to zero in the expansion \eqref{pert(NDp)} have to come with a zero coefficient.
The expansion \eqref{pert(NDp)} is dominated by the term with $m=n=1$:
\begin{align}  \frac{5(a_2^3+2a_0a_3^2 - 2a_0a_2a_4)}{8a_2^3} & = \# N_{F1}^{(1)} \frac{{\alpha'}^3\sqrt{g_s}}{L^6}  + \mathrm{subleading}
\end{align}
As in section \ref{sub:coefE} we assume that the parameters $N_{D3}^{(1)} \frac{{\alpha'}^2}{L^4}$ and $N_{D5}^{(1)}\frac{\sqrt{g_s}{\alpha'}} {L^2}$ are smaller than one.
The subleading terms contain higher powers of the D-brane numbers $N_{D3}^{(1)} $ and $N_{D5}^{(1)} $ that are suppressed by additional powers of $g_s$ and/or $1/N$.
We deduce that the asymptotic metric reads:
\be\label{asympMetric} ds^2 = L^2 \frac{dz^2 + \left(1+  \# \frac{N_{F1}^{(1)} \sqrt{g_s}}{N^{\frac{3}{2}}} \frac{z^2}{r^2}\right)\eta_{\mu \nu} dx^\mu dx^\nu }{z^2} + ... \ee
where the $\#$ is a numerical coefficient.
Using numerics we find that this coefficient is negative and of order $\sim - 0.0016$.

We can rewrite the linearized metric \eqref{asympMetric} as:
\be ds^2 = L^2 \frac{dz^2 + \left(1+  \# \frac{{\alpha'}^4 g_s^2}{g_s L^4} \frac{N_{F1}^{(1)} }{\alpha' \sqrt{g_s}L^2} \frac{z^2}{r^2}\right)\eta_{\mu \nu} dx^\mu dx^\nu }{z^2} + ... \ee
where $g_s^{1/4}L$ is the string frame radius and $ \frac{{\alpha'}^4 g_s^2}{g_s L^4} $ can be identified with a 6-dimensional Newton's constant.
This result matches perfectly with the computation done in \cite{Benichou:2010sj} for the linear backreaction of a string in the six-dimensional spacetime $AdS_5 \times S^1$, in agreement with the comment we made at the end of the previous paragraph.

%%%%%%%%%%%%%%%%%%%%%%%%%

\section{Radial geodesics}\label{app:geodesic}

In this appendix we discuss the spacelike geodesics in the genus-one geometry.
In particular our goal is to construct the geodesics that emanate from the AdS boundary and extend up to the horizon.

The affine parameter along these geodesics provides a good radial coordinate to parametrize the geometry.
This is similar in spirit to the construction of the Fefferman-Graham coordinates, that are the Gaussian normal coordinates emanating from the boundary. The difference is that we do not demand the geodesic to be normal to the boundary, but rather to reach the horizon. In this way the coordinate system we obtain covers the entire spacetime and its range of validity is not limited to a neighborhood of the boundary. Close to the boundary and far away from the backreacting stack of strings, the geodesics that we discuss reduce to the normal geodesics used to define the  Fefferman-Graham coordinates.

The metric is:
\be ds^2 = f_1^2 \frac{dw^2-dt^2}{w^2} + f_2^2 ds_{S^2}^2 + f_4^2 ds_{S^4}^2 + 4\rho^2(dx^2+dy^2) \ee
where $x$ and $y$ parametrize the rectangle of Figure \ref{genus1}, \emph{right}.
We are interested in spacelike geodesics, lying at constant time $t$. We can consistently assume that the geodesics lie at a fixed point of the two- and four-spheres. So the data we have to find is $w(\lambda)$, $x(\lambda)$ and $y(\lambda)$, where $\lambda$ is the affine parameter of the geodesic.
Assuming affine parametrization, the equations of motions can be derived from the simplified Lagrangian:
\be \tilde{L} = f_1^2 \frac{\dot w^2}{w^2} + 4 \rho^2(\dot x^2 + \dot y^2) \ee
where a dot stands for the derivative with respect to the affine parameter $\lambda$.
The affine parametrization condition implies that this Lagrangian evaluates to 1 on the solution.
The equation of motion for $w(\lambda)$ reads:
\be \partial_\lambda \left( f_1^2 \frac{\dot w}{w^2} \right) = - f_1^2 \frac{\dot w^2}{w^3} \ee
This is equivalent to:
\be \partial_\lambda \left(f_1^2 \frac{\dot w}{w} \right) = 0 \ee
Consequently there is a constant $C$ fixed by initial conditions such that:
\be\label{EOMw} \frac{\dot w}{w}  = \frac{C}{f_1^2} \ee
The equation of motion for $x(\lambda)$ reads:
\be \partial_x(f_1^2)\frac{\dot w^2}{w^2} + 4 \partial_x(\rho^2)(\dot x^2 + \dot y^2) = 8 \partial_\lambda(\rho^2 \dot x) \ee
and similarly for $y(\lambda)$:
\be\label{EOMy} \partial_y(f_1^2)\frac{\dot w^2}{w^2} + 4 \partial_y(\rho^2)(\dot x^2 + \dot y^2) = 8 \partial_\lambda(\rho^2 \dot y) \ee
These equations are difficult to solve in general.

\paragraph{On the boundary of the Riemann surface.}
In the following we limit ourselves to the study of geodesics living on the boundary of the Riemann surface $\Sigma$.
We are interested in geodesics going from the asymptotic region up to the horizon.
They extend from the AdS boundary at $x=1$ up to $x=\omega_1$ where they reach the horizon at $w=\infty$ (see Figure \ref{choicePath}, \emph{bottom}).
It is possible to find the analytic expression of these geodesics.
Looking at the behavior of the harmonic function $h_1$ and $h_2$ in the neighborhood of the boundary of $\Sigma$, we find that the functions $f_1$ and $\rho$ that appear in the metric behave like:
\be f_1^2(x,y) = f_1^2(x) + \mathcal{O}(y^2) \ee
\be \rho^2(x,y) = \rho^2(x) + \mathcal{O}(y^2) \ee
This implies that $\partial_y(f_1^2)|_{y=0}=0=\partial_y(\rho^2)|_{y=0}$. So setting $y=\dot y = 0$ solves the equation of motion for $y$ \eqref{EOMy}. This proves the existence of a geodesic living uniquely on the boundary of $\Sigma$.
Next instead of solving the equation of motion for $x$, we can use the affine parametrization condition:
\be 1 = 4 \rho^2 \dot x^2 + f_1^2 \frac{\dot w^2}{w^2} \ee
From which we deduce using equation \eqref{EOMw}:
\be\label{defV} \dot x^2 = \frac{1}{4 \rho^2} \left(1-\frac{C^2}{f_1^2} \right)  \equiv V(x) \ee
We can deduce $w$ as a function of $x$:
\be\label{w(x)}  w(x) =  w_0 \exp \left(\int_{x_0}^x dx' \frac{C}{f_1^2(x')} \frac{1}{\sqrt{V(x')}} \right)\ee
For the geodesic to reach the horizon, we need $w$ to go to infinity at $x=\omega_1$. So the constant $C$ introduced in \eqref{EOMw} has to satisfy:
\be\label{valueC} C = f_1(\omega_1) \ee
Indeed the functions $f_1$ and $\rho$ are everywhere regular and non-vanishing. Moreover $\p_x f_1(\omega_1) = 0$. Consequently the choice \eqref{valueC} implies that $V(x) \propto (x-\omega_1)^2$ for $x$ close to $\omega_1$. We deduce from equation \eqref{w(x)} that $w(\omega_1)=\infty$.

\bibliography{openStringsBackreactingD3}
\bibliographystyle{JHEP}

\end{document}